\ifdefined\pdfminorversion
\fi
\documentclass[aps,prl,twocolumn,superscriptaddress,floatfix]{revtex4-2}

\usepackage{amsfonts}
\usepackage{amsmath}
\usepackage{graphicx}
\usepackage[percent]{overpic}
\usepackage{dsfont}
\usepackage{array}
\usepackage{amssymb}
\usepackage{bbm}
\usepackage{bm}
\usepackage{booktabs}
\usepackage{tikz}
\usetikzlibrary{arrows.meta,decorations.markings}
\usepackage{xcolor}
\definecolor{darkblue}{HTML}{004D6B}
\definecolor{brightblue}{HTML}{0072BD}
\definecolor{darkred}{HTML}{8c1515}
\definecolor{darkgreen}{HTML}{006400}
\usepackage{hyperref}
\hypersetup{
    pdftitle={Measurement-induced criticality for non-Abelian anyons: The monitored golden chain},
	colorlinks=true,
	urlcolor=darkred,
	citecolor=darkblue,
	linkcolor=darkred,
	breaklinks
}
\usepackage{physics}
\usepackage{times}
\usepackage{soul}
\setstcolor{red}

\usepackage{cleveref}
\crefname{figure}{Fig.}{Figs.}
\Crefname{figure}{Figure}{Figures}
\crefname{section}{Sec.}{Secs.}
\Crefname{section}{Section}{Sections}
\crefname{section*}{Sec.}{Secs.}
\Crefname{section*}{Section}{Sections}
\crefname{equation}{Eq.}{Eqs.}
\Crefname{equation}{Equation}{Equations}
\crefname{table}{Table}{Tables}
\Crefname{table}{Table}{Tables}

\soulregister{\cite}{7}
\soulregister{\ref}{7}
\soulregister{\pageref}{7}
\soulregister{\cref}{7}
\soulregister{\Cref}{7}

\begin{document}

\title{Measurement-induced criticality for non-Abelian anyons: The monitored golden chain}

 \author{Zhaohui Zhi}
 \affiliation{The Hong Kong University of Science and Technology (Guangzhou), Nansha, Guangzhou, 511453, Guangdong, China}
 \author{Bo Han}
 \affiliation{Institute for Theoretical Physics, University of Cologne, Z\"ulpicher Straße 77, 50937 Cologne, Germany}
 \author{Romain Vasseur}
 \affiliation{Department of Theoretical Physics, University of Geneva, 24 quai Ernest-Ansermet, 1211 Gen\`eve, Switzerland}
 \affiliation{Geneva Quantum Center, University of Geneva}
 \author{Andreas W. W. Ludwig}
 \affiliation{Department of Physics, University of California, Santa Barbara, CA 93106, USA}
 \author{Simon Trebst}
 \affiliation{Institute for Theoretical Physics, University of Cologne, Z\"ulpicher Straße 77, 50937 Cologne, Germany}
 \author{Guo-Yi Zhu}
 \email{guoyizhu@hkust-gz.edu.cn}
 \affiliation{The Hong Kong University of Science and Technology (Guangzhou), Nansha, Guangzhou, 511453, Guangdong, China}

\date{\today}

\begin{abstract}

Non-Abelian anyon chains whose interactions favor a single fusion channel host quantum critical ground states described by minimal-model conformal field theories (CFTs) protected by a topological symmetry. We show that their {\sl measurement-only} counterpart, where non-commuting fusion charge measurements act locally on the anyons of a chain of $L$ sites, provides a fundamental example of a non-unitary CFT controlling a non-Abelian learning problem. Mapping the monitored dynamics to a
random tensor network, we numerically find $1/L$ scaling Lyapunov gaps, logarithmic entanglement scaling with the chord length, and an effective central charge in the Shannon entropy of the measurement record. The conserved topological charge sharpens with dynamical exponent $z= 1$ and is learnable by a decoder on an $O(L)$ time scale. As the topological symmetry is a non-invertible symmetry of the $SU(2)_k$ anyon theories, we establish the monitored golden chain $(k=3)$ as a non-Abelian generalization of the weakly (Kramers-Wannier) self-dual quantum critical point of the toric code $(k=2)$, and the first of a family for $k\geq 3$.
\end{abstract}

\maketitle


In quantum topology, Fibonacci anyons are one of the simplest types of non-Abelian anyons~\cite{FreedmanLarsenWang02,ReadRezayi99,Nayak08RMP,Trebst08introfibonacci}: they carry topological charge $\tau$ and obey a fusion rule $\tau\times\tau=\mathbf{1}+\tau$, which yields their characteristic   
quantum dimension $\phi= (1+\sqrt5)/2$  
matching the golden ratio. Despite their simplicity, their braiding is sufficient for universal topological quantum computation~\cite{FreedmanLarsenWang02, Simon05fibocompute}, and can in principle be realized in fractional quantum Hall systems~\cite{ReadRezayi99,Mong17fibonacci125}.  
Previously, some of us introduced the {\sl golden chain}~\cite{Trebst07goldenchain}, a one-dimensional model of {\sl interacting} Fibonacci anyons,
which for every pair of neighboring anyons {\sl energetically} prefers fusion into the trivial channel (akin to an antiferromagnetic spin-spin interaction~\cite{Trebst08introfibonacci}). This model is exactly solvable~\cite{Trebst07goldenchain} ---
the interacting anyon chain is conformally critical and, for `antiferromagnetic' couplings,  
described by the tricritical Ising conformal field theory (CFT), with central charge $c=7/10$. 
Remarkably, this critical state of the anyon chain does not require any fine-tuning as a multicritical point would in, say, a one-dimensional (1D) quantum spin chain or two-dimensional (2D) classical statistical mechanics model, but is instead protected by a {\sl topological symmetry}~\cite{Trebst07goldenchain}, giving the critical state a natural meaning in terms of a topological edge 
mode~\cite{Gils2009,Gils09topology}. 
This topological symmetry is an early example of a {\sl generalized non-invertible symmetry}~\cite{Frohlich2004KWdefects,Trebst07goldenchain,Gaiotto15generalized,
Buican_2017_anyonichains,Chang19defectlines,ThorngrenWang2019fusion,Ji20categorical,
Kong20categorical,aasen2020topologicaldefectslatticedualities,Chatterjee2025emergent,
SeibergShao2024Majorana,Shao2023TASI,Seiberg2024LSM,Okada2024quantumoperations,
Cao2024generating,chen2026SSBnonInvertible}. 
Going beyond the familiar setting of Kramers--Wannier (KW) duality,
this topological symmetry obeys a fusion algebra rather than the group structure of an ordinary symmetry in the Ginzburg--Landau--Wilson paradigm; consequently, it has no inverse as a symmetry operation.

\begin{figure*}[tb!]
    \centering
    \includegraphics[width=\textwidth]{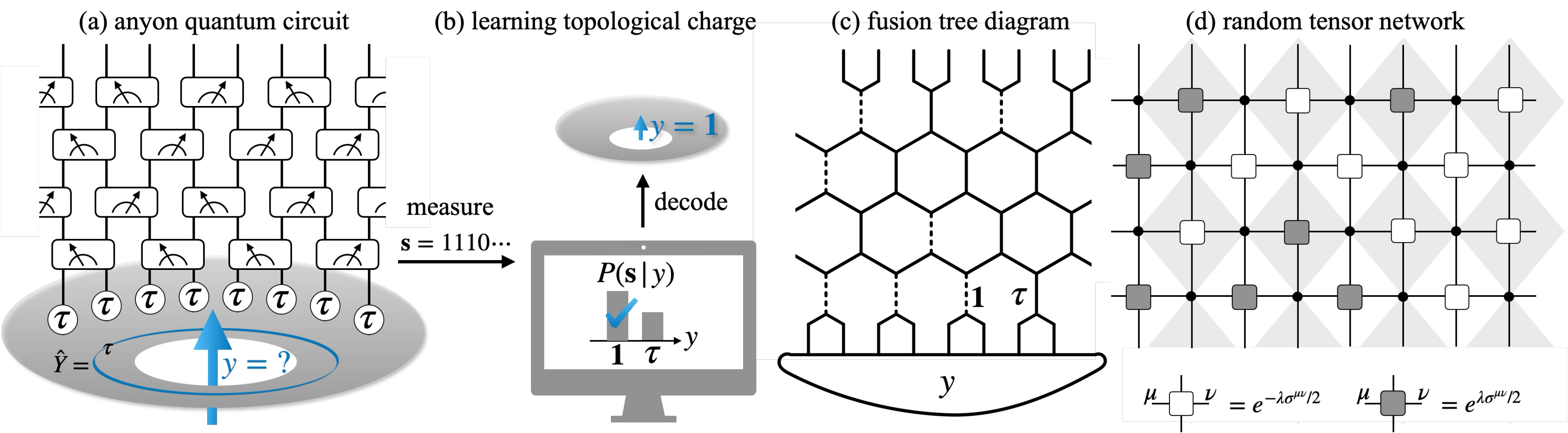}
    \caption{\textbf{Schematic of the model: quantum circuit and classical simulation.}
    (a) A ring of Fibonacci anyons is created from a parent 2D topological quantum liquid on a torus (gray shading). 
    The topological symmetry operator $\hat{Y}$ threads an auxiliary $\tau$ line encircling the hole. Its eigenvalue diagnoses the global topological charge $y$ at the center of the ring, or equivalently, the anyonic flux through the hole of the torus.
    A brickwork circuit with periodic boundary conditions repeatedly measures the local total charge, or fusion channel, of neighboring anyons, with a tunable measurement strength $\gamma\in[0,1]$ interpolating between weak and projective measurement.
    (b) One can input the measurement record into a classical decoder with which one can infer the most likely topological charge. The simulation of the decoder is performed in basis shown in (c)(d). 
    (c) The collective state of the anyon chain following the Born-rule trajectory in (a) is represented by a fusion-tree diagram, where solid and dashed lines denote $\tau$ and $\mathbf{1}$, respectively. The diagram can be viewed as a spacetime string-net configuration.
    (d) Constrained random tensor network representation of the fusion tree in (c), where the building blocks are Kraus operators $M_s$ explained in the main text. 
    Here each vertical leg labels a corresponding spine index of the sequential fusion tree and  $\mu, \nu \in \{{\bf 1}, \tau\}$. 
    This tensor network is a random generalization of the RSOS model: each black dot represents a
    binary variable.
     }
    \label{fig:fusion_protocol}
\end{figure*}

In its Hamiltonian description, the golden chain describes the formation of a critical state in a closed system, where one can think of the imaginary-time 
path integral dynamics of competing projectors (onto the trivial fusion channel) as a purification mechanism that leads, in the long-time
limit, to a unique (critical) ground state.
This closed-system setting can be contrasted with an open-system setting where {\sl measurements} provide a qualitatively different route to the formation
of a late-time many-body quantum state --
each measurement outcome selects a quantum trajectory with its {\sl Born probability}. 
Such monitoring can suppress volume-law entanglement in random circuits~\cite{Skinner2019,Altman2020weak,Jian2020,Huse2020qec,Vedika2021measure} and efficiently prepare gapped topological states~\cite{Bravyi2022,Google22surfacecode,Roushan21sycamoretoriccode,Iqbal23iontoriccode}. 
But whether local measurements can also stabilize criticality in a non-Abelian anyon system, and what role its topological symmetry plays, are open questions
that have motivated us to study a {\sl monitored golden chain} model in this Letter.
Specifically, we study a brickwork circuit of Born measurements of the total topological charge of neighboring Fibonacci anyons, which we refer to as fusion measurements, as illustrated in \cref{fig:fusion_protocol}(a).
Such measurements are the elementary operations of measurement-only topological quantum computation~\cite{Nayak2008measurement}, in which sequences of fusion measurements replace braiding. Importantly, every Kraus operator preserves the topological symmetry, making the measurement protocol a natural candidate
for a {\sl measurement-only} generalization of the golden chain Hamiltonian model~\cite{Trebst07goldenchain}. 

While the presence of topological symmetry conserves the topological charge in the quantum system, one might ask how quickly  a classical observer can infer this {\it global} topological charge of a non-Abelian anyon chain from a record of {\it local} non-commuting fusion measurements [\cref{fig:fusion_protocol}(b)]?
For a chain of $L$ Fibonacci anyons we find that the charge sharpens with dynamical exponent $z=1$
and becomes learnable by a decoder on an $O(L)$ time scale. By mapping the monitored dynamics to a random tensor network, we numerically demonstrate that this learning problem is controlled by a {\sl non-unitary} CFT, whose effective (Casimir) central charge and scaling dimensions we calculate.
One might expect that the {\sl measurement-only} critical theory is
protected by the same topological symmetry that renders the ground states of interacting anyon chains quantum critical.
Indeed, we find that monitoring renders the golden chain always critical, stable and independent of the strength of measurement.
Recalling
the topological symmetry as a non-invertible symmetry of $SU(2)_k$ anyon theories, we establish the monitored golden chain $(k=3)$ as 
the first of the non-Abelian family with $k\geq 3$.


{\it Model}.---We study a system of $L$ Fibonacci anyons whose Hilbert space dimension grows asymptotically as $\phi^L$ stipulated by the fusion rule $\tau \times \tau = \mathbf{1} + \tau$.
The `antiferromagnetic' golden-chain Hamiltonian is $H=-\sum_j\Pi^{\mathbf{1}}_{j,j+1}$, where $\Pi^s_{j,j+1}$ {\sl projects} neighboring anyons at sites $j$ and $j+1$ onto fusion channel $s\in\{\mathbf{1},\tau\}$ which are the anyon chain analogs of the spin-singlet or triplet projectors in the conventional Heisenberg spin-$1/2$ chain~\cite{Trebst08introfibonacci}.
When multiplied by the golden ratio $\phi$, the resulting operators
$\phi \Pi^{\mathbf{1}}_{j,j+1}$ satisfy the Temperley--Lieb algebra~\cite{TemperleyLiebPRC1971} describing minimal model conformal field theories
and statistical mechanics models, related to the Jones polynomial~\cite{VJonesCRASP1984,VJonesAnnMath1987}
, see Refs.~\cite{VJonesCRASP1984,WenzlInventMath1988,KunibaAkutsuWadatiJPSJ1986,PasquierNPB1987, FendleyJPhysA2006}.%

Here we propose to replace each Hamiltonian interaction by a {\sl weak measurement} of the total charge of the corresponding neighboring pair, equivalently its local fusion channel, with Kraus operators
\begin{equation}
M_{\mathbf{1}} = \frac{e^{\lambda \Pi^{\mathbf{1}}}}{\sqrt{1+e^{2\lambda}}} \ ,\quad \quad
M_\tau = \frac{e^{\lambda \Pi^{\tau}}}{\sqrt{1+e^{2\lambda}}}\ ,
\label{LabelEqDefKrausOperators}
\end{equation}
where $M_{\mathbf{1}}$ ($M_\tau$) favors fusion to $\mathbf{1}$ ($\tau$). They form a two-outcome positive operator-valued measure (POVM), satisfying the completeness relation $\sum_{s}M_s^\dagger M_s=\mathbbm{1}$, where a {\sl measurement strength} $\gamma=\tanh\lambda\in[0,1]$ interpolates between no measurement ($\gamma=0$, $M_s\propto \mathbbm{1}$) and a projective measurement of the fusion channel ($\gamma=1$, $M_s=\Pi^s$). In the projective limit $\gamma=1$, the local measurement collapses the pair into a definite fusion channel $s=\mathbf{1}$ or $\tau$~\footnote{
Note that the paired anyons whose total charge is measured into $\mathbf{1}$ do not annihilate as in Ref.~\cite{Zou21fibonacci}, but instead remain spatially separated as in measurement-only topological quantum computation~\cite{Nayak2008measurement}}.
Repeatedly measuring the odd and even bonds defines discrete-time monitoring dynamics on the brickwork spacetime lattice in \cref{fig:fusion_protocol}(a), which is labeled as a (random) transfer matrix $\mathcal{T}(\mathbf{s})=\prod_{j,t}M_{s_{j,t}}$ where the product follows the time ordering.
Starting from a given initial state $\ket{\psi_0}$, the unnormalized quantum state following a trajectory $\mathbf{s}$ (the set of binary measurement outcomes) is $\mathcal{T}(\mathbf{s})\ket{\psi_0}$
whose probability is given by the Born rule
$
    P(\mathbf{s})
    =\lVert\mathcal{T}(\mathbf{s})\ket{\psi_0}\rVert^2 \, .
$

\begin{figure*}[t!]
    \centering
    \begin{overpic}[width=0.32\textwidth]{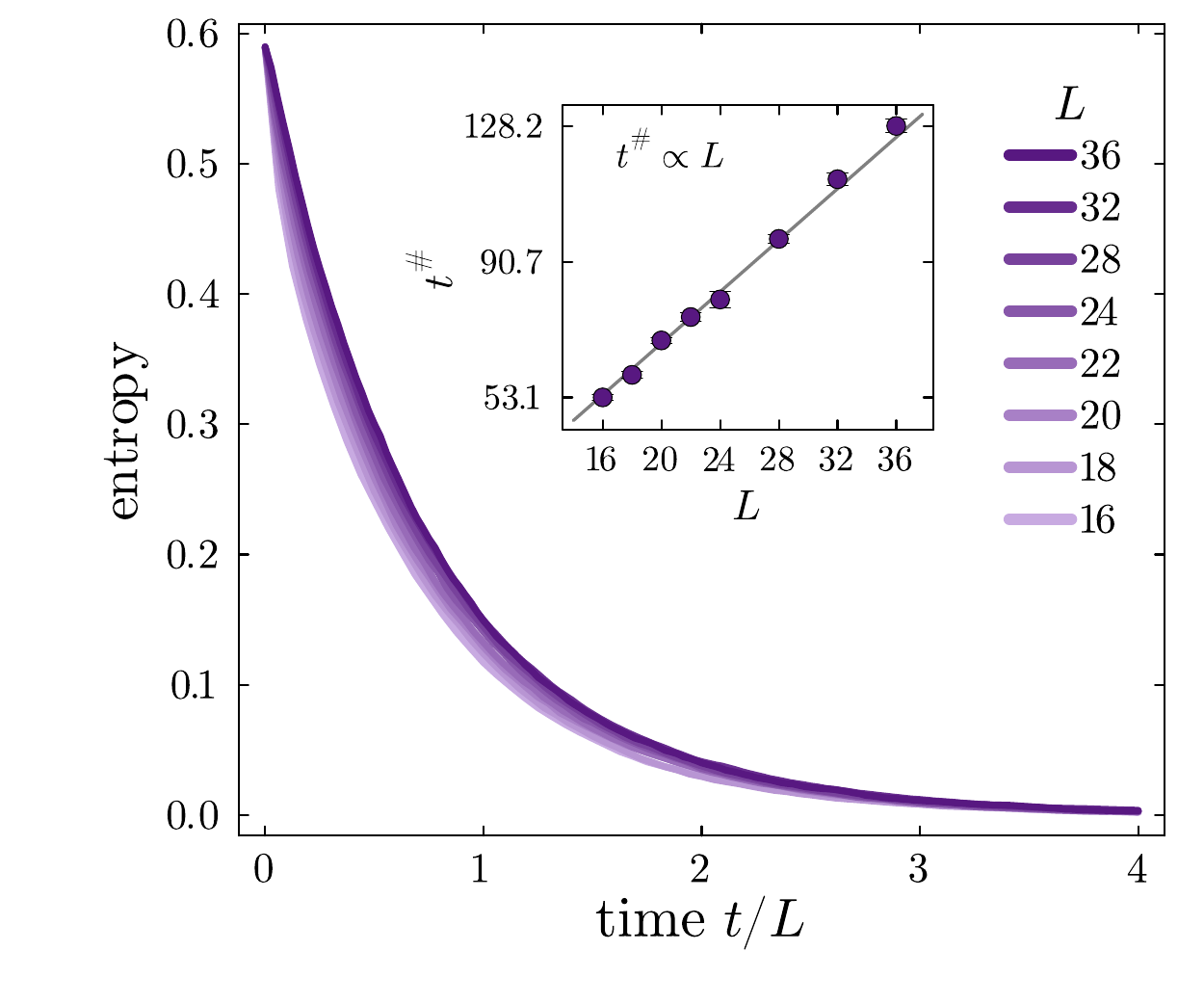}
        \put(-2,81){(a)}
    \end{overpic}
    \hfill
    \begin{overpic}[width=0.32\textwidth]{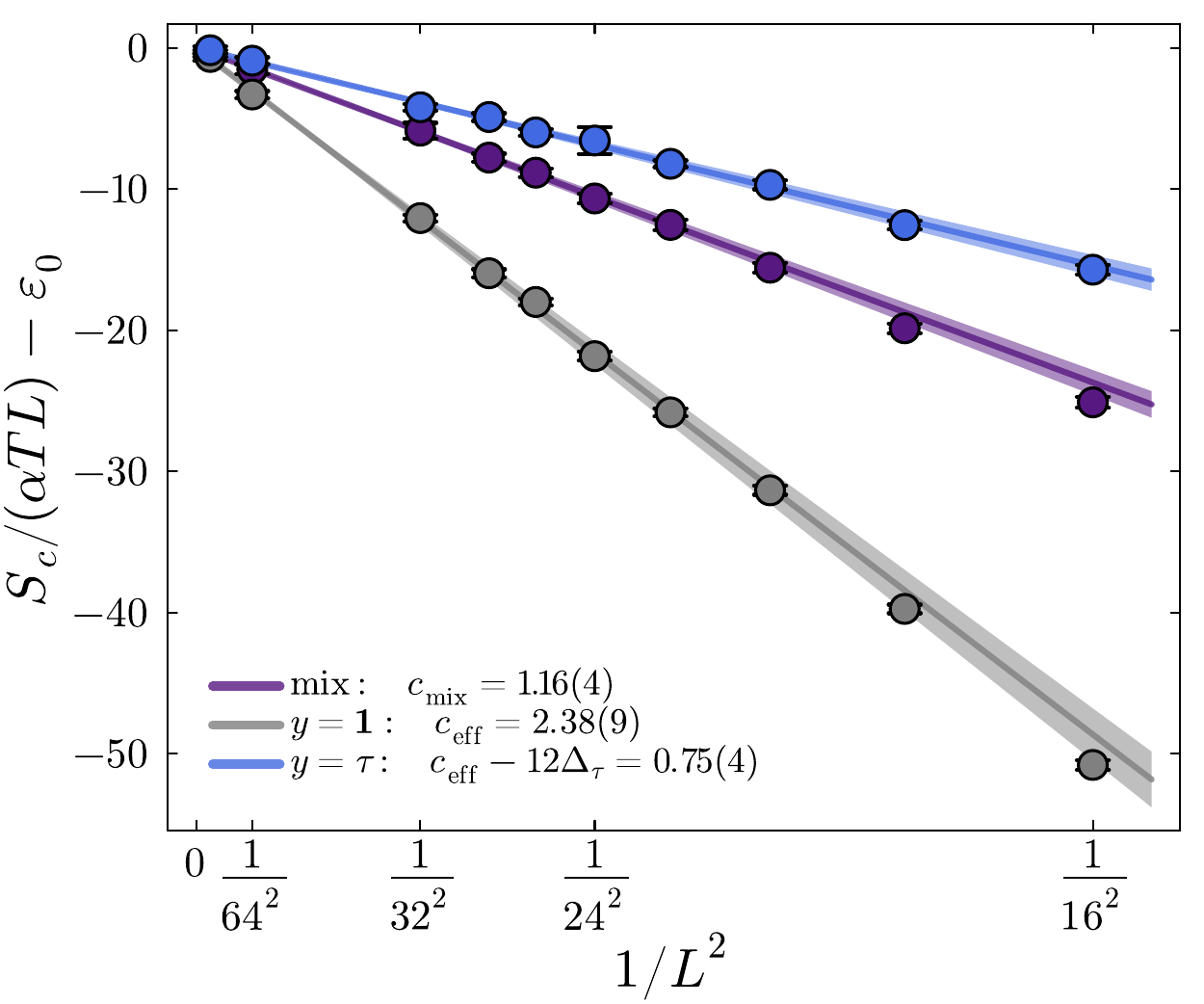}
        \put(-2,81){(b)}
        \put(7,82){\scalebox{0.72}{$\times 10^{-4}$}}
    \end{overpic}
    \hfill
    \begin{overpic}[width=0.32\textwidth]{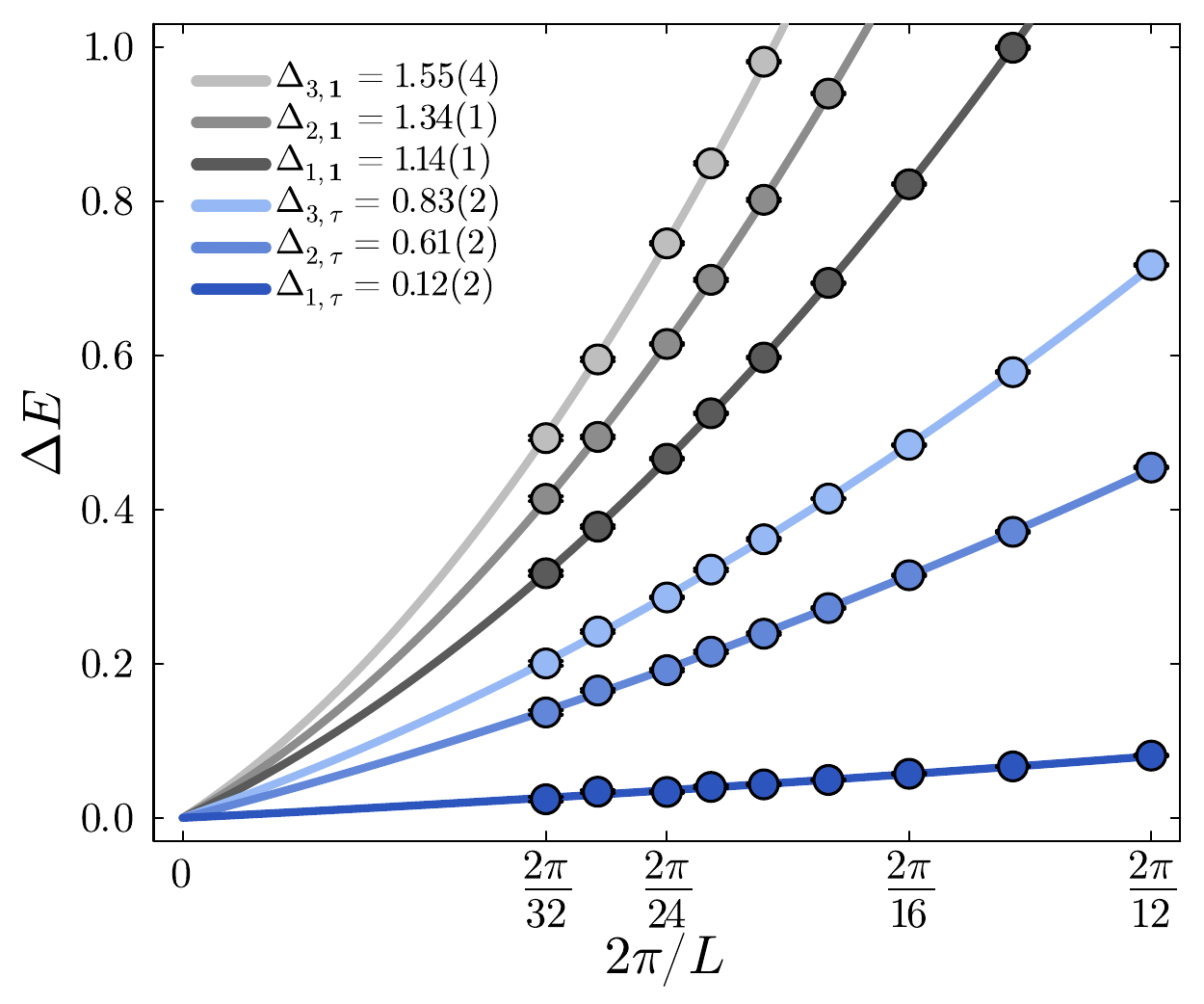}
        \put(-2,81){(c)}
    \end{overpic}
    \caption{
        (a)
        {\bf Purification dynamics from an initial incoherent mixture.}
        The entropy is averaged over Born-rule trajectories at measurement strength $\gamma=0.95$, and is found to decay exponentially with $t/L$.
        The inset extracts the sharpening time $t^\#$ at which $S_R(t^\#)=0.01S_R(0)$,
        which grows nearly linearly with  system size $L$: a power-law fit gives $t^\#\propto L^z$ with $z=1.09(2)$, while the gray line in the inset shows $z=1$ as a guide to the eyes.
        (b) {\bf Topological-charge-resolved Shannon entropy scaling}.
        The purple data show the mix sector initiated from the fusion-tree basis state $\ket{\psi_0}$, 
        while gray data shows the $y=\mathbf{1}$ sector obtained by setting the initial state to be antiferromagnetic Hamiltonian ground state 
        and blue data shows the $y=\tau$ sector by setting ferromagnetic Hamiltonian ground state (only for $L\notin 6\mathbb Z_{>0}$) as initial state. 
        For $L\in 6\mathbb Z_{>0}$ data, we use projected $\ket{\psi_0}$ to $y=\tau$ sector as initial state.
        The slopes of the curves versus $\pi / (6L^2)$ are fit (for $L\geq22$) giving $\Delta_\tau=0.136(8)$ and $c_{\rm eff}=2.38(9)$.
        The purple line is equivalent to a golden-ratio-weighted average of the blue and the gray line. 
        (c) {\bf  Scaling of topological-sector-resolved Lyapunov spectrum}.
        Rescaled low-lying Lyapunov gaps in each topological sector are measured relative to the true ground-state level: $E_{n,y}-E_{0,\mathbf{1}}$ for $n=1,2,3$ in $y=\mathbf{1}$ (gray) and $y=\tau$ (blue) sectors.
        Each curve is fitted to $\Delta E_{n,y}=\Delta_{n,y}x+b_{n,y}x^2$, with $x=2\pi/L$, using $L=16,18,20,22,24,28,32$.
        The $1/L$ term dominates in the long-wavelength limit. 
     }
    \label{fig:conformal}
\end{figure*}

We describe a state of the anyon chain  in its fusion-tree basis~\cite{Trebst07goldenchain} as $\ket{\boldsymbol{x}}=\ket{x_0,\ldots,x_{L-1}}$ 
corresponding to
\begin{equation}
\begin{tikzpicture}[
    baseline=-0.4ex,
    x=1.0cm,
    y=0.72cm,
    spine/.style={line width=0.9pt},%
    anyon/.style={line width=0.8pt},
    vertex/.style={circle,fill=black,inner sep=1.1pt}
]
    \node at (-0.8,0) {$\cdots$};
    \draw[spine] (-0.5,0) -- (4.5,0);
    \node at (4.85,0) {$\cdots$};
    \foreach \j in {0,...,4}{
        \draw[anyon] (\j,0) -- (\j,0.62);
        \node[vertex] at (\j,0) {};
        \node[above] at (\j,0.62) {\small$\tau$};
    }
    \node[below] at (0.5,0) {\small$x_{j-2}$};
    \node[below] at (1.5,0) {\small$x_{j-1}$};
    \node[below] at (2.5,0) {\small$x_j$};
    \node[below] at (3.5,0) {\small$x_{j+1}$};
\end{tikzpicture} \,,
\label{eq:fusion_path}
\end{equation} 
where the vertical legs represent physical Fibonacci anyons and the horizontal spine index $x_i\in\{\mathbf{1},\tau\}$. Under periodic boundary conditions~\cite{Trebst07goldenchain,Pfeifer12fibo}
we force $x_L=x_0$ which closes the spine forming a loop as shown in \cref{fig:fusion_protocol}(c). 
The label set $\{x_i\}$ describes an effective spin chain subject to the Fibonacci fusion rule constraint, enabling classical simulation. 
We next map the measurement gates to this basis in the following. 


\emph{Tensor network and partition function}.--
A Born-sampled trajectory is shown as a fusion-tree diagram in \cref{fig:fusion_protocol}(c), where solid and dashed vertical legs denote outcomes $\tau$ and $\mathbf{1}$, respectively. 
This fusion tree
can be mapped to a random 2D tensor network.
To see this one can use the $F$-move to map the neighboring anyon fusion operator to a gate acting on the spine index $x_j$ but conditioned upon $x_{j-1}$ and $x_{j+1}$. Thus one layer of measurements is a matrix product operator (MPO), composed by each local rank-4 tensor as shown in \cref{fig:fusion_protocol}(d). The detailed derivation and its tensor elements are technical details left to the Supplemental Material (SM)~\cite{SM}\nocite{Friedan1984,Belavin1984,BianchiniErcolessiPearceRavanini15,KimPearce87anisotropy}.
Assembling all such measurement gates yields a circuit as shown in \cref{fig:fusion_protocol}(d), where each vertical line represents the worldline of a spine index. As a 2D tensor network with randomness {\sl conditioned} upon the measurement outcomes, it can be interpreted as a partition function of a statistical model for binary
spin variables on the black dots,
where each measurement generates an interaction between the four neighboring dots (indicated by the diamond shades). 
For a uniformly (in space and time) post-selected trajectory $\mathbf{s}=\mathbf{1}$, this is exactly the restricted-solid-on-solid (RSOS) model, which leads to the tricritical Ising CFT of the Hamiltonian ground state~\cite{Trebst07goldenchain}.
Without any post-selection, Born sampling instead makes the local weights {\sl stochastic} and {\sl correlated}, 
as the probabilities reflect any existing correlations of the underlying many-body quantum state (prior to measurement).
We find that, after long time evolution, the average probability of the local fusion charge projective measurement settles to universal values $P(\tau)=1/\phi$ and $P(\mathbf{1})=1/\phi^2$ regardless of the initial state, see the End Matter (EM) for more details. However, the whole trajectory actually encodes the information of the topological symmetry charge as detailed below. 


{\it Topological symmetry: charge sharpening and learning}.--
Threading an auxiliary $\tau$ line around the ring defines the topological symmetry $\hat Y$~\cite{Trebst07goldenchain}, see \cref{fig:fusion_protocol}(a). Its matrix elements are $\bra{\boldsymbol{x}'}\hat Y\ket{\boldsymbol{x}}=\prod_{i=0}^{L-1}\bigl(F_{\,x'_{i+1}}^{\tau x_i \tau}\bigr)^{x_{i+1}}_{\,x'_i}$, with $x'_L=x'_0$ and $F$ the fusion matrix of the Fibonacci anyon theory, see the SM~\cite{SM} for its explicit derivation and the matrix product operator (MPO) representation. Every local two-anyon charge measurement commutes with $\hat Y$, which itself obeys the Fibonacci fusion algebra $\hat Y^2=\mathbbm{1}+\hat Y$. Thus, the local measurements do not change the global topological sector. Instead, $\hat Y$ acts on the sectors $y=\mathbf{1}$ and $y=\tau$ with eigenvalues $S_{y\tau}/S_{y\mathbf{1}}=\phi$ and $-\phi^{-1}$, respectively. These
eigenvalues, determined by the modular $S$-matrix $S_{y, y'}$, encode the mutual braiding statistics between the auxiliary $\tau$ line and the global anyonic flux.

For simplicity we first consider a translationally invariant fusion-tree basis state as the initial state
\begin{equation}
\ket{\psi_0}\equiv
\ket{\tau\tau\cdots\tau\tau}
=\sqrt{w_{\mathbf{1}}}\ket{\psi_{0,\mathbf{1}}}
+\sqrt{w_\tau}\ket{\psi_{0,\tau}}\,,
\label{eq:initial_fusion_path}
\end{equation}
where $\ket{\psi_{0,y}}$ is its normalized projection onto the topological charge sector $y$.
The amplitudes exponentially converge to the quantum dimensions of the two sectors $\sqrt{w_{y}}=d_y/D$, 
where $d_\mathbf{1}=1$, $d_\tau=\phi$, and $D=\sqrt{1+\phi^2}$ is the total quantum dimension.
Therefore $\bra{\psi_0}\hat{Y}\ket{\psi_0}\to 0$, which also holds for any other fusion-tree basis state, see SM~\cite{SM}.
Thus, in the thermodynamic limit, the sector populations approach $w_\mathbf{1}\to(d_\mathbf{1}/D)^2\simeq28\%$ and $w_\tau\to(d_\tau/D)^2\simeq72\%$.
The resulting spacetime configuration in \cref{fig:fusion_protocol}(c) resembles a Fibonacci string net~\cite{LevinWen05stringnetcondensation}, except that the horizontal zigzag worldlines remain fixed to $\tau$ (indicating the physical Fibonacci anyons).

Although the protocol contains no direct measurement of $\hat Y$, the accumulating record of local fusion measurements enables an observer to infer the global topological charge, thereby purifying the post-measurement state into a definite charge sector. This process is known as charge (or spin) sharpening in monitored systems with $U(1)$~\cite{Agrawal22chargesharpening,Vasseur22learnability,Ippoliti24learnability,Agrawal24measurementcollapse,Barratt22chargesharpening} or $SU(2)$ symmetries~\cite{Majidy23spinsharpening,Feng25chargespinsharpening,Ma26nonabelianlearnability}, while here our protocol infers the non-Abelian anyon charge.
Following a given quantum trajectory, the coherent superposition \cref{eq:initial_fusion_path}
evolves, and collapses into a definite topological charge sector after a sharpening time scale $t^\#$, as 
shown in \cref{fig:Y_sharpening} of the EM.
To determine the sharpening time
and how it scales with the system size, we modify the initial state to be entangled with a reference qubit $R$, $\sqrt{w_{\mathbf{1}}}\ket{\psi_{0,\mathbf{1}}}\ket{1}_R+\sqrt{w_\tau}\ket{\psi_{0,\tau}}\ket{0}_R$. Tracing out the reference qubit removes the inter-sector coherences of \cref{eq:initial_fusion_path} and leaves a mixed initial state. The decay of the reference qubit entropy then measures how rapidly the measurement record reveals the conserved topological charge [\cref{fig:conformal}(a)];  we find that the sharpening time for the non-Abelian
topological charge scales {\sl linearly} with the system size, $t^\# \propto L$, which contrasts with the $t^\#=O(1)$ of $U(1)$ Abelian charges~\cite{Agrawal22chargesharpening,Vasseur22learnability,Agrawal24measurementcollapse,Barratt22chargesharpening} and $t^\#\propto L^2$ for $SU(2)$ charges~\cite{Majidy23spinsharpening,Feng25chargespinsharpening,Ma26nonabelianlearnability} at their measurement-only limits. 

In order to numerically probe learnability explicitly, we 
design a classical maximum-likelihood decoder (see SM~\cite{SM}) \\
to process the monitored trajectory $\mathbf{s}$ of an {\sl unknown} quantum state, 
and infer its most likely topological charge by maximizing the conditional probability $P(\mathbf{s}|y)$ over $y$
, see \cref{fig:fusion_protocol}(b).
\Cref{fig:topological_charge_decoding_dynamics} in the End Matter shows the decoding error probability decreases exponentially with $t/L$ of the trajectory, i.e., the success probability is $1-O(e^{-t/L})$, which implies an $O(L)$ learning time scale consistent with $t^\#$.
The linear scaling of the sharpening and learning time scale implies {\sl spacetime-symmetric} critical dynamics (dynamical exponent $z=1$), whose underlying critical theory is expanded below. 


\emph{Critical scaling}.--
For the full record $\mathbf{s}$
we define, following the logic presented in 
Ref.~\cite{Pixley22MIPTCFT} for a generic monitored quantum circuit, the free energy $F(\mathbf{s})=-\ln P(\mathbf{s})$.
The Born average of the free energy is 
the measurement record's Shannon entropy, $S_c=-\sum_{\mathbf{s}}P(\mathbf{s})\ln P(\mathbf{s})$. For a 2D 
critical system on a long cylinder 
whose averaged observables obey conformal symmetry, the averaged
free energy density obeys conformal scaling with a universal $1/L$ Casimir correction,
\begin{equation}
\frac{S_c}{\alpha T}
\equiv E_0=\epsilon_0 L-\frac{\pi c_{\rm eff}}{6L}+\ldots \, ,
\label{eq:casimir}
\end{equation}
which defines the (universal) effective (Casimir) central charge $c_{\rm eff}$ of the critical system, while the prefactor $\epsilon_0$ of the extensive term is 
non-universal. The spacetime anisotropy factor (sound velocity) $\alpha$ is extracted by matching spatial and temporal correlation lengths~\cite{Pixley22MIPTCFT,Zabalo_2020,Gullans20scalabledecoder} (see the 
EM and SM~\cite{SM}). 
We use the notation $E_0$ for the negative 
of the largest Lyapunov exponent in analogy with the ground-state energy of a non-random $(1{+}1)$D quantum chain; 
the analogous excited energy levels 
$\{E_{n,y}=-\lambda_{n,y}/\alpha\}$, with $y\in\{\mathbf{1},\tau\}$ denoting the topological charge sector,
can 
be extracted from the Lyapunov spectrum $\{\lambda_{n,y}\}$ of the (random) circuit transfer matrix $\mathcal{T}(\mathbf{s})^{\otimes 2}$.
The level index starts from $n=0$ for the unique ground state, which lies in the vacuum sector $y=\mathbf{1}$; consequently, the levels of the $y=\tau$ sector are labeled by $n=1,2,\ldots\,$.
They encode a set of {\it typical} scaling dimensions $\Delta_{n,y}$ of the underlying non-unitary
CFT~\cite{Ludwig90hierarchies,Pixley22MIPTCFT,Ludwig24stat,Ludwig24tricriticalising}
as
\begin{equation}
     E_{n,y}=\frac{2\pi}{L}\left(\Delta_{n,y}-\frac{c_{\rm eff}}{12}\right)+\ldots \ ,
     \label{LabelEquationCFTEnergies}
\end{equation}
where subleading non-universal $1/L^2$ terms are not displayed. We caution that the scaling is sensitive to the topological sector;
while a general imaginary time evolution with a {\it constant} Hermitian Hamiltonian ``cools'' the evolved state to the unique ground state of that Hamiltonian in the `vacuum sector', we find that here the Born ensemble sampled by the monitored circuit dynamics turns out to
include both
topological sectors, depending on the initial state. This is a direct consequence of the charge sharpening discussed above, where each trajectory collapses into a definite $\hat Y$ sector according to the measurement outcomes $\mathbf{s}$, and the charge is conserved upon averaging all the trajectories.
In our numerical results for the critical scaling shown in Figs.~\ref{fig:conformal}(b),(c), we employ a color coding to represent the topological sectors:
gray color for topological charge $y=\mathbf{1}$, blue color for the $y=\tau$ sector. 
In (b), the purple color indicates a mixture of both sectors with probabilities 
according to their quantum dimensions $d_y^2/D^2$, a combination which naturally appears in any dynamics initiated from an arbitrary fusion-tree basis state with periodic boundary conditions, such as \cref{eq:initial_fusion_path}. Here the lowest average energy is a weighted average of the $y=\mathbf{1}$ and $y=\tau$ sectors: $E_{0,\rm mix}=w_\mathbf{1}E_{0,\mathbf{1}} + w_{\tau} E_{1,\tau}$, where $\Delta_\tau\equiv\Delta_{1,\tau}$ is the scaling dimension of the lowest-lying level in the $y=\tau$ sector, defined through its gap $E_{1,\tau}-E_{0,\mathbf{1}}=\frac{2\pi}{L}\Delta_\tau$ above the vacuum-sector ground state, since $\Delta_{0, \mathbf{1}}=0$ by definition. Correspondingly, the Casimir form $E_{0,\rm mix}=\epsilon_0L-\frac{\pi c_{\rm mix}}{6L}+\ldots$ defines the mixed-sector $c_{\rm mix}=c_{\rm eff}-12w_\tau\Delta_\tau$.
The dominant linear scaling terms of the energy gaps with $1/L$ in (c) are a reflection of emergent conformal
symmetry in the long-wavelength limit. 
Each individual trajectory lacks of course spacetime translational symmetry, but because the Lyapunov exponents are self-averaging, they characterize (typical) observables
of the underlying trajectory-averaged
theory~\cite{Ludwig90hierarchies,Pixley22MIPTCFT,Ludwig24stat,Ludwig24tricriticalising}.
While the latter exhibits conformal invariance, note that the ``low-energy'' states here do not 
correspond to relevant operators of the CFT  that can be added as perturbations to the Hamiltonian or action 
(as it would be in a unitary CFT setting).

\begin{figure}[tb!]
    \includegraphics[width=\columnwidth]{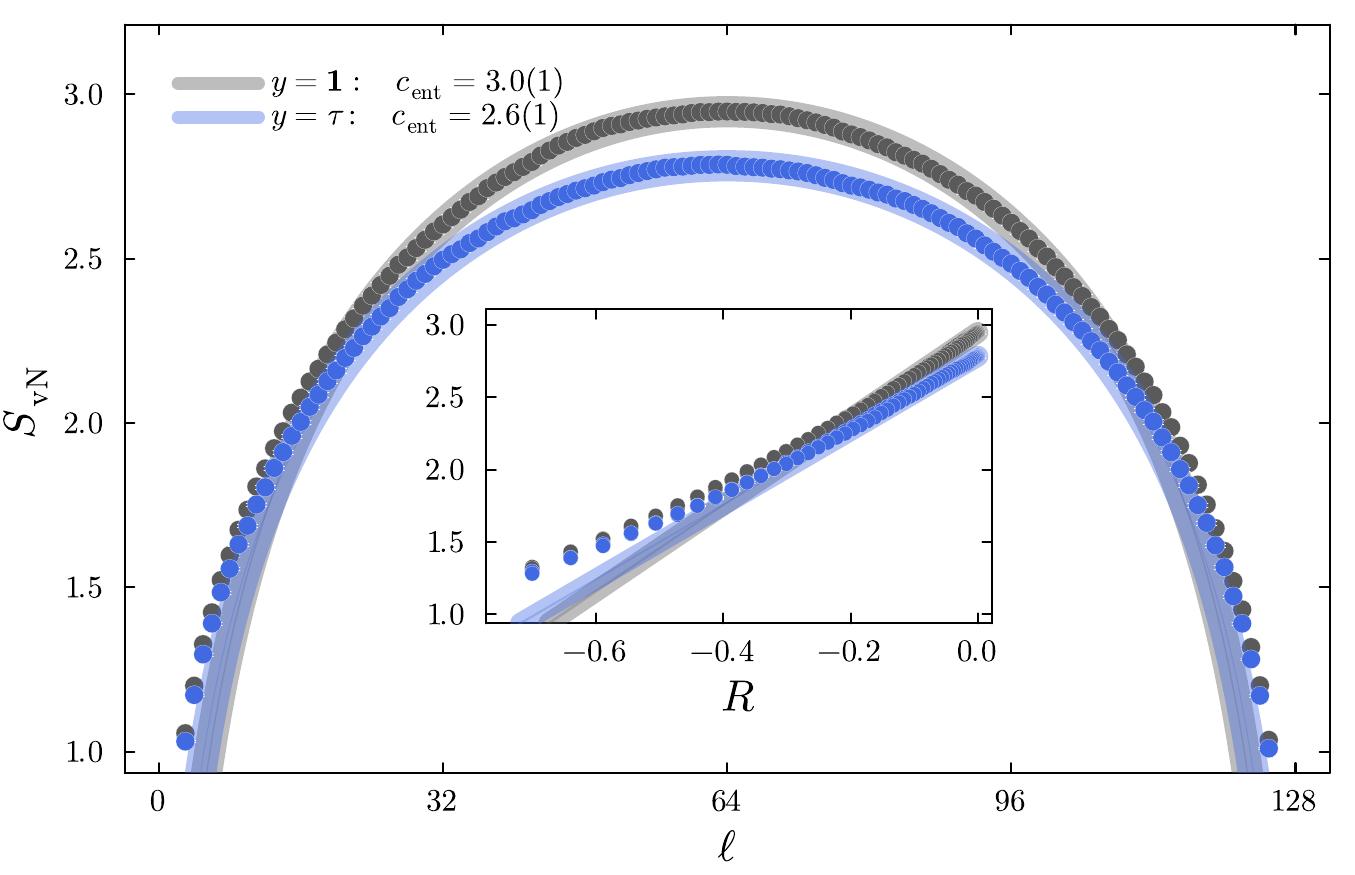}
    \caption{
    {\bf Critical entanglement entropy scaling} of the late-time steady state for the monitored golden chain at measurement strength $\gamma=0.95$.
    Data are shown for a chain of fixed size $L=128$ and varying bipartite subsystem size $\ell$ in the $y=\mathbf{1}$ sector (gray) and the $y=\tau$ sector (blue).
    Each curve is fitted independently over the bulk window $L/4\le \ell \le3L/4$ ($32\le \ell \le96$), giving $c_{\rm ent}=3.0(1)$ and $2.6(1)$ for the $y=\mathbf{1}$ and $y=\tau$ sectors, respectively.
    Inset: $S_{\rm vN}$ versus the rescaled logarithmic chord distance $R=\frac{1}{3}\ln[\sin(\pi\ell/L)]$.
    The MPS simulations use bond dimension $\chi=400$ and evolve up to $512$ layers ($T=256$) of fusion measurements; the $y=\mathbf{1}$ and $y=\tau$ data are averaged over $2000$ Born-random trajectories, respectively.
    }
    \label{fig:monitoredee}
\end{figure}

The quantum state emerging at late time is found to be critically entangled, whose Born-averaged bipartite von Neumann entropy obeys the scaling in \cref{fig:monitoredee}
\begin{equation}
    \overline{S_{\rm vN}}=\sum_\mathbf{s} P(\mathbf{s})\,
    S_{\rm vN}\!\left[\rho_A(\mathbf{s})\right] =
    c_{\rm ent} R + \ldots \, ,
\label{eq:cent}
\end{equation}
where $R=\frac{1}{3}\ln[\sin(\pi\ell/L)]$ for a subsystem of $\ell$ consecutive anyons; at the fixed $L$ used in \cref{fig:monitoredee}, the factor $L/\pi$ in the chord length is absorbed into the nonuniversal intercept.
The prefactor $c_{\rm ent}$ reveals the scaling dimension of a boundary-condition-changing (bcc) operator~\cite{Vasseur2019,Jian2020} of the underlying non-unitary CFT at the {\sl temporal} boundary.
Its value is clearly distinct from the effective central charge, $c_{\rm eff}\neq c_{\rm ent}$, underlining the non-unitary character of the CFT which arises from the lack of translational invariance (in space and time) in each quantum trajectory. As is well known~\cite{Vasseur2019,Jian2020,Pixley22MIPTCFT}, for this reason the two universal numbers $c_{\rm eff}$ and $c_{\rm ent}$ are not constrained to coincide, unlike in non-random, thus translationally invariant unitary CFTs. 


{\it Discussion}.--
The Fibonacci-anyon chain is part of a larger family of $SU(2)_k$ (anyonic) quantum spin chains~\cite{Trebst07goldenchain,Gils2009} 
where it corresponds to the quantization level $k=3$.
Its sister model at $k=2$ is a chain of {\sl Ising} anyons $\sigma$ where one
repeatedly weakly measures their fusion channel $\sigma\times\sigma=\mathbf{1}+\psi$, 
resulting in two possible outcomes, either the vacuum or a fermion. 
But while both the Ising and Fibonacci anyon theories are  non-Abelian theories of the  $SU(2)_k$ family, 
their monitored chains exhibit {\sl qualitatively different} outcome dependence, which is rooted in the fact that
their fusion channels carry different topological information.
Specifically, for Ising anyons, both possible measurement outcomes $\mathbf{1}$ and $\psi$ are {\sl Abelian}. 
By contrast, for Fibonacci anyons, the `singlet' outcome $\mathbf{1}$ is Abelian, while the `triplet' outcome $\tau$ is {\sl non-Abelian}.
As a result, the two measurement outcomes produce {\sl inequivalent} topological-charge-resolved states and can lead to distinct subsequent fusion and entanglement dynamics. 
In the fusion-tree basis, the weakly monitored Ising-anyon chain maps to a weakly monitored (1+1)D Ising (or Majorana) chain, where fermion parities of neighboring Majorana fermions are weakly measured. Its critical behavior was established by some of us~\cite{Wang25selfdual} via the dual problem of decohering or measuring a (2+0)D $\mathbb{Z}_2$ toric-code state protected by Kramers--Wannier self-duality, yielding $c_{\rm eff}=0.447(1)$ and $c_{\rm ent}=0.795(1)$, and a description by a non-linear sigma model~\footnote{In agreement with independent work by Fava et al.~\cite{Nahum23nlsmFreeFermion}.}.
More generally, for SU(2)$_k$, the quantum dimension of the fundamental anyon is $d_k = 2 \cos{\frac{\pi}{k+2}}$. As a result, for the fusion channel into singlet or triplet, we have the respective quantum dimensions $d_{\text{singlet}} = 1$ and $d_{\text{triplet}} = d^2_k - d_{\text{singlet}} = 1+ 2 \cos{\frac{2\pi}{k+2}}$, which for $k\geq3$ is larger than 1, making the Fibonacci-anyon chain the first member of the family to have non-Abelian fusion outcomes. 
We leave the systematic study of higher monitored SU(2)$_k$ anyon chains (with $k\geq4$) to future work.

Motivated by the above-mentioned duality between (1+1)D monitored Ising anyon circuits and (2+0)D decohered topologically ordered state~\cite{Wang25selfdual}, the (1+1)D monitored Fibonacci-anyon criticality we uncover in this paper could potentially describe the decoherence of a certain 2D topological quantum order associated with Fibonacci anyons, a non-Abelian generalization of the quantum critical point of decohering the toric code for $k=2$~\cite{Eckstein2024,Eckstein2025,Wang25selfdual,Beri24coherenterror,Yan2026sigma,Yang2026decoding,Negari2024,Putz25nishimori,Yoshida2026,Learning_transition_guoyi,patil2026highernishimoricriticalityexact, patil2026pottsZq,Mukherjee26zqthreshold,VijayLee26infoCritical}. Similar to the Ising anyon/toric code setting~\cite{Wang25selfdual}, such a 2D state could be formulated by lifting our spacetime string-net configuration (obeying the fusion rules) to a projected entangled pair state (as a string-net condensation wave function~\cite{LevinWen05stringnetcondensation} under certain {\it deformations}~\cite{Gils2009, Isakov_2011, Troyer10topocrit, Zhu19, Zhang20deformFib}), which under decoherence turns into a mixture of classical snapshots (see \cref{fig:from_1d_to_2d}). 
The learnability perspective on the monitored anyon chain naturally leads to an open question:
if one inserts local {\sl unitary} gates (such as braiding operations) that compete with the non-unitary dynamics of measurement,
does one observe a sharpening / learnability phase transition that generalizes the $U(1)$~\cite{Agrawal22chargesharpening,Vasseur22learnability,Ippoliti24learnability,Agrawal24measurementcollapse,Barratt22chargesharpening} and $SU(2)$~\cite{Majidy23spinsharpening,Feng25chargespinsharpening,Ma26nonabelianlearnability} cases? We will explore this in future studies. 
\\

\emph{Data availability}.-- 
The numerical data shown in the figures are available on Zenodo~\cite{zenodo}.\\

\begin{acknowledgments}
{\it Acknowledgments}.--
We thank M.~Buchhold for fruitful discussions in the early stages of this project, and R.~Patil for collaboration on closely related projects. 
GYZ thanks Nat Tantivasadakarn and Shang-Qiang Ning for discussions about the topological symmetry. 
GYZ and ZHZ thank Jin-Guo Liu
for significantly improving the efficiency of the numerical simulation code. 
The MPS functionality in our code
is built on the ITensor software library~\cite{itensor}.
All authors acknowledge hospitality of Nordita, the Nordic Institute for Theoretical Physics, during the summer 2026 workshop 
``Entanglement dynamics: Open quantum systems, monitored circuits, and topological order" during the final stages of this work. 
BH acknowledges the hospitality of KITP at UC Santa Barbara, during the program ``Engineering Fractionalization in Quantum Many-Body Systems: From New Quantum Materials to Quantum Simulators'', where this work has been finalized.
GYZ acknowledges the support of NSFC-Young Scientists Fund (grant no.~12504181) and Start-up Fund of HKUST(GZ) (grant no.~G0101000221) 
and Guangdong provincial project (grant no.~2024QN11X201) and the Guangdong Basic and Applied Basic Research Foundation (grant no.~2026A1515010965). 
The Cologne research group is supported, in part, by the Deutsche Forschungsgemeinschaft (DFG, German Research Foundation) 
under Germany's Excellence Strategy—Cluster of Excellence Matter
and Light for Quantum Computing (ML4Q) EXC 2004/1 -- 390534769 as well as within the CRC network TR 183 (Project Grant
No. 277101999) as part of subproject B01.
RV acknowledges support from the Swiss National Science Foundation (grant 10008234) and the Foundation for the University of Geneva.
\end{acknowledgments}

\clearpage
\section{End Matter}
\suppressfloats[t]

\subsection{Average measurement outcome probability}
\begin{figure}[th!]
    \centering
    \includegraphics[width=0.9\columnwidth]{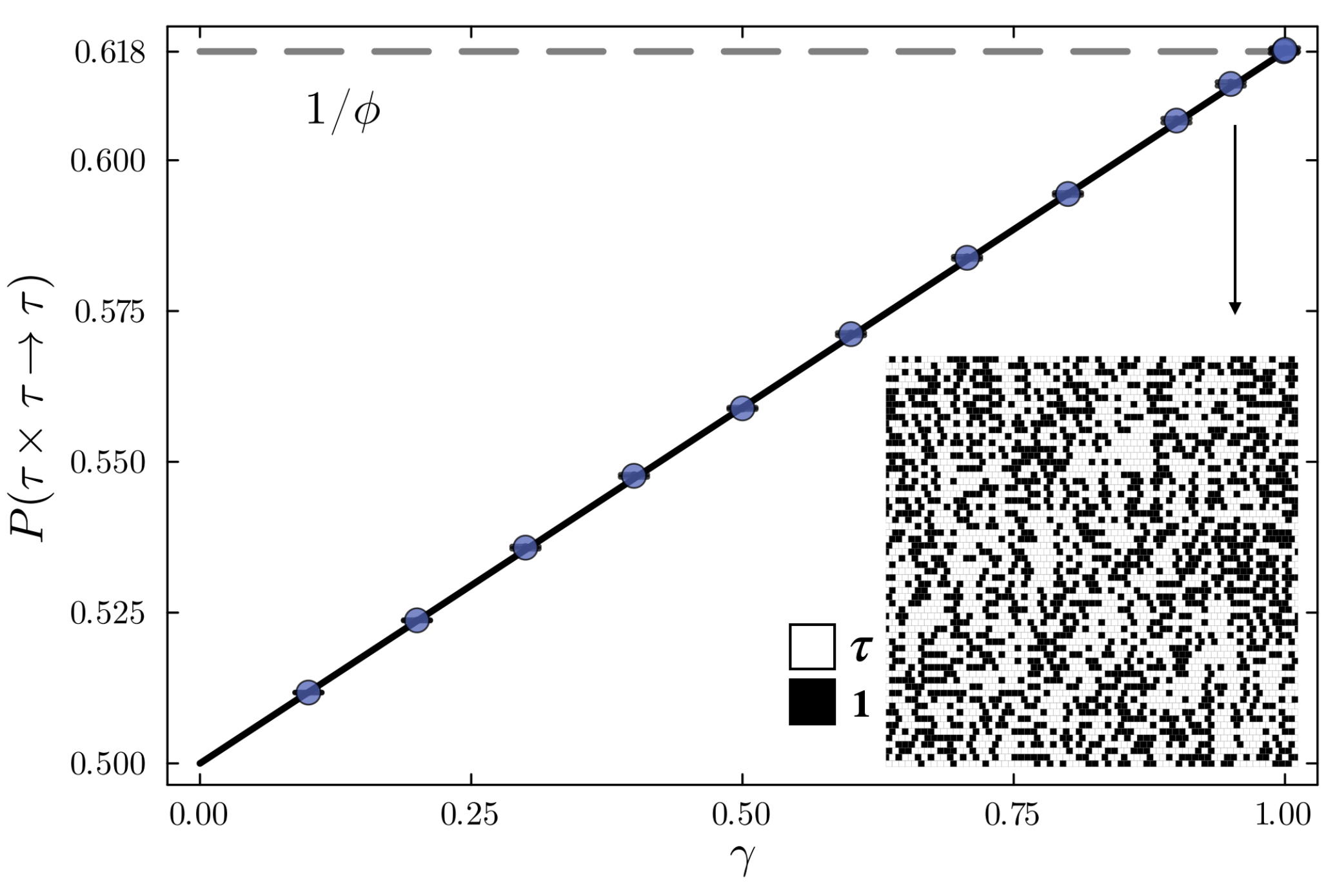}
    \caption{
        {\bf Average probability of the $\tau\times\tau\to\tau$ outcome.}
        Inset: a late-time Born trajectory at $\gamma=0.95$, $L=128$, and $T=256$; black (white) squares denote $\mathbf{1}$ ($\tau$).
    }
    \label{fig:outcome_probability}
\end{figure}

At late times, the local fusion probabilities approach
\[
	P(\tau\times \tau \to \tau) = 1/\phi \,,\qquad P(\tau\times \tau \to \mathbf{1}) = 1/\phi^2 \,.
\]
Finite strength reduces the readout fidelity to $(1+\gamma)/2$, giving
$P(\tau\times\tau\to \tau)=(1+\gamma)/(2\phi)+(1-\gamma)/(2\phi^2)$,
in agreement with
the numerical sampling results in \cref{fig:outcome_probability}. Both topological sectors have the same local measurement outcome probability, so their charge is encoded in spacetime correlations rather than the mean outcome density. The SM derives these results, including finite-size corrections~\cite{SM}.

\subsection{Charge sharpening and learning topological charge}

\begin{figure}[h!]
    \centering
    \includegraphics[width=0.9\columnwidth]{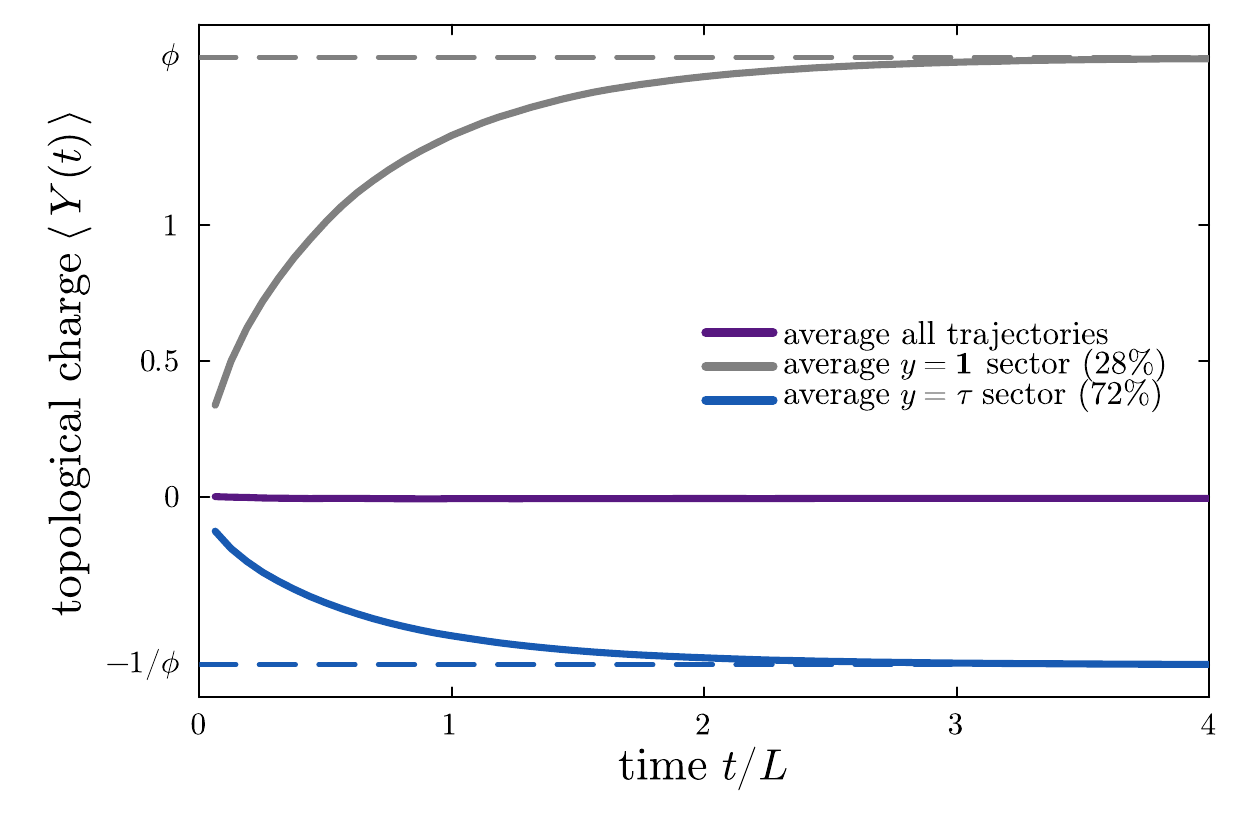}
    \caption{
        {\bf Topological charge sharpening.}
        The unconditional average (purple) splits into trajectories that sharpen to
        $y=\mathbf{1}$ (gray) or $\tau$ (blue), with eigenvalues $\phi$ and $-\phi^{-1}$.
        Here $L=16$ and $\gamma=0.95$.
    }
    \label{fig:Y_sharpening}
\end{figure}

\begin{figure}[h!]
    \centering
    \includegraphics[width=0.9\columnwidth]{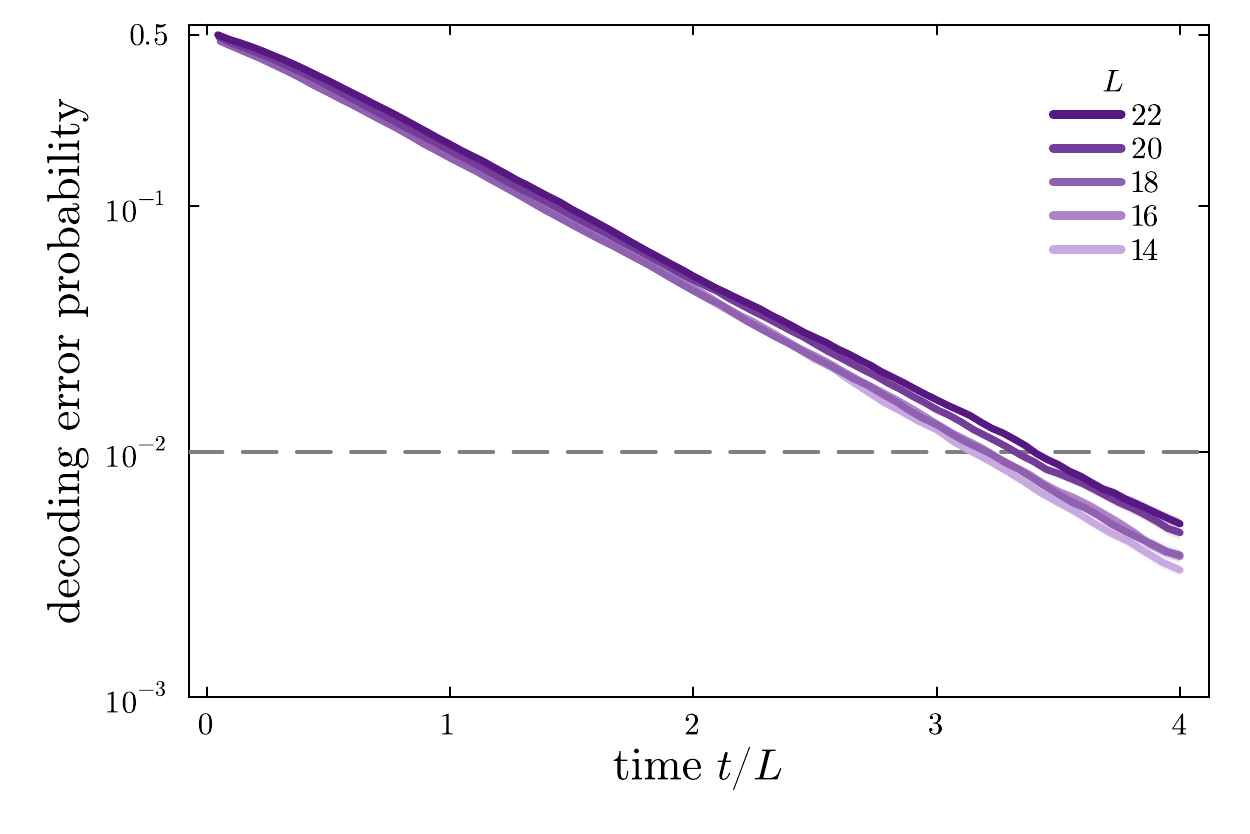}
    \caption{
        {\bf Learning the global charge.}
        At $\gamma=0.95$, the maximum-likelihood error decays exponentially with $t/L$ and falls below $1\%$ by $t=4L$.
        Each size for each sector tests $10^4$ trajectories.
}
    \label{fig:topological_charge_decoding_dynamics}
\end{figure}

An initial fusion-tree basis state spans both charge sectors, but each trajectory sharpens to
one of them [\cref{fig:Y_sharpening}]. If the initial charge $y^*$ is fixed but unknown,
the same record can instead be decoded by comparing its likelihood in the two sectors:
$\tilde y=\arg\max_y\|\mathcal{T}(\mathbf{s})\ket{\psi_y}\|^2$.
\Cref{fig:topological_charge_decoding_dynamics} shows that sharpening and classical
learning both require an $O(L)$ record. The SM gives the Bayesian sharpening description,
decoder construction, sector averaging, and sampling protocol~\cite{SM}.

\subsection{Varying measurement strength}

\begin{figure}[h!]
    \centering
    \includegraphics[width=0.9\columnwidth]{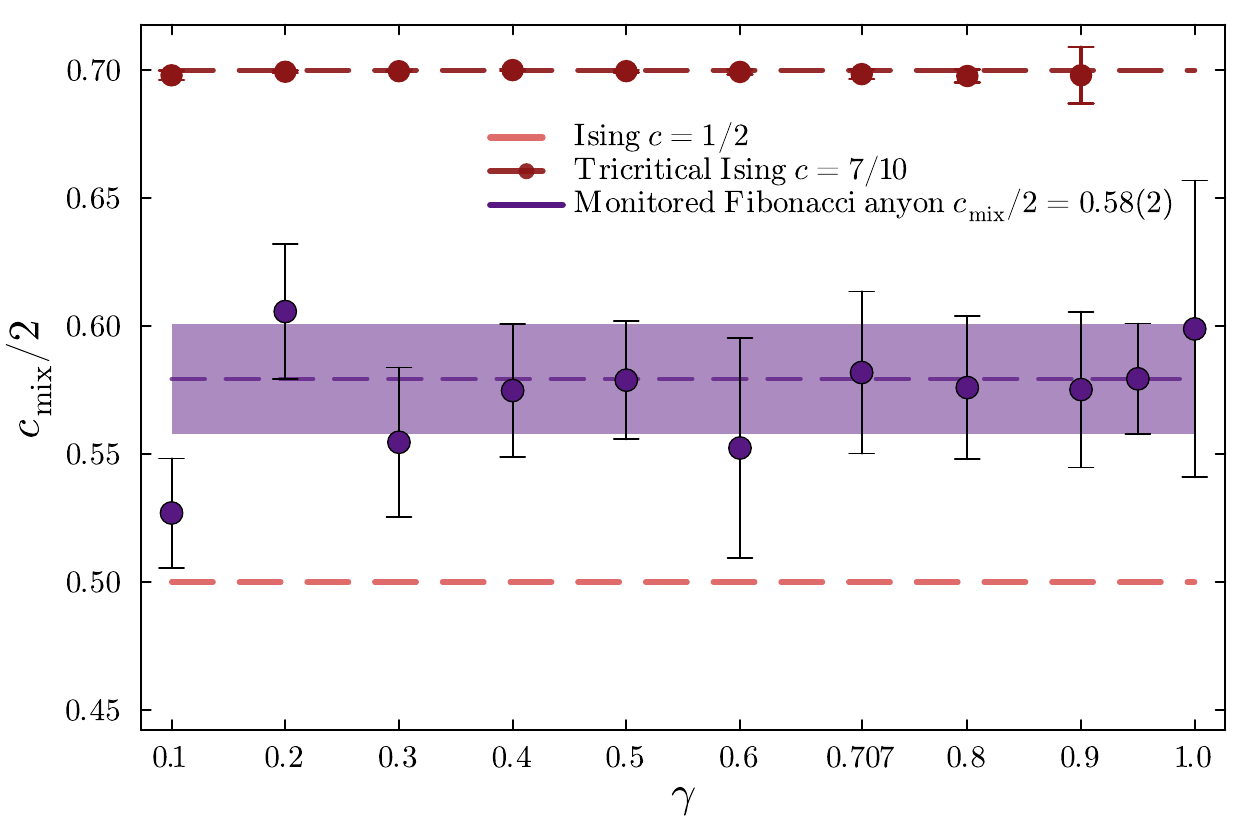}
    \caption{
        {\bf Casimir central charge versus measurement strength.}
        Purple: $c_{\rm mix}(\gamma)/2$ (ribbon: uncertainty at $\gamma=0.95$).
        Dark red: uniform $\mathbf{s}=\mathbf{1}$ post-selection. Dashed lines mark tricritical Ising $c=7/10$ and Ising $c=1/2$ for comparison. 
    }
    \label{fig:cCasimir}
\end{figure}

The extracted universality is intrinsic to the monitored golden chain rather than a single measurement strength $\gamma$: repeating the Casimir analysis over the entire interval $\gamma\in[0,1]$ (\cref{fig:cCasimir}), $c_{\rm mix}$ is 
consistent (within error bars) with being constant
from weak to projective measurement, including $\gamma=0.95$ and $\gamma=1$.
This suggests that weak and projective measurements of the Fibonacci chain belong to the same universality class: weakening the measurement strength away from the projective limit does not appear to be
a relevant perturbation of the measurement-only brickwork circuit. This is in contrast to the monitored Ising chain ($k=2$), when one measures  $Z_jZ_{j+1}$ parity and $X_j$ with uniform measurement strength (weak or projective); in this case, weak measurements are critical 
(described by the class-D nonlinear sigma model), but the projective limit is gapped,
corresponding to a paramagnet and a GHZ state as a product of Majorana fermion dimers, on even and odd bonds, respectively~\cite{Wang25selfdual}. 
This is also to be distinguished from the well-known setup
of {\it non-uniform random} projective measurements of $Z_jZ_{j+1}$ with probability $p$ and of $X_j$ with probability $(1-p)$; in that case the Ising or Majorana chain exhibits criticality in the 2D percolation universality class at $p=1/2$, where it possesses statistical (``weak'') translational and (Kramers-Wannier) self-dual symmetry~\cite{Nahum20freefermion,
Lang_2020,Vasseur24transfer}. 
Relaxing, at 
$p=1/2$, the projective limit to weak measurements leads to the generic class-D non-linear sigma model universality class mentioned above.
Coming back to the Fibonacci case, we note that  for  uniformly post-selected antiferromagnetic  measurements  the chain is  critical (conformal)  for finite weak measurement strength, but the projective limit dimerizes into a valence-bond solid for the (antiferro) $\tau \times \tau \to {\bf 1}$ fusion channel.
Clean Ising and tricritical-Ising values are shown for comparison in Fig.~\ref{fig:cCasimir}.

\subsection{Spacetime correlation matching}

\begin{figure}[tbh!]
    \centering
    \includegraphics[width=0.9\columnwidth]{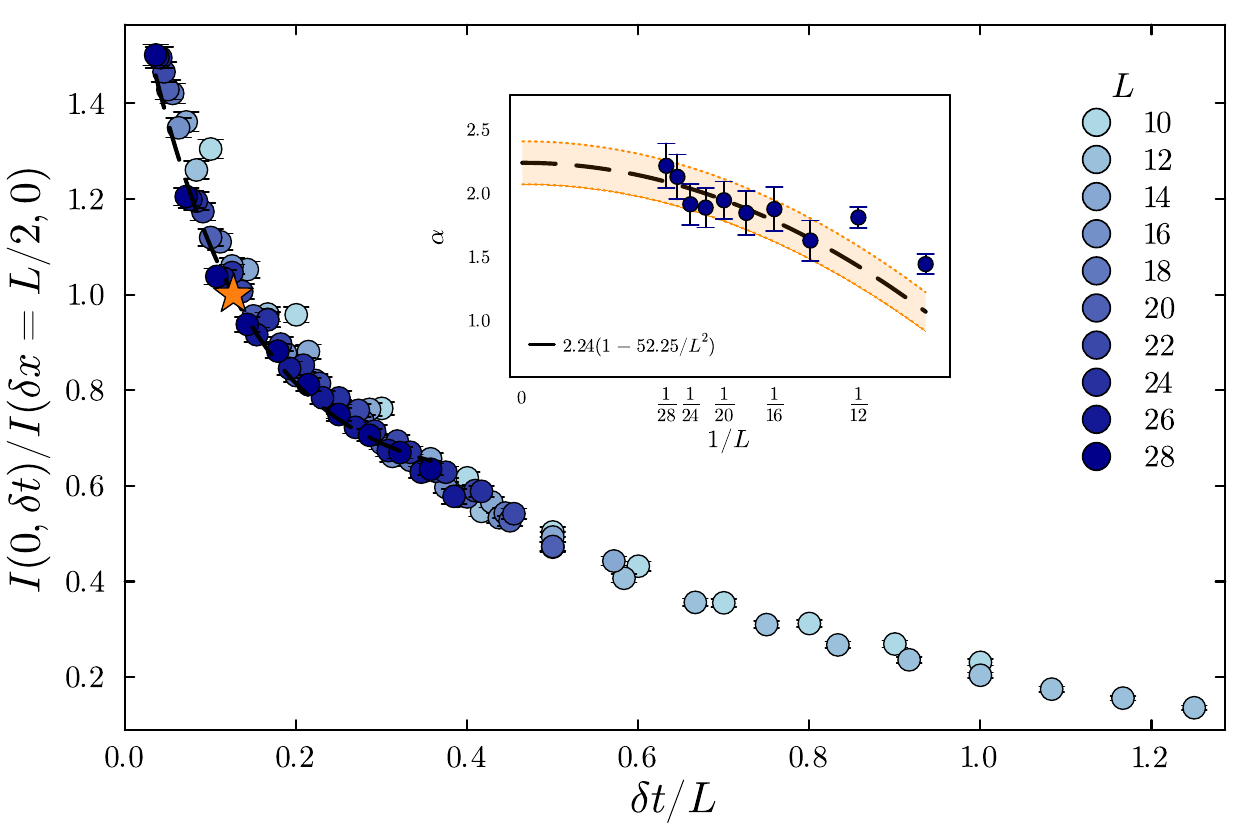}
    \caption{{\bf Spatiotemporal correlation matching} at $\gamma=0.95$.
    For $L=10$--$28$, $I(0,\delta t)/I(L/2,0)$ crosses unity at $\delta t^*$.
    Inset: extrapolation using $L=14$--$28$ gives $\alpha_\infty=2.24(8)$.
    }
    \label{fig:spacetimecorr}
\end{figure}

\begin{figure}[tb!]
    \centering
    \includegraphics[width=0.9\columnwidth]{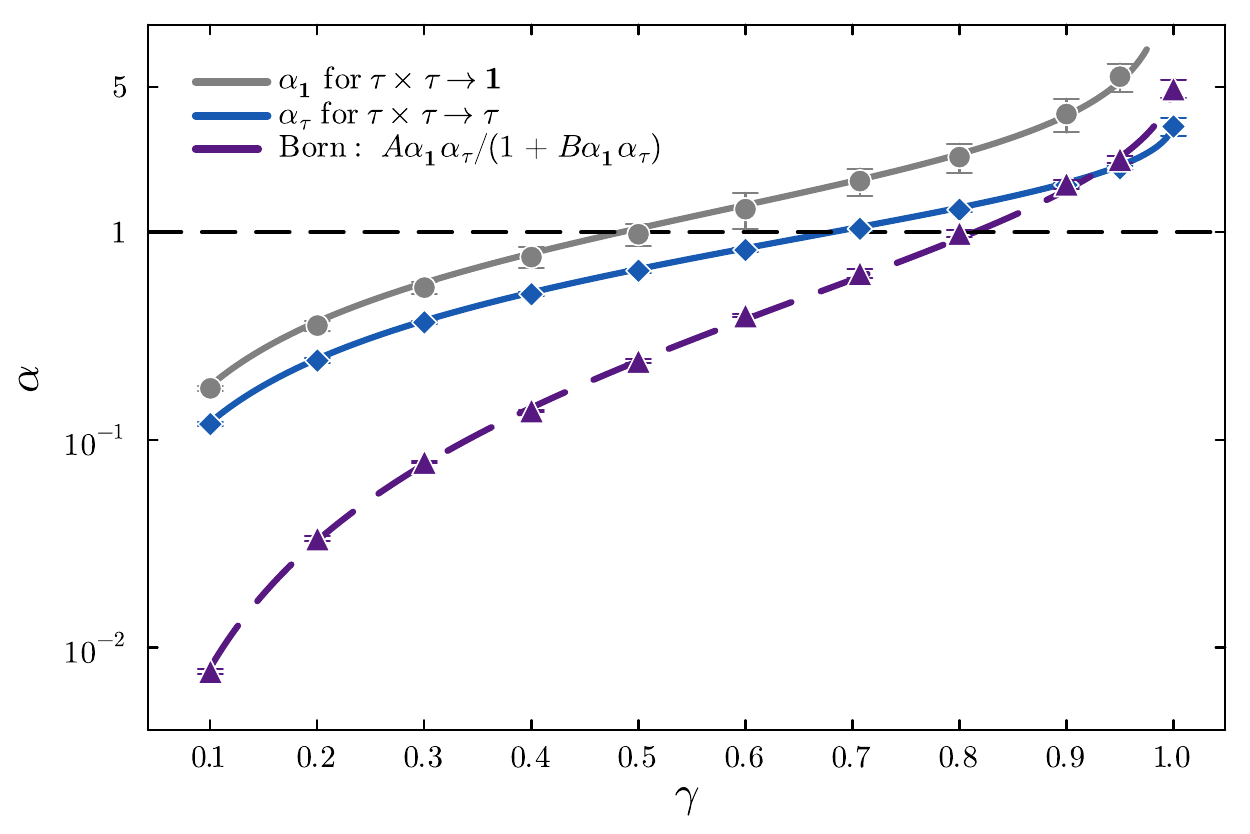}
    \caption{\textbf{Spacetime anisotropy versus $\gamma$.}
    Solid curves: exact analytic AFM/FM RSOS results; markers: numerical AFM (gray), FM
    (blue), and Born (purple) estimates. Dashed purple: phenomenological Born fit.
    }
    \label{fig:anisotropy_comparison}
\end{figure}

Following Refs.~\cite{Pixley22MIPTCFT,Zabalo_2020,Gullans20scalabledecoder,Vasseur24transfer},
we match the mutual information $I(\delta x, \delta t)$ of bulk probes separated in space or time.
Equating $I(0,\delta t^*)=I(L/2,0)$ on the cylinder gives
$\alpha=\ln(1+\sqrt2)L/(\pi\delta t^*)$ [\cref{fig:spacetimecorr}].
\Cref{fig:anisotropy_comparison} compares the Born result with the exact uniformly
post-selected RSOS values. The derivation and benchmarks are in the SM~\cite{SM}.

\subsection{From 1+1D to 2D}

\begin{figure}[h!]
    \centering
    \includegraphics[width=0.9\columnwidth]{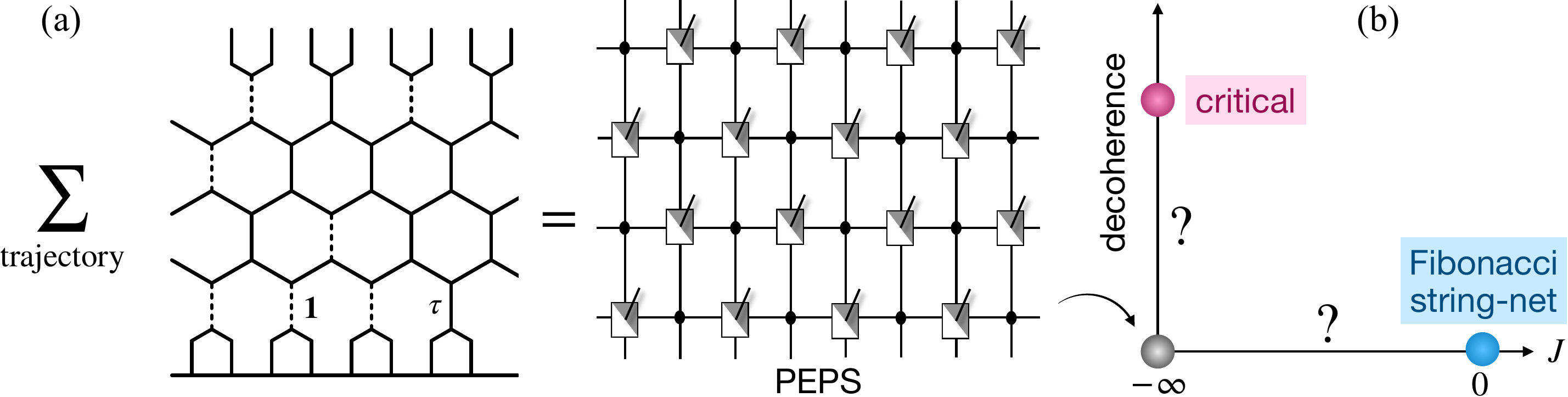}
    \caption{
        (a) {\bf From monitored trajectories to a 2D PEPS.}
        The coherent trajectory superposition (left; solid/dashed lines denote
        $\tau$/$\mathbf{1}$) has the PEPS representation shown at right.
        (b) {\bf Schematic phase diagram} connecting the monitored critical state,
        its parent PEPS, and the Fibonacci string-net through deformation $J$ and decoherence.
    }
    \label{fig:from_1d_to_2d}
\end{figure}

Local fusion measurements can be implemented by coupling each measured pair to an ancilla and then projectively measuring it.
If the ancillas remain unmeasured and coherent, the anyon chain and the 2D ancilla array form a pure 2D PEPS shown in \cref{fig:from_1d_to_2d}~\cite{Wang25selfdual}. Equivalently, the 1+1D monitored circuit becomes lifted to a 2+1D sequential unitary circuit preparing this PEPS.
This PEPS can be compared with the Fibonacci string-net condensate~\cite{LevinWen05stringnetcondensation}. Its horizontal zigzag fluxes are fixed to $\tau$, as in the Levin-Wen model Hamiltonian under a strong anisotropic string tension~\cite{Gils2009},
$
H_{\text{ Levin-Wen}} - J \sum_{e\in \text{ zig-zag}} \delta_{s(e),\mathbf{1}} + \cdots \ ,
$
and under certain boundary condition.
Alternatively, one might as well add a Rokhsar-Kivelson-like term to deform the wave function~\cite{Isakov_2011, Troyer10topocrit, Zhu19, Zhang20deformFib} directly, i.e. applying $e^{J  \delta_{s(e),\mathbf{1}}}$ to the horizontal zig-zag edges for the wave function. 
The string-net ground state, our PEPS and the monitored 1+1D criticality can be embedded in a schematic phase diagram shown in \cref{fig:from_1d_to_2d}(b). 
Note that the wave function amplitudes of this PEPS are generated by 
Born's rule, while the wave function amplitudes of the string-net condensate are given by the chromatic polynomial or the Boltzmann  weight of the $(\phi+2)$-state Potts model~\cite{Fidkowski09nonabelions,Fendley08quantumloopsnets}.
We leave it to future work to study the possible pure state and mixed state phase transitions (denoted by question marks in \cref{fig:from_1d_to_2d}(b)). 

\clearpage
\bibliography{measurements}

\begin{thebibliography}{92}%
\makeatletter
\providecommand \@ifxundefined [1]{%
 \@ifx{#1\undefined}
}%
\providecommand \@ifnum [1]{%
 \ifnum #1\expandafter \@firstoftwo
 \else \expandafter \@secondoftwo
 \fi
}%
\providecommand \@ifx [1]{%
 \ifx #1\expandafter \@firstoftwo
 \else \expandafter \@secondoftwo
 \fi
}%
\providecommand \natexlab [1]{#1}%
\providecommand \enquote  [1]{``#1''}%
\providecommand \bibnamefont  [1]{#1}%
\providecommand \bibfnamefont [1]{#1}%
\providecommand \citenamefont [1]{#1}%
\providecommand \href@noop [0]{\@secondoftwo}%
\providecommand \href [0]{\begingroup \@sanitize@url \@href}%
\providecommand \@href[1]{\@@startlink{#1}\@@href}%
\providecommand \@@href[1]{\endgroup#1\@@endlink}%
\providecommand \@sanitize@url [0]{\catcode `\\12\catcode `\$12\catcode `\&12\catcode `\#12\catcode `\^12\catcode `\_12\catcode `\%12\relax}%
\providecommand \@@startlink[1]{}%
\providecommand \@@endlink[0]{}%
\providecommand \url  [0]{\begingroup\@sanitize@url \@url }%
\providecommand \@url [1]{\endgroup\@href {#1}{\urlprefix }}%
\providecommand \urlprefix  [0]{URL }%
\providecommand \Eprint [0]{\href }%
\providecommand \doibase [0]{https://doi.org/}%
\providecommand \selectlanguage [0]{\@gobble}%
\providecommand \bibinfo  [0]{\@secondoftwo}%
\providecommand \bibfield  [0]{\@secondoftwo}%
\providecommand \translation [1]{[#1]}%
\providecommand \BibitemOpen [0]{}%
\providecommand \bibitemStop [0]{}%
\providecommand \bibitemNoStop [0]{.\EOS\space}%
\providecommand \EOS [0]{\spacefactor3000\relax}%
\providecommand \BibitemShut  [1]{\csname bibitem#1\endcsname}%
\let\auto@bib@innerbib\@empty
\bibitem [{\citenamefont {Freedman}\ \emph {et~al.}(2002)\citenamefont {Freedman}, \citenamefont {Larsen},\ and\ \citenamefont {Wang}}]{FreedmanLarsenWang02}%
  \BibitemOpen
  \bibfield  {author} {\bibinfo {author} {\bibfnamefont {M.}~\bibnamefont {Freedman}}, \bibinfo {author} {\bibfnamefont {M.}~\bibnamefont {Larsen}},\ and\ \bibinfo {author} {\bibfnamefont {Z.}~\bibnamefont {Wang}},\ }\bibfield  {title} {\bibinfo {title} {{A modular functor which is universal for quantum computation}},\ }\href {https://doi.org/10.1007/s002200200645} {\bibfield  {journal} {\bibinfo  {journal} {Communications in Mathematical Physics}\ }\textbf {\bibinfo {volume} {227}},\ \bibinfo {pages} {605} (\bibinfo {year} {2002})}\BibitemShut {NoStop}%
\bibitem [{\citenamefont {Read}\ and\ \citenamefont {Rezayi}(1999)}]{ReadRezayi99}%
  \BibitemOpen
  \bibfield  {author} {\bibinfo {author} {\bibfnamefont {N.}~\bibnamefont {Read}}\ and\ \bibinfo {author} {\bibfnamefont {E.}~\bibnamefont {Rezayi}},\ }\bibfield  {title} {\bibinfo {title} {{Beyond paired quantum {H}all states: {P}arafermions and incompressible states in the first excited {L}andau level}},\ }\href {https://doi.org/10.1103/PhysRevB.59.8084} {\bibfield  {journal} {\bibinfo  {journal} {Physical Review B}\ }\textbf {\bibinfo {volume} {59}},\ \bibinfo {pages} {8084} (\bibinfo {year} {1999})}\BibitemShut {NoStop}%
\bibitem [{\citenamefont {Nayak}\ \emph {et~al.}(2008)\citenamefont {Nayak}, \citenamefont {Simon}, \citenamefont {Stern}, \citenamefont {Freedman},\ and\ \citenamefont {Das~Sarma}}]{Nayak08RMP}%
  \BibitemOpen
  \bibfield  {author} {\bibinfo {author} {\bibfnamefont {C.}~\bibnamefont {Nayak}}, \bibinfo {author} {\bibfnamefont {S.~H.}\ \bibnamefont {Simon}}, \bibinfo {author} {\bibfnamefont {A.}~\bibnamefont {Stern}}, \bibinfo {author} {\bibfnamefont {M.}~\bibnamefont {Freedman}},\ and\ \bibinfo {author} {\bibfnamefont {S.}~\bibnamefont {Das~Sarma}},\ }\bibfield  {title} {\bibinfo {title} {{Non-{A}belian anyons and topological quantum computation}},\ }\href {https://doi.org/10.1103/RevModPhys.80.1083} {\bibfield  {journal} {\bibinfo  {journal} {Reviews of Modern Physics}\ }\textbf {\bibinfo {volume} {80}},\ \bibinfo {pages} {1083} (\bibinfo {year} {2008})}\BibitemShut {NoStop}%
\bibitem [{\citenamefont {Trebst}\ \emph {et~al.}(2008)\citenamefont {Trebst}, \citenamefont {Troyer}, \citenamefont {Wang},\ and\ \citenamefont {Ludwig}}]{Trebst08introfibonacci}%
  \BibitemOpen
  \bibfield  {author} {\bibinfo {author} {\bibfnamefont {S.}~\bibnamefont {Trebst}}, \bibinfo {author} {\bibfnamefont {M.}~\bibnamefont {Troyer}}, \bibinfo {author} {\bibfnamefont {Z.}~\bibnamefont {Wang}},\ and\ \bibinfo {author} {\bibfnamefont {A.~W.~W.}\ \bibnamefont {Ludwig}},\ }\bibfield  {title} {\bibinfo {title} {{A Short Introduction to Fibonacci Anyon Models}},\ }\href {https://doi.org/10.1143/ptps.176.384} {\bibfield  {journal} {\bibinfo  {journal} {Progress of Theoretical Physics Supplement}\ }\textbf {\bibinfo {volume} {176}},\ \bibinfo {pages} {384} (\bibinfo {year} {2008})}\BibitemShut {NoStop}%
\bibitem [{\citenamefont {Bonesteel}\ \emph {et~al.}(2005)\citenamefont {Bonesteel}, \citenamefont {Hormozi}, \citenamefont {Zikos},\ and\ \citenamefont {Simon}}]{Simon05fibocompute}%
  \BibitemOpen
  \bibfield  {author} {\bibinfo {author} {\bibfnamefont {N.~E.}\ \bibnamefont {Bonesteel}}, \bibinfo {author} {\bibfnamefont {L.}~\bibnamefont {Hormozi}}, \bibinfo {author} {\bibfnamefont {G.}~\bibnamefont {Zikos}},\ and\ \bibinfo {author} {\bibfnamefont {S.~H.}\ \bibnamefont {Simon}},\ }\bibfield  {title} {\bibinfo {title} {{Braid Topologies for Quantum Computation}},\ }\href {https://doi.org/10.1103/PhysRevLett.95.140503} {\bibfield  {journal} {\bibinfo  {journal} {Phys. Rev. Lett.}\ }\textbf {\bibinfo {volume} {95}},\ \bibinfo {pages} {140503} (\bibinfo {year} {2005})}\BibitemShut {NoStop}%
\bibitem [{\citenamefont {Mong}\ \emph {et~al.}(2017)\citenamefont {Mong}, \citenamefont {Zaletel}, \citenamefont {Pollmann},\ and\ \citenamefont {Papi{\'c}}}]{Mong17fibonacci125}%
  \BibitemOpen
  \bibfield  {author} {\bibinfo {author} {\bibfnamefont {R.~S.~K.}\ \bibnamefont {Mong}}, \bibinfo {author} {\bibfnamefont {M.~P.}\ \bibnamefont {Zaletel}}, \bibinfo {author} {\bibfnamefont {F.}~\bibnamefont {Pollmann}},\ and\ \bibinfo {author} {\bibfnamefont {Z.}~\bibnamefont {Papi{\'c}}},\ }\bibfield  {title} {\bibinfo {title} {{Fibonacci anyons and charge density order in the $12/5$ and $13/5$ quantum {H}all plateaus}},\ }\href {https://doi.org/10.1103/PhysRevB.95.115136} {\bibfield  {journal} {\bibinfo  {journal} {Physical Review B}\ }\textbf {\bibinfo {volume} {95}},\ \bibinfo {pages} {115136} (\bibinfo {year} {2017})}\BibitemShut {NoStop}%
\bibitem [{\citenamefont {Feiguin}\ \emph {et~al.}(2007)\citenamefont {Feiguin}, \citenamefont {Trebst}, \citenamefont {Ludwig}, \citenamefont {Troyer}, \citenamefont {Kitaev}, \citenamefont {Wang},\ and\ \citenamefont {Freedman}}]{Trebst07goldenchain}%
  \BibitemOpen
  \bibfield  {author} {\bibinfo {author} {\bibfnamefont {A.}~\bibnamefont {Feiguin}}, \bibinfo {author} {\bibfnamefont {S.}~\bibnamefont {Trebst}}, \bibinfo {author} {\bibfnamefont {A.~W.~W.}\ \bibnamefont {Ludwig}}, \bibinfo {author} {\bibfnamefont {M.}~\bibnamefont {Troyer}}, \bibinfo {author} {\bibfnamefont {A.}~\bibnamefont {Kitaev}}, \bibinfo {author} {\bibfnamefont {Z.}~\bibnamefont {Wang}},\ and\ \bibinfo {author} {\bibfnamefont {M.~H.}\ \bibnamefont {Freedman}},\ }\bibfield  {title} {\bibinfo {title} {{Interacting Anyons in Topological Quantum Liquids: The Golden Chain}},\ }\href {https://doi.org/10.1103/PhysRevLett.98.160409} {\bibfield  {journal} {\bibinfo  {journal} {Phys. Rev. Lett.}\ }\textbf {\bibinfo {volume} {98}},\ \bibinfo {pages} {160409} (\bibinfo {year} {2007})}\BibitemShut {NoStop}%
\bibitem [{\citenamefont {Gils}\ \emph {et~al.}(2009{\natexlab{a}})\citenamefont {Gils}, \citenamefont {Ardonne}, \citenamefont {Trebst}, \citenamefont {Ludwig}, \citenamefont {Troyer},\ and\ \citenamefont {Wang}}]{Gils2009}%
  \BibitemOpen
  \bibfield  {author} {\bibinfo {author} {\bibfnamefont {C.}~\bibnamefont {Gils}}, \bibinfo {author} {\bibfnamefont {E.}~\bibnamefont {Ardonne}}, \bibinfo {author} {\bibfnamefont {S.}~\bibnamefont {Trebst}}, \bibinfo {author} {\bibfnamefont {A.~W.~W.}\ \bibnamefont {Ludwig}}, \bibinfo {author} {\bibfnamefont {M.}~\bibnamefont {Troyer}},\ and\ \bibinfo {author} {\bibfnamefont {Z.}~\bibnamefont {Wang}},\ }\bibfield  {title} {\bibinfo {title} {{Collective States of Interacting Anyons, Edge States, and the Nucleation of Topological Liquids}},\ }\href {https://doi.org/10.1103/PhysRevLett.103.070401} {\bibfield  {journal} {\bibinfo  {journal} {Phys. Rev. Lett.}\ }\textbf {\bibinfo {volume} {103}},\ \bibinfo {pages} {070401} (\bibinfo {year} {2009}{\natexlab{a}})}\BibitemShut {NoStop}%
\bibitem [{\citenamefont {Gils}\ \emph {et~al.}(2009{\natexlab{b}})\citenamefont {Gils}, \citenamefont {Trebst}, \citenamefont {Kitaev}, \citenamefont {Ludwig}, \citenamefont {Troyer},\ and\ \citenamefont {Wang}}]{Gils09topology}%
  \BibitemOpen
  \bibfield  {author} {\bibinfo {author} {\bibfnamefont {C.}~\bibnamefont {Gils}}, \bibinfo {author} {\bibfnamefont {S.}~\bibnamefont {Trebst}}, \bibinfo {author} {\bibfnamefont {A.}~\bibnamefont {Kitaev}}, \bibinfo {author} {\bibfnamefont {A.~W.~W.}\ \bibnamefont {Ludwig}}, \bibinfo {author} {\bibfnamefont {M.}~\bibnamefont {Troyer}},\ and\ \bibinfo {author} {\bibfnamefont {Z.}~\bibnamefont {Wang}},\ }\bibfield  {title} {\bibinfo {title} {{Topology-driven quantum phase transitions in time-reversal-invariant anyonic quantum liquids}},\ }\href {https://doi.org/10.1038/nphys1396} {\bibfield  {journal} {\bibinfo  {journal} {Nature Physics}\ }\textbf {\bibinfo {volume} {5}},\ \bibinfo {pages} {834} (\bibinfo {year} {2009}{\natexlab{b}})}\BibitemShut {NoStop}%
\bibitem [{\citenamefont {Fr{\"o}hlich}\ \emph {et~al.}(2004)\citenamefont {Fr{\"o}hlich}, \citenamefont {Fuchs}, \citenamefont {Runkel},\ and\ \citenamefont {Schweigert}}]{Frohlich2004KWdefects}%
  \BibitemOpen
  \bibfield  {author} {\bibinfo {author} {\bibfnamefont {J.}~\bibnamefont {Fr{\"o}hlich}}, \bibinfo {author} {\bibfnamefont {J.}~\bibnamefont {Fuchs}}, \bibinfo {author} {\bibfnamefont {I.}~\bibnamefont {Runkel}},\ and\ \bibinfo {author} {\bibfnamefont {C.}~\bibnamefont {Schweigert}},\ }\bibfield  {title} {\bibinfo {title} {{Kramers--Wannier duality from conformal defects}},\ }\href {https://doi.org/10.1103/PhysRevLett.93.070601} {\bibfield  {journal} {\bibinfo  {journal} {Phys. Rev. Lett.}\ }\textbf {\bibinfo {volume} {93}},\ \bibinfo {pages} {070601} (\bibinfo {year} {2004})}\BibitemShut {NoStop}%
\bibitem [{\citenamefont {Gaiotto}\ \emph {et~al.}(2015)\citenamefont {Gaiotto}, \citenamefont {Kapustin}, \citenamefont {Seiberg},\ and\ \citenamefont {Willett}}]{Gaiotto15generalized}%
  \BibitemOpen
  \bibfield  {author} {\bibinfo {author} {\bibfnamefont {D.}~\bibnamefont {Gaiotto}}, \bibinfo {author} {\bibfnamefont {A.}~\bibnamefont {Kapustin}}, \bibinfo {author} {\bibfnamefont {N.}~\bibnamefont {Seiberg}},\ and\ \bibinfo {author} {\bibfnamefont {B.}~\bibnamefont {Willett}},\ }\bibfield  {title} {\bibinfo {title} {{Generalized Global Symmetries}},\ }\href {https://doi.org/10.1007/JHEP02(2015)172} {\bibfield  {journal} {\bibinfo  {journal} {Journal of High Energy Physics}\ }\textbf {\bibinfo {volume} {2015}},\ \bibinfo {pages} {172} (\bibinfo {year} {2015})}\BibitemShut {NoStop}%
\bibitem [{\citenamefont {Buican}\ and\ \citenamefont {Gromov}(2017)}]{Buican_2017_anyonichains}%
  \BibitemOpen
  \bibfield  {author} {\bibinfo {author} {\bibfnamefont {M.}~\bibnamefont {Buican}}\ and\ \bibinfo {author} {\bibfnamefont {A.}~\bibnamefont {Gromov}},\ }\bibfield  {title} {\bibinfo {title} {{Anyonic Chains, Topological Defects, and Conformal Field Theory}},\ }\href {https://doi.org/10.1007/s00220-017-2995-6} {\bibfield  {journal} {\bibinfo  {journal} {Communications in Mathematical Physics}\ }\textbf {\bibinfo {volume} {356}},\ \bibinfo {pages} {1017–1056} (\bibinfo {year} {2017})}\BibitemShut {NoStop}%
\bibitem [{\citenamefont {Chang}\ \emph {et~al.}(2019)\citenamefont {Chang}, \citenamefont {Lin}, \citenamefont {Shao}, \citenamefont {Wang},\ and\ \citenamefont {Yin}}]{Chang19defectlines}%
  \BibitemOpen
  \bibfield  {author} {\bibinfo {author} {\bibfnamefont {C.-M.}\ \bibnamefont {Chang}}, \bibinfo {author} {\bibfnamefont {Y.-H.}\ \bibnamefont {Lin}}, \bibinfo {author} {\bibfnamefont {S.-H.}\ \bibnamefont {Shao}}, \bibinfo {author} {\bibfnamefont {Y.}~\bibnamefont {Wang}},\ and\ \bibinfo {author} {\bibfnamefont {X.}~\bibnamefont {Yin}},\ }\bibfield  {title} {\bibinfo {title} {{Topological Defect Lines and Renormalization Group Flows in Two Dimensions}},\ }\href {https://doi.org/10.1007/JHEP01(2019)026} {\bibfield  {journal} {\bibinfo  {journal} {Journal of High Energy Physics}\ }\textbf {\bibinfo {volume} {2019}},\ \bibinfo {pages} {26} (\bibinfo {year} {2019})}\BibitemShut {NoStop}%
\bibitem [{\citenamefont {Thorngren}\ and\ \citenamefont {Wang}(2024)}]{ThorngrenWang2019fusion}%
  \BibitemOpen
  \bibfield  {author} {\bibinfo {author} {\bibfnamefont {R.}~\bibnamefont {Thorngren}}\ and\ \bibinfo {author} {\bibfnamefont {Y.}~\bibnamefont {Wang}},\ }\bibfield  {title} {\bibinfo {title} {{Fusion category symmetry. Part I. Anomaly in-flow and gapped phases}},\ }\href {https://doi.org/10.1007/JHEP04(2024)132} {\bibfield  {journal} {\bibinfo  {journal} {Journal of High Energy Physics}\ }\textbf {\bibinfo {volume} {2024}},\ \bibinfo {pages} {132} (\bibinfo {year} {2024})}\BibitemShut {NoStop}%
\bibitem [{\citenamefont {Ji}\ and\ \citenamefont {Wen}(2020)}]{Ji20categorical}%
  \BibitemOpen
  \bibfield  {author} {\bibinfo {author} {\bibfnamefont {W.}~\bibnamefont {Ji}}\ and\ \bibinfo {author} {\bibfnamefont {X.-G.}\ \bibnamefont {Wen}},\ }\bibfield  {title} {\bibinfo {title} {{Categorical symmetry and noninvertible anomaly in symmetry-breaking and topological phase transitions}},\ }\href {https://doi.org/10.1103/PhysRevResearch.2.033417} {\bibfield  {journal} {\bibinfo  {journal} {Phys. Rev. Res.}\ }\textbf {\bibinfo {volume} {2}},\ \bibinfo {pages} {033417} (\bibinfo {year} {2020})}\BibitemShut {NoStop}%
\bibitem [{\citenamefont {Kong}\ \emph {et~al.}(2020)\citenamefont {Kong}, \citenamefont {Lan}, \citenamefont {Wen}, \citenamefont {Zhang},\ and\ \citenamefont {Zheng}}]{Kong20categorical}%
  \BibitemOpen
  \bibfield  {author} {\bibinfo {author} {\bibfnamefont {L.}~\bibnamefont {Kong}}, \bibinfo {author} {\bibfnamefont {T.}~\bibnamefont {Lan}}, \bibinfo {author} {\bibfnamefont {X.-G.}\ \bibnamefont {Wen}}, \bibinfo {author} {\bibfnamefont {Z.-H.}\ \bibnamefont {Zhang}},\ and\ \bibinfo {author} {\bibfnamefont {H.}~\bibnamefont {Zheng}},\ }\bibfield  {title} {\bibinfo {title} {{Algebraic higher symmetry and categorical symmetry: A holographic and entanglement view of symmetry}},\ }\href {https://doi.org/10.1103/PhysRevResearch.2.043086} {\bibfield  {journal} {\bibinfo  {journal} {Phys. Rev. Res.}\ }\textbf {\bibinfo {volume} {2}},\ \bibinfo {pages} {043086} (\bibinfo {year} {2020})}\BibitemShut {NoStop}%
\bibitem [{\citenamefont {Aasen}\ \emph {et~al.}(2020)\citenamefont {Aasen}, \citenamefont {Fendley},\ and\ \citenamefont {Mong}}]{aasen2020topologicaldefectslatticedualities}%
  \BibitemOpen
  \bibfield  {author} {\bibinfo {author} {\bibfnamefont {D.}~\bibnamefont {Aasen}}, \bibinfo {author} {\bibfnamefont {P.}~\bibnamefont {Fendley}},\ and\ \bibinfo {author} {\bibfnamefont {R.~S.~K.}\ \bibnamefont {Mong}},\ }\href@noop {} {\bibinfo {title} {{Topological Defects on the Lattice: Dualities and Degeneracies}}} (\bibinfo {year} {2020}),\ \Eprint {https://arxiv.org/abs/2008.08598} {arXiv:2008.08598 [cond-mat.stat-mech]} \BibitemShut {NoStop}%
\bibitem [{\citenamefont {Chatterjee}\ \emph {et~al.}(2025)\citenamefont {Chatterjee}, \citenamefont {Ji},\ and\ \citenamefont {Wen}}]{Chatterjee2025emergent}%
  \BibitemOpen
  \bibfield  {author} {\bibinfo {author} {\bibfnamefont {A.}~\bibnamefont {Chatterjee}}, \bibinfo {author} {\bibfnamefont {W.}~\bibnamefont {Ji}},\ and\ \bibinfo {author} {\bibfnamefont {X.-G.}\ \bibnamefont {Wen}},\ }\bibfield  {title} {\bibinfo {title} {{Emergent generalized symmetry and maximal symmetry topological order}},\ }\href {https://doi.org/10.1103/jzfv-ygmr} {\bibfield  {journal} {\bibinfo  {journal} {Phys. Rev. B}\ }\textbf {\bibinfo {volume} {112}},\ \bibinfo {pages} {115142} (\bibinfo {year} {2025})}\BibitemShut {NoStop}%
\bibitem [{\citenamefont {Seiberg}\ and\ \citenamefont {Shao}(2024)}]{SeibergShao2024Majorana}%
  \BibitemOpen
  \bibfield  {author} {\bibinfo {author} {\bibfnamefont {N.}~\bibnamefont {Seiberg}}\ and\ \bibinfo {author} {\bibfnamefont {S.-H.}\ \bibnamefont {Shao}},\ }\bibfield  {title} {\bibinfo {title} {{Majorana chain and Ising model: (non-invertible) translations, anomalies, and emanant symmetries}},\ }\href {https://doi.org/10.21468/SciPostPhys.16.3.064} {\bibfield  {journal} {\bibinfo  {journal} {SciPost Phys.}\ }\textbf {\bibinfo {volume} {16}},\ \bibinfo {pages} {064} (\bibinfo {year} {2024})}\BibitemShut {NoStop}%
\bibitem [{\citenamefont {Shao}(2023)}]{Shao2023TASI}%
  \BibitemOpen
  \bibfield  {author} {\bibinfo {author} {\bibfnamefont {S.-H.}\ \bibnamefont {Shao}},\ }\href@noop {} {\bibinfo {title} {{What's Done Cannot Be Undone: TASI Lectures on Non-Invertible Symmetries}}} (\bibinfo {year} {2023}),\ \Eprint {https://arxiv.org/abs/2308.00747} {arXiv:2308.00747 [hep-th]} \BibitemShut {NoStop}%
\bibitem [{\citenamefont {Seiberg}\ \emph {et~al.}(2024)\citenamefont {Seiberg}, \citenamefont {Seifnashri},\ and\ \citenamefont {Shao}}]{Seiberg2024LSM}%
  \BibitemOpen
  \bibfield  {author} {\bibinfo {author} {\bibfnamefont {N.}~\bibnamefont {Seiberg}}, \bibinfo {author} {\bibfnamefont {S.}~\bibnamefont {Seifnashri}},\ and\ \bibinfo {author} {\bibfnamefont {S.-H.}\ \bibnamefont {Shao}},\ }\bibfield  {title} {\bibinfo {title} {{Non-invertible symmetries and LSM-type constraints on a tensor product Hilbert space}},\ }\href {https://doi.org/10.21468/SciPostPhys.16.6.154} {\bibfield  {journal} {\bibinfo  {journal} {SciPost Phys.}\ }\textbf {\bibinfo {volume} {16}},\ \bibinfo {pages} {154} (\bibinfo {year} {2024})}\BibitemShut {NoStop}%
\bibitem [{\citenamefont {Okada}\ and\ \citenamefont {Tachikawa}(2024)}]{Okada2024quantumoperations}%
  \BibitemOpen
  \bibfield  {author} {\bibinfo {author} {\bibfnamefont {M.}~\bibnamefont {Okada}}\ and\ \bibinfo {author} {\bibfnamefont {Y.}~\bibnamefont {Tachikawa}},\ }\bibfield  {title} {\bibinfo {title} {{Noninvertible Symmetries Act Locally by Quantum Operations}},\ }\href {https://doi.org/10.1103/PhysRevLett.133.191602} {\bibfield  {journal} {\bibinfo  {journal} {Phys. Rev. Lett.}\ }\textbf {\bibinfo {volume} {133}},\ \bibinfo {pages} {191602} (\bibinfo {year} {2024})}\BibitemShut {NoStop}%
\bibitem [{\citenamefont {Cao}\ \emph {et~al.}(2024)\citenamefont {Cao}, \citenamefont {Li},\ and\ \citenamefont {Yamazaki}}]{Cao2024generating}%
  \BibitemOpen
  \bibfield  {author} {\bibinfo {author} {\bibfnamefont {W.}~\bibnamefont {Cao}}, \bibinfo {author} {\bibfnamefont {L.}~\bibnamefont {Li}},\ and\ \bibinfo {author} {\bibfnamefont {M.}~\bibnamefont {Yamazaki}},\ }\bibfield  {title} {\bibinfo {title} {{Generating lattice non-invertible symmetries}},\ }\href {https://doi.org/10.21468/SciPostPhys.17.4.104} {\bibfield  {journal} {\bibinfo  {journal} {SciPost Phys.}\ }\textbf {\bibinfo {volume} {17}},\ \bibinfo {pages} {104} (\bibinfo {year} {2024})}\BibitemShut {NoStop}%
\bibitem [{\citenamefont {Chen}\ \emph {et~al.}(2026)\citenamefont {Chen}, \citenamefont {Liu}, \citenamefont {Lu},\ and\ \citenamefont {Tantivasadakarn}}]{chen2026SSBnonInvertible}%
  \BibitemOpen
  \bibfield  {author} {\bibinfo {author} {\bibfnamefont {X.}~\bibnamefont {Chen}}, \bibinfo {author} {\bibfnamefont {S.}~\bibnamefont {Liu}}, \bibinfo {author} {\bibfnamefont {D.-C.}\ \bibnamefont {Lu}},\ and\ \bibinfo {author} {\bibfnamefont {N.}~\bibnamefont {Tantivasadakarn}},\ }\href@noop {} {\bibinfo {title} {{Spontaneous breaking of non-invertible symmetries and duality to beyond-Landau transitions}}} (\bibinfo {year} {2026}),\ \Eprint {https://arxiv.org/abs/2605.27672} {arXiv:2605.27672 [cond-mat.str-el]} \BibitemShut {NoStop}%
\bibitem [{\citenamefont {Skinner}\ \emph {et~al.}(2019)\citenamefont {Skinner}, \citenamefont {Ruhman},\ and\ \citenamefont {Nahum}}]{Skinner2019}%
  \BibitemOpen
  \bibfield  {author} {\bibinfo {author} {\bibfnamefont {B.}~\bibnamefont {Skinner}}, \bibinfo {author} {\bibfnamefont {J.}~\bibnamefont {Ruhman}},\ and\ \bibinfo {author} {\bibfnamefont {A.}~\bibnamefont {Nahum}},\ }\bibfield  {title} {\bibinfo {title} {{Measurement-Induced Phase Transitions in the Dynamics of Entanglement}},\ }\href {https://doi.org/10.1103/PhysRevX.9.031009} {\bibfield  {journal} {\bibinfo  {journal} {Phys. Rev. X}\ }\textbf {\bibinfo {volume} {9}},\ \bibinfo {pages} {031009} (\bibinfo {year} {2019})}\BibitemShut {NoStop}%
\bibitem [{\citenamefont {Bao}\ \emph {et~al.}(2020)\citenamefont {Bao}, \citenamefont {Choi},\ and\ \citenamefont {Altman}}]{Altman2020weak}%
  \BibitemOpen
  \bibfield  {author} {\bibinfo {author} {\bibfnamefont {Y.}~\bibnamefont {Bao}}, \bibinfo {author} {\bibfnamefont {S.}~\bibnamefont {Choi}},\ and\ \bibinfo {author} {\bibfnamefont {E.}~\bibnamefont {Altman}},\ }\bibfield  {title} {\bibinfo {title} {{Theory of the phase transition in random unitary circuits with measurements}},\ }\href {https://doi.org/10.1103/PhysRevB.101.104301} {\bibfield  {journal} {\bibinfo  {journal} {Phys. Rev. B}\ }\textbf {\bibinfo {volume} {101}},\ \bibinfo {pages} {104301} (\bibinfo {year} {2020})}\BibitemShut {NoStop}%
\bibitem [{\citenamefont {Jian}\ \emph {et~al.}(2020)\citenamefont {Jian}, \citenamefont {You}, \citenamefont {Vasseur},\ and\ \citenamefont {Ludwig}}]{Jian2020}%
  \BibitemOpen
  \bibfield  {author} {\bibinfo {author} {\bibfnamefont {C.-M.}\ \bibnamefont {Jian}}, \bibinfo {author} {\bibfnamefont {Y.-Z.}\ \bibnamefont {You}}, \bibinfo {author} {\bibfnamefont {R.}~\bibnamefont {Vasseur}},\ and\ \bibinfo {author} {\bibfnamefont {A.~W.~W.}\ \bibnamefont {Ludwig}},\ }\bibfield  {title} {\bibinfo {title} {{Measurement-induced criticality in random quantum circuits}},\ }\href {https://doi.org/10.1103/PhysRevB.101.104302} {\bibfield  {journal} {\bibinfo  {journal} {Phys. Rev. B}\ }\textbf {\bibinfo {volume} {101}},\ \bibinfo {pages} {104302} (\bibinfo {year} {2020})}\BibitemShut {NoStop}%
\bibitem [{\citenamefont {Gullans}\ and\ \citenamefont {Huse}(2020{\natexlab{a}})}]{Huse2020qec}%
  \BibitemOpen
  \bibfield  {author} {\bibinfo {author} {\bibfnamefont {M.~J.}\ \bibnamefont {Gullans}}\ and\ \bibinfo {author} {\bibfnamefont {D.~A.}\ \bibnamefont {Huse}},\ }\bibfield  {title} {\bibinfo {title} {{Dynamical Purification Phase Transition Induced by Quantum Measurements}},\ }\href {https://doi.org/10.1103/PhysRevX.10.041020} {\bibfield  {journal} {\bibinfo  {journal} {Phys. Rev. X}\ }\textbf {\bibinfo {volume} {10}},\ \bibinfo {pages} {041020} (\bibinfo {year} {2020}{\natexlab{a}})}\BibitemShut {NoStop}%
\bibitem [{\citenamefont {Ippoliti}\ \emph {et~al.}(2021)\citenamefont {Ippoliti}, \citenamefont {Gullans}, \citenamefont {Gopalakrishnan}, \citenamefont {Huse},\ and\ \citenamefont {Khemani}}]{Vedika2021measure}%
  \BibitemOpen
  \bibfield  {author} {\bibinfo {author} {\bibfnamefont {M.}~\bibnamefont {Ippoliti}}, \bibinfo {author} {\bibfnamefont {M.~J.}\ \bibnamefont {Gullans}}, \bibinfo {author} {\bibfnamefont {S.}~\bibnamefont {Gopalakrishnan}}, \bibinfo {author} {\bibfnamefont {D.~A.}\ \bibnamefont {Huse}},\ and\ \bibinfo {author} {\bibfnamefont {V.}~\bibnamefont {Khemani}},\ }\bibfield  {title} {\bibinfo {title} {{Entanglement Phase Transitions in Measurement-Only Dynamics}},\ }\href {https://doi.org/10.1103/PhysRevX.11.011030} {\bibfield  {journal} {\bibinfo  {journal} {Phys. Rev. X}\ }\textbf {\bibinfo {volume} {11}},\ \bibinfo {pages} {011030} (\bibinfo {year} {2021})}\BibitemShut {NoStop}%
\bibitem [{\citenamefont {Bravyi}\ \emph {et~al.}(2022)\citenamefont {Bravyi}, \citenamefont {Kim}, \citenamefont {Kliesch},\ and\ \citenamefont {Koenig}}]{Bravyi2022}%
  \BibitemOpen
  \bibfield  {author} {\bibinfo {author} {\bibfnamefont {S.}~\bibnamefont {Bravyi}}, \bibinfo {author} {\bibfnamefont {I.}~\bibnamefont {Kim}}, \bibinfo {author} {\bibfnamefont {A.}~\bibnamefont {Kliesch}},\ and\ \bibinfo {author} {\bibfnamefont {R.}~\bibnamefont {Koenig}},\ }\href@noop {} {\bibinfo {title} {{Adaptive constant-depth circuits for manipulating non-abelian anyons}}} (\bibinfo {year} {2022}),\ \Eprint {https://arxiv.org/abs/2205.01933} {arXiv:2205.01933} \BibitemShut {NoStop}%
\bibitem [{\citenamefont {{Google Quantum AI}}(2023)}]{Google22surfacecode}%
  \BibitemOpen
  \bibfield  {author} {\bibinfo {author} {\bibnamefont {{Google Quantum AI}}},\ }\bibfield  {title} {\bibinfo {title} {{Suppressing quantum errors by scaling a surface code logical qubit}},\ }\href {https://doi.org/10.1038/s41586-022-05434-1} {\bibfield  {journal} {\bibinfo  {journal} {Nature}\ }\textbf {\bibinfo {volume} {614}},\ \bibinfo {pages} {676} (\bibinfo {year} {2023})}\BibitemShut {NoStop}%
\bibitem [{\citenamefont {{Google Quantum AI}}(2021)}]{Roushan21sycamoretoriccode}%
  \BibitemOpen
  \bibfield  {author} {\bibinfo {author} {\bibnamefont {{Google Quantum AI}}},\ }\bibfield  {title} {\bibinfo {title} {{Realizing topologically ordered states on a quantum processor}},\ }\href {https://doi.org/10.1126/science.abi8378} {\bibfield  {journal} {\bibinfo  {journal} {Science}\ }\textbf {\bibinfo {volume} {374}},\ \bibinfo {pages} {1237} (\bibinfo {year} {2021})}\BibitemShut {NoStop}%
\bibitem [{\citenamefont {{Iqbal}}\ \emph {et~al.}(2024)\citenamefont {{Iqbal}}, \citenamefont {{Tantivasadakarn}}, \citenamefont {{Gatterman}}, \citenamefont {{Gerber}}, \citenamefont {{Gilmore}}, \citenamefont {{Gresh}}, \citenamefont {{Hankin}}, \citenamefont {{Hewitt}}, \citenamefont {{Horst}}, \citenamefont {{Matheny}}, \citenamefont {{Mengle}}, \citenamefont {{Neyenhuis}}, \citenamefont {{Vishwanath}}, \citenamefont {{Foss-Feig}}, \citenamefont {{Verresen}},\ and\ \citenamefont {{Dreyer}}}]{Iqbal23iontoriccode}%
  \BibitemOpen
  \bibfield  {author} {\bibinfo {author} {\bibfnamefont {M.}~\bibnamefont {{Iqbal}}}, \bibinfo {author} {\bibfnamefont {N.}~\bibnamefont {{Tantivasadakarn}}}, \bibinfo {author} {\bibfnamefont {T.~M.}\ \bibnamefont {{Gatterman}}}, \bibinfo {author} {\bibfnamefont {J.~A.}\ \bibnamefont {{Gerber}}}, \bibinfo {author} {\bibfnamefont {K.}~\bibnamefont {{Gilmore}}}, \bibinfo {author} {\bibfnamefont {D.}~\bibnamefont {{Gresh}}}, \bibinfo {author} {\bibfnamefont {A.}~\bibnamefont {{Hankin}}}, \bibinfo {author} {\bibfnamefont {N.}~\bibnamefont {{Hewitt}}}, \bibinfo {author} {\bibfnamefont {C.~V.}\ \bibnamefont {{Horst}}}, \bibinfo {author} {\bibfnamefont {M.}~\bibnamefont {{Matheny}}}, \bibinfo {author} {\bibfnamefont {T.}~\bibnamefont {{Mengle}}}, \bibinfo {author} {\bibfnamefont {B.}~\bibnamefont {{Neyenhuis}}}, \bibinfo {author} {\bibfnamefont {A.}~\bibnamefont {{Vishwanath}}}, \bibinfo {author} {\bibfnamefont {M.}~\bibnamefont {{Foss-Feig}}}, \bibinfo {author} {\bibfnamefont {R.}~\bibnamefont {{Verresen}}},\ and\ \bibinfo {author} {\bibfnamefont {H.}~\bibnamefont {{Dreyer}}},\ }\bibfield  {title} {\bibinfo {title} {{Topological Order from Measurements and Feed-Forward on a Trapped Ion Quantum Computer}},\ }\href {https://doi.org/10.1038/s42005-024-01698-3} {\bibfield  {journal} {\bibinfo  {journal} {Communications Physics}\ }\textbf {\bibinfo {volume} {7}},\ \bibinfo {pages} {205} (\bibinfo {year} {2024})}\BibitemShut {NoStop}%
\bibitem [{\citenamefont {Bonderson}\ \emph {et~al.}(2008)\citenamefont {Bonderson}, \citenamefont {Freedman},\ and\ \citenamefont {Nayak}}]{Nayak2008measurement}%
  \BibitemOpen
  \bibfield  {author} {\bibinfo {author} {\bibfnamefont {P.}~\bibnamefont {Bonderson}}, \bibinfo {author} {\bibfnamefont {M.}~\bibnamefont {Freedman}},\ and\ \bibinfo {author} {\bibfnamefont {C.}~\bibnamefont {Nayak}},\ }\bibfield  {title} {\bibinfo {title} {{Measurement-Only Topological Quantum Computation}},\ }\href {https://doi.org/10.1103/PhysRevLett.101.010501} {\bibfield  {journal} {\bibinfo  {journal} {Phys. Rev. Lett.}\ }\textbf {\bibinfo {volume} {101}},\ \bibinfo {pages} {010501} (\bibinfo {year} {2008})}\BibitemShut {NoStop}%
\bibitem [{\citenamefont {Temperley}\ and\ \citenamefont {Lieb}(1971)}]{TemperleyLiebPRC1971}%
  \BibitemOpen
  \bibfield  {author} {\bibinfo {author} {\bibfnamefont {H.~N.~V.}\ \bibnamefont {Temperley}}\ and\ \bibinfo {author} {\bibfnamefont {E.~H.}\ \bibnamefont {Lieb}},\ }\bibfield  {title} {\bibinfo {title} {{Relations between the ‘percolation’ and ‘colouring’ problem and other graph-theoretical problems associated with regular planar lattices: some exact results for the ‘percolation’ problem}},\ }\href {https://doi.org/10.1098/rspa.1971.0067} {\bibfield  {journal} {\bibinfo  {journal} {Proceedings of the Royal Society of London. A. Mathematical and Physical Sciences}\ }\textbf {\bibinfo {volume} {322}},\ \bibinfo {pages} {251} (\bibinfo {year} {1971})}\BibitemShut {NoStop}%
\bibitem [{\citenamefont {Jones}(1984)}]{VJonesCRASP1984}%
  \BibitemOpen
  \bibfield  {author} {\bibinfo {author} {\bibfnamefont {V.~F.~R.}\ \bibnamefont {Jones}},\ }\bibfield  {title} {\bibinfo {title} {{Groupes de tresses, alg\`ebres de Hecke et facteurs de type $\mathrm{II}_1$}},\ }\href@noop {} {\bibfield  {journal} {\bibinfo  {journal} {C. R. Acad. Sci. Paris S\'er. I Math.}\ }\textbf {\bibinfo {volume} {298}},\ \bibinfo {pages} {505} (\bibinfo {year} {1984})}\BibitemShut {NoStop}%
\bibitem [{\citenamefont {Jones}(1987)}]{VJonesAnnMath1987}%
  \BibitemOpen
  \bibfield  {author} {\bibinfo {author} {\bibfnamefont {V.~F.~R.}\ \bibnamefont {Jones}},\ }\bibfield  {title} {\bibinfo {title} {Hecke algebra representations of braid groups and link polynomials},\ }\href {https://doi.org/10.2307/1971403} {\bibfield  {journal} {\bibinfo  {journal} {Annals of Mathematics}\ }\textbf {\bibinfo {volume} {126}},\ \bibinfo {pages} {335} (\bibinfo {year} {1987})}\BibitemShut {NoStop}%
\bibitem [{\citenamefont {Wenzl}(1988)}]{WenzlInventMath1988}%
  \BibitemOpen
  \bibfield  {author} {\bibinfo {author} {\bibfnamefont {H.}~\bibnamefont {Wenzl}},\ }\bibfield  {title} {\bibinfo {title} {{Hecke algebras of type $A_n$ and subfactors}},\ }\href {https://doi.org/10.1007/BF01404457} {\bibfield  {journal} {\bibinfo  {journal} {Inventiones Mathematicae}\ }\textbf {\bibinfo {volume} {92}},\ \bibinfo {pages} {349} (\bibinfo {year} {1988})}\BibitemShut {NoStop}%
\bibitem [{\citenamefont {Kuniba}\ \emph {et~al.}(1986)\citenamefont {Kuniba}, \citenamefont {Akutsu},\ and\ \citenamefont {Wadati}}]{KunibaAkutsuWadatiJPSJ1986}%
  \BibitemOpen
  \bibfield  {author} {\bibinfo {author} {\bibfnamefont {A.}~\bibnamefont {Kuniba}}, \bibinfo {author} {\bibfnamefont {Y.}~\bibnamefont {Akutsu}},\ and\ \bibinfo {author} {\bibfnamefont {M.}~\bibnamefont {Wadati}},\ }\bibfield  {title} {\bibinfo {title} {{Virasoro Algebra, von Neumann Algebra and Critical Eight-Vertex SOS Models}},\ }\href {https://doi.org/10.1143/JPSJ.55.3285} {\bibfield  {journal} {\bibinfo  {journal} {Journal of the Physical Society of Japan}\ }\textbf {\bibinfo {volume} {55}},\ \bibinfo {pages} {3285} (\bibinfo {year} {1986})}\BibitemShut {NoStop}%
\bibitem [{\citenamefont {Pasquier}(1987)}]{PasquierNPB1987}%
  \BibitemOpen
  \bibfield  {author} {\bibinfo {author} {\bibfnamefont {V.}~\bibnamefont {Pasquier}},\ }\bibfield  {title} {\bibinfo {title} {{Two-dimensional critical systems labelled by Dynkin diagrams}},\ }\href {https://doi.org/10.1016/0550-3213(87)90332-4} {\bibfield  {journal} {\bibinfo  {journal} {Nuclear Physics B}\ }\textbf {\bibinfo {volume} {285}},\ \bibinfo {pages} {162} (\bibinfo {year} {1987})}\BibitemShut {NoStop}%
\bibitem [{\citenamefont {Fendley}(2006)}]{FendleyJPhysA2006}%
  \BibitemOpen
  \bibfield  {author} {\bibinfo {author} {\bibfnamefont {P.}~\bibnamefont {Fendley}},\ }\bibfield  {title} {\bibinfo {title} {{Loop models and their critical points}},\ }\href {https://doi.org/10.1088/0305-4470/39/50/011} {\bibfield  {journal} {\bibinfo  {journal} {Journal of Physics A: Mathematical and General}\ }\textbf {\bibinfo {volume} {39}},\ \bibinfo {pages} {15445} (\bibinfo {year} {2006})}\BibitemShut {NoStop}%
\bibitem [{Note1()}]{Note1}%
  \BibitemOpen
  \bibinfo {note} {Note that the paired anyons whose total charge is measured into $\protect \mathbf {1}$ do not annihilate as in Ref.~\cite {Zou21fibonacci}, but instead remain spatially separated as in measurement-only topological quantum computation~\cite {Nayak2008measurement}}\BibitemShut {NoStop}%
\bibitem [{\citenamefont {Pfeifer}\ \emph {et~al.}(2012)\citenamefont {Pfeifer}, \citenamefont {Buerschaper}, \citenamefont {Trebst}, \citenamefont {Ludwig}, \citenamefont {Troyer},\ and\ \citenamefont {Vidal}}]{Pfeifer12fibo}%
  \BibitemOpen
  \bibfield  {author} {\bibinfo {author} {\bibfnamefont {R.~N.~C.}\ \bibnamefont {Pfeifer}}, \bibinfo {author} {\bibfnamefont {O.}~\bibnamefont {Buerschaper}}, \bibinfo {author} {\bibfnamefont {S.}~\bibnamefont {Trebst}}, \bibinfo {author} {\bibfnamefont {A.~W.~W.}\ \bibnamefont {Ludwig}}, \bibinfo {author} {\bibfnamefont {M.}~\bibnamefont {Troyer}},\ and\ \bibinfo {author} {\bibfnamefont {G.}~\bibnamefont {Vidal}},\ }\bibfield  {title} {\bibinfo {title} {{Translation invariance, topology, and protection of criticality in chains of interacting anyons}},\ }\href {https://doi.org/10.1103/PhysRevB.86.155111} {\bibfield  {journal} {\bibinfo  {journal} {Phys. Rev. B}\ }\textbf {\bibinfo {volume} {86}},\ \bibinfo {pages} {155111} (\bibinfo {year} {2012})}\BibitemShut {NoStop}%
\bibitem [{SM()}]{SM}%
  \BibitemOpen
  \href@noop {} {}\bibinfo {note} {See Supplemental Material at [URL will be inserted by publisher] for (i) From a measurement trajectory to the RSOS tensor network, (ii) Topological Symmetry, (iii) Measurement statistics, (iv) Spacetime anisotropy, (v) Shannon entropy density, (vi) Casimir fits across measurement strengths, and (vii) Computational resources, which includes Refs.~\cite{Friedan1984,Belavin1984,BianchiniErcolessiPearceRavanini15,KimPearce87anisotropy}.}\BibitemShut {Stop}%
\bibitem [{\citenamefont {Friedan}\ \emph {et~al.}(1985)\citenamefont {Friedan}, \citenamefont {Qiu},\ and\ \citenamefont {Shenker}}]{Friedan1984}%
  \BibitemOpen
  \bibfield  {author} {\bibinfo {author} {\bibfnamefont {D.}~\bibnamefont {Friedan}}, \bibinfo {author} {\bibfnamefont {Z.}~\bibnamefont {Qiu}},\ and\ \bibinfo {author} {\bibfnamefont {S.~H.}\ \bibnamefont {Shenker}},\ }\bibfield  {title} {\bibinfo {title} {{Superconformal Invariance in Two-Dimensions and the Tricritical Ising Model}},\ }\href {https://doi.org/10.1016/0370-2693(85)90819-6} {\bibfield  {journal} {\bibinfo  {journal} {Phys. Lett. B}\ }\textbf {\bibinfo {volume} {151}},\ \bibinfo {pages} {37} (\bibinfo {year} {1985})}\BibitemShut {NoStop}%
\bibitem [{\citenamefont {Belavin}\ \emph {et~al.}(1984)\citenamefont {Belavin}, \citenamefont {Polyakov},\ and\ \citenamefont {Zamolodchikov}}]{Belavin1984}%
  \BibitemOpen
  \bibfield  {author} {\bibinfo {author} {\bibfnamefont {A.~A.}\ \bibnamefont {Belavin}}, \bibinfo {author} {\bibfnamefont {A.~M.}\ \bibnamefont {Polyakov}},\ and\ \bibinfo {author} {\bibfnamefont {A.~B.}\ \bibnamefont {Zamolodchikov}},\ }\bibfield  {title} {\bibinfo {title} {{Infinite Conformal Symmetry in Two-Dimensional Quantum Field Theory}},\ }\href {https://doi.org/10.1016/0550-3213(84)90052-X} {\bibfield  {journal} {\bibinfo  {journal} {Nucl. Phys. B}\ }\textbf {\bibinfo {volume} {241}},\ \bibinfo {pages} {333} (\bibinfo {year} {1984})}\BibitemShut {NoStop}%
\bibitem [{\citenamefont {Bianchini}\ \emph {et~al.}(2015)\citenamefont {Bianchini}, \citenamefont {Ercolessi}, \citenamefont {Pearce},\ and\ \citenamefont {Ravanini}}]{BianchiniErcolessiPearceRavanini15}%
  \BibitemOpen
  \bibfield  {author} {\bibinfo {author} {\bibfnamefont {D.}~\bibnamefont {Bianchini}}, \bibinfo {author} {\bibfnamefont {E.}~\bibnamefont {Ercolessi}}, \bibinfo {author} {\bibfnamefont {P.~A.}\ \bibnamefont {Pearce}},\ and\ \bibinfo {author} {\bibfnamefont {F.}~\bibnamefont {Ravanini}},\ }\bibfield  {title} {\bibinfo {title} {{RSOS Quantum Chains Associated with Off-Critical Minimal Models and $\mathbb{Z}_n$ Parafermions}},\ }\href {https://doi.org/10.1088/1742-5468/2015/03/P03010} {\bibfield  {journal} {\bibinfo  {journal} {J. Stat. Mech.}\ }\textbf {\bibinfo {volume} {2015}},\ \bibinfo {pages} {P03010} (\bibinfo {year} {2015})}\BibitemShut {NoStop}%
\bibitem [{\citenamefont {Kim}\ and\ \citenamefont {Pearce}(1987)}]{KimPearce87anisotropy}%
  \BibitemOpen
  \bibfield  {author} {\bibinfo {author} {\bibfnamefont {D.}~\bibnamefont {Kim}}\ and\ \bibinfo {author} {\bibfnamefont {P.~A.}\ \bibnamefont {Pearce}},\ }\bibfield  {title} {\bibinfo {title} {{Scaling Dimensions and Conformal Anomaly in Anisotropic Lattice Spin Models}},\ }\href {https://doi.org/10.1088/0305-4470/20/7/006} {\bibfield  {journal} {\bibinfo  {journal} {J. Phys. A: Math. Gen.}\ }\textbf {\bibinfo {volume} {20}},\ \bibinfo {pages} {L451} (\bibinfo {year} {1987})}\BibitemShut {NoStop}%
\bibitem [{\citenamefont {Levin}\ and\ \citenamefont {Wen}(2005)}]{LevinWen05stringnetcondensation}%
  \BibitemOpen
  \bibfield  {author} {\bibinfo {author} {\bibfnamefont {M.~A.}\ \bibnamefont {Levin}}\ and\ \bibinfo {author} {\bibfnamefont {X.-G.}\ \bibnamefont {Wen}},\ }\bibfield  {title} {\bibinfo {title} {{String-net condensation: A physical mechanism for topological phases}},\ }\href {https://doi.org/10.1103/PhysRevB.71.045110} {\bibfield  {journal} {\bibinfo  {journal} {Phys. Rev. B}\ }\textbf {\bibinfo {volume} {71}},\ \bibinfo {pages} {045110} (\bibinfo {year} {2005})}\BibitemShut {NoStop}%
\bibitem [{\citenamefont {Agrawal}\ \emph {et~al.}(2022)\citenamefont {Agrawal}, \citenamefont {Zabalo}, \citenamefont {Chen}, \citenamefont {Wilson}, \citenamefont {Potter}, \citenamefont {Pixley}, \citenamefont {Gopalakrishnan},\ and\ \citenamefont {Vasseur}}]{Agrawal22chargesharpening}%
  \BibitemOpen
  \bibfield  {author} {\bibinfo {author} {\bibfnamefont {U.}~\bibnamefont {Agrawal}}, \bibinfo {author} {\bibfnamefont {A.}~\bibnamefont {Zabalo}}, \bibinfo {author} {\bibfnamefont {K.}~\bibnamefont {Chen}}, \bibinfo {author} {\bibfnamefont {J.~H.}\ \bibnamefont {Wilson}}, \bibinfo {author} {\bibfnamefont {A.~C.}\ \bibnamefont {Potter}}, \bibinfo {author} {\bibfnamefont {J.~H.}\ \bibnamefont {Pixley}}, \bibinfo {author} {\bibfnamefont {S.}~\bibnamefont {Gopalakrishnan}},\ and\ \bibinfo {author} {\bibfnamefont {R.}~\bibnamefont {Vasseur}},\ }\bibfield  {title} {\bibinfo {title} {{Entanglement and Charge-Sharpening Transitions in U(1) Symmetric Monitored Quantum Circuits}},\ }\href {https://doi.org/10.1103/PhysRevX.12.041002} {\bibfield  {journal} {\bibinfo  {journal} {Phys. Rev. X}\ }\textbf {\bibinfo {volume} {12}},\ \bibinfo {pages} {041002} (\bibinfo {year} {2022})}\BibitemShut {NoStop}%
\bibitem [{\citenamefont {Barratt}\ \emph {et~al.}(2022{\natexlab{a}})\citenamefont {Barratt}, \citenamefont {Agrawal}, \citenamefont {Potter}, \citenamefont {Gopalakrishnan},\ and\ \citenamefont {Vasseur}}]{Vasseur22learnability}%
  \BibitemOpen
  \bibfield  {author} {\bibinfo {author} {\bibfnamefont {F.}~\bibnamefont {Barratt}}, \bibinfo {author} {\bibfnamefont {U.}~\bibnamefont {Agrawal}}, \bibinfo {author} {\bibfnamefont {A.~C.}\ \bibnamefont {Potter}}, \bibinfo {author} {\bibfnamefont {S.}~\bibnamefont {Gopalakrishnan}},\ and\ \bibinfo {author} {\bibfnamefont {R.}~\bibnamefont {Vasseur}},\ }\bibfield  {title} {\bibinfo {title} {{Transitions in the Learnability of Global Charges from Local Measurements}},\ }\href {https://doi.org/10.1103/PhysRevLett.129.200602} {\bibfield  {journal} {\bibinfo  {journal} {Phys. Rev. Lett.}\ }\textbf {\bibinfo {volume} {129}},\ \bibinfo {pages} {200602} (\bibinfo {year} {2022}{\natexlab{a}})}\BibitemShut {NoStop}%
\bibitem [{\citenamefont {Ippoliti}\ and\ \citenamefont {Khemani}(2024)}]{Ippoliti24learnability}%
  \BibitemOpen
  \bibfield  {author} {\bibinfo {author} {\bibfnamefont {M.}~\bibnamefont {Ippoliti}}\ and\ \bibinfo {author} {\bibfnamefont {V.}~\bibnamefont {Khemani}},\ }\bibfield  {title} {\bibinfo {title} {{Learnability Transitions in Monitored Quantum Dynamics via Eavesdropper's Classical Shadows}},\ }\href {https://doi.org/10.1103/PRXQuantum.5.020304} {\bibfield  {journal} {\bibinfo  {journal} {PRX Quantum}\ }\textbf {\bibinfo {volume} {5}},\ \bibinfo {pages} {020304} (\bibinfo {year} {2024})}\BibitemShut {NoStop}%
\bibitem [{\citenamefont {Agrawal}\ \emph {et~al.}(2024)\citenamefont {Agrawal}, \citenamefont {Lopez-Piqueres}, \citenamefont {Vasseur}, \citenamefont {Gopalakrishnan},\ and\ \citenamefont {Potter}}]{Agrawal24measurementcollapse}%
  \BibitemOpen
  \bibfield  {author} {\bibinfo {author} {\bibfnamefont {U.}~\bibnamefont {Agrawal}}, \bibinfo {author} {\bibfnamefont {J.}~\bibnamefont {Lopez-Piqueres}}, \bibinfo {author} {\bibfnamefont {R.}~\bibnamefont {Vasseur}}, \bibinfo {author} {\bibfnamefont {S.}~\bibnamefont {Gopalakrishnan}},\ and\ \bibinfo {author} {\bibfnamefont {A.~C.}\ \bibnamefont {Potter}},\ }\bibfield  {title} {\bibinfo {title} {{Observing Quantum Measurement Collapse as a Learnability Phase Transition}},\ }\href {https://doi.org/10.1103/PhysRevX.14.041012} {\bibfield  {journal} {\bibinfo  {journal} {Phys. Rev. X}\ }\textbf {\bibinfo {volume} {14}},\ \bibinfo {pages} {041012} (\bibinfo {year} {2024})}\BibitemShut {NoStop}%
\bibitem [{\citenamefont {Barratt}\ \emph {et~al.}(2022{\natexlab{b}})\citenamefont {Barratt}, \citenamefont {Agrawal}, \citenamefont {Gopalakrishnan}, \citenamefont {Huse}, \citenamefont {Vasseur},\ and\ \citenamefont {Potter}}]{Barratt22chargesharpening}%
  \BibitemOpen
  \bibfield  {author} {\bibinfo {author} {\bibfnamefont {F.}~\bibnamefont {Barratt}}, \bibinfo {author} {\bibfnamefont {U.}~\bibnamefont {Agrawal}}, \bibinfo {author} {\bibfnamefont {S.}~\bibnamefont {Gopalakrishnan}}, \bibinfo {author} {\bibfnamefont {D.~A.}\ \bibnamefont {Huse}}, \bibinfo {author} {\bibfnamefont {R.}~\bibnamefont {Vasseur}},\ and\ \bibinfo {author} {\bibfnamefont {A.~C.}\ \bibnamefont {Potter}},\ }\bibfield  {title} {\bibinfo {title} {{Field Theory of Charge Sharpening in Symmetric Monitored Quantum Circuits}},\ }\href {https://doi.org/10.1103/PhysRevLett.129.120604} {\bibfield  {journal} {\bibinfo  {journal} {Phys. Rev. Lett.}\ }\textbf {\bibinfo {volume} {129}},\ \bibinfo {pages} {120604} (\bibinfo {year} {2022}{\natexlab{b}})}\BibitemShut {NoStop}%
\bibitem [{\citenamefont {Majidy}\ \emph {et~al.}(2023)\citenamefont {Majidy}, \citenamefont {Agrawal}, \citenamefont {Gopalakrishnan}, \citenamefont {Potter}, \citenamefont {Vasseur},\ and\ \citenamefont {Yunger~Halpern}}]{Majidy23spinsharpening}%
  \BibitemOpen
  \bibfield  {author} {\bibinfo {author} {\bibfnamefont {S.}~\bibnamefont {Majidy}}, \bibinfo {author} {\bibfnamefont {U.}~\bibnamefont {Agrawal}}, \bibinfo {author} {\bibfnamefont {S.}~\bibnamefont {Gopalakrishnan}}, \bibinfo {author} {\bibfnamefont {A.~C.}\ \bibnamefont {Potter}}, \bibinfo {author} {\bibfnamefont {R.}~\bibnamefont {Vasseur}},\ and\ \bibinfo {author} {\bibfnamefont {N.}~\bibnamefont {Yunger~Halpern}},\ }\bibfield  {title} {\bibinfo {title} {{Critical Phase and Spin Sharpening in SU(2)-Symmetric Monitored Quantum Circuits}},\ }\href {https://doi.org/10.1103/PhysRevB.108.054307} {\bibfield  {journal} {\bibinfo  {journal} {Phys. Rev. B}\ }\textbf {\bibinfo {volume} {108}},\ \bibinfo {pages} {054307} (\bibinfo {year} {2023})}\BibitemShut {NoStop}%
\bibitem [{\citenamefont {Feng}\ \emph {et~al.}(2025)\citenamefont {Feng}, \citenamefont {Fishchenko}, \citenamefont {Gopalakrishnan},\ and\ \citenamefont {Ippoliti}}]{Feng25chargespinsharpening}%
  \BibitemOpen
  \bibfield  {author} {\bibinfo {author} {\bibfnamefont {X.}~\bibnamefont {Feng}}, \bibinfo {author} {\bibfnamefont {N.}~\bibnamefont {Fishchenko}}, \bibinfo {author} {\bibfnamefont {S.}~\bibnamefont {Gopalakrishnan}},\ and\ \bibinfo {author} {\bibfnamefont {M.}~\bibnamefont {Ippoliti}},\ }\bibfield  {title} {\bibinfo {title} {{Charge and Spin Sharpening Transitions on Dynamical Quantum Trees}},\ }\href {https://doi.org/10.22331/q-2025-04-07-1692} {\bibfield  {journal} {\bibinfo  {journal} {Quantum}\ }\textbf {\bibinfo {volume} {9}},\ \bibinfo {pages} {1692} (\bibinfo {year} {2025})}\BibitemShut {NoStop}%
\bibitem [{\citenamefont {Ma}\ and\ \citenamefont {Vasseur}(2026)}]{Ma26nonabelianlearnability}%
  \BibitemOpen
  \bibfield  {author} {\bibinfo {author} {\bibfnamefont {R.}~\bibnamefont {Ma}}\ and\ \bibinfo {author} {\bibfnamefont {R.}~\bibnamefont {Vasseur}},\ }\href@noop {} {\bibinfo {title} {{Statistical Mechanics of Non-Abelian Learnability Transitions}}} (\bibinfo {year} {2026}),\ \Eprint {https://arxiv.org/abs/2608.19325} {arXiv:2608.19325 [quant-ph]} \BibitemShut {NoStop}%
\bibitem [{\citenamefont {Zabalo}\ \emph {et~al.}(2022)\citenamefont {Zabalo}, \citenamefont {Gullans}, \citenamefont {Wilson}, \citenamefont {Vasseur}, \citenamefont {Ludwig}, \citenamefont {Gopalakrishnan}, \citenamefont {Huse},\ and\ \citenamefont {Pixley}}]{Pixley22MIPTCFT}%
  \BibitemOpen
  \bibfield  {author} {\bibinfo {author} {\bibfnamefont {A.}~\bibnamefont {Zabalo}}, \bibinfo {author} {\bibfnamefont {M.~J.}\ \bibnamefont {Gullans}}, \bibinfo {author} {\bibfnamefont {J.~H.}\ \bibnamefont {Wilson}}, \bibinfo {author} {\bibfnamefont {R.}~\bibnamefont {Vasseur}}, \bibinfo {author} {\bibfnamefont {A.~W.~W.}\ \bibnamefont {Ludwig}}, \bibinfo {author} {\bibfnamefont {S.}~\bibnamefont {Gopalakrishnan}}, \bibinfo {author} {\bibfnamefont {D.~A.}\ \bibnamefont {Huse}},\ and\ \bibinfo {author} {\bibfnamefont {J.~H.}\ \bibnamefont {Pixley}},\ }\bibfield  {title} {\bibinfo {title} {{Operator Scaling Dimensions and Multifractality at Measurement-Induced Transitions}},\ }\href {https://doi.org/10.1103/PhysRevLett.128.050602} {\bibfield  {journal} {\bibinfo  {journal} {Phys. Rev. Lett.}\ }\textbf {\bibinfo {volume} {128}},\ \bibinfo {pages} {050602} (\bibinfo {year} {2022})}\BibitemShut {NoStop}%
\bibitem [{\citenamefont {Zabalo}\ \emph {et~al.}(2020)\citenamefont {Zabalo}, \citenamefont {Gullans}, \citenamefont {Wilson}, \citenamefont {Gopalakrishnan}, \citenamefont {Huse},\ and\ \citenamefont {Pixley}}]{Zabalo_2020}%
  \BibitemOpen
  \bibfield  {author} {\bibinfo {author} {\bibfnamefont {A.}~\bibnamefont {Zabalo}}, \bibinfo {author} {\bibfnamefont {M.~J.}\ \bibnamefont {Gullans}}, \bibinfo {author} {\bibfnamefont {J.~H.}\ \bibnamefont {Wilson}}, \bibinfo {author} {\bibfnamefont {S.}~\bibnamefont {Gopalakrishnan}}, \bibinfo {author} {\bibfnamefont {D.~A.}\ \bibnamefont {Huse}},\ and\ \bibinfo {author} {\bibfnamefont {J.~H.}\ \bibnamefont {Pixley}},\ }\bibfield  {title} {\bibinfo {title} {{Critical properties of the measurement-induced transition in random quantum circuits}},\ }\href {https://doi.org/10.1103/PhysRevB.101.060301} {\bibfield  {journal} {\bibinfo  {journal} {Phys. Rev. B}\ }\textbf {\bibinfo {volume} {101}},\ \bibinfo {pages} {060301(R)} (\bibinfo {year} {2020})}\BibitemShut {NoStop}%
\bibitem [{\citenamefont {Gullans}\ and\ \citenamefont {Huse}(2020{\natexlab{b}})}]{Gullans20scalabledecoder}%
  \BibitemOpen
  \bibfield  {author} {\bibinfo {author} {\bibfnamefont {M.~J.}\ \bibnamefont {Gullans}}\ and\ \bibinfo {author} {\bibfnamefont {D.~A.}\ \bibnamefont {Huse}},\ }\bibfield  {title} {\bibinfo {title} {{Scalable Probes of Measurement-Induced Criticality}},\ }\href {https://doi.org/10.1103/PhysRevLett.125.070606} {\bibfield  {journal} {\bibinfo  {journal} {Phys. Rev. Lett.}\ }\textbf {\bibinfo {volume} {125}},\ \bibinfo {pages} {070606} (\bibinfo {year} {2020}{\natexlab{b}})}\BibitemShut {NoStop}%
\bibitem [{\citenamefont {Ludwig}(1990)}]{Ludwig90hierarchies}%
  \BibitemOpen
  \bibfield  {author} {\bibinfo {author} {\bibfnamefont {A.~W.~W.}\ \bibnamefont {Ludwig}},\ }\bibfield  {title} {\bibinfo {title} {{Infinite hierarchies of exponents in a diluted ferromagnet and their interpretation}},\ }\href {https://doi.org/10.1016/0550-3213(90)90126-X} {\bibfield  {journal} {\bibinfo  {journal} {Nucl. Phys. B}\ }\textbf {\bibinfo {volume} {330}},\ \bibinfo {pages} {639} (\bibinfo {year} {1990})}\BibitemShut {NoStop}%
\bibitem [{\citenamefont {Li}\ \emph {et~al.}(2024)\citenamefont {Li}, \citenamefont {Vasseur}, \citenamefont {Fisher},\ and\ \citenamefont {Ludwig}}]{Ludwig24stat}%
  \BibitemOpen
  \bibfield  {author} {\bibinfo {author} {\bibfnamefont {Y.}~\bibnamefont {Li}}, \bibinfo {author} {\bibfnamefont {R.}~\bibnamefont {Vasseur}}, \bibinfo {author} {\bibfnamefont {M.~P.~A.}\ \bibnamefont {Fisher}},\ and\ \bibinfo {author} {\bibfnamefont {A.~W.~W.}\ \bibnamefont {Ludwig}},\ }\bibfield  {title} {\bibinfo {title} {{Statistical mechanics model for Clifford random tensor networks and monitored quantum circuits}},\ }\href {https://doi.org/10.1103/PhysRevB.109.174307} {\bibfield  {journal} {\bibinfo  {journal} {Phys. Rev. B}\ }\textbf {\bibinfo {volume} {109}},\ \bibinfo {pages} {174307} (\bibinfo {year} {2024})}\BibitemShut {NoStop}%
\bibitem [{\citenamefont {{Patil}}\ and\ \citenamefont {{Ludwig}}(2024)}]{Ludwig24tricriticalising}%
  \BibitemOpen
  \bibfield  {author} {\bibinfo {author} {\bibfnamefont {R.~A.}\ \bibnamefont {{Patil}}}\ and\ \bibinfo {author} {\bibfnamefont {A.~W.~W.}\ \bibnamefont {{Ludwig}}},\ }\href@noop {} {\bibinfo {title} {{Highly complex novel critical behavior from the intrinsic randomness of quantum mechanical measurements on critical ground states -- a controlled renormalization group analysis}}} (\bibinfo {year} {2024}),\ \Eprint {https://arxiv.org/abs/2409.02107} {arXiv:2409.02107 [cond-mat.stat-mech]} \BibitemShut {NoStop}%
\bibitem [{\citenamefont {Vasseur}\ \emph {et~al.}(2019)\citenamefont {Vasseur}, \citenamefont {Potter}, \citenamefont {You},\ and\ \citenamefont {Ludwig}}]{Vasseur2019}%
  \BibitemOpen
  \bibfield  {author} {\bibinfo {author} {\bibfnamefont {R.}~\bibnamefont {Vasseur}}, \bibinfo {author} {\bibfnamefont {A.~C.}\ \bibnamefont {Potter}}, \bibinfo {author} {\bibfnamefont {Y.-Z.}\ \bibnamefont {You}},\ and\ \bibinfo {author} {\bibfnamefont {A.~W.~W.}\ \bibnamefont {Ludwig}},\ }\bibfield  {title} {\bibinfo {title} {{Entanglement transitions from holographic random tensor networks}},\ }\href {https://doi.org/10.1103/PhysRevB.100.134203} {\bibfield  {journal} {\bibinfo  {journal} {Phys. Rev. B}\ }\textbf {\bibinfo {volume} {100}},\ \bibinfo {pages} {134203} (\bibinfo {year} {2019})}\BibitemShut {NoStop}%
\bibitem [{\citenamefont {Wang}\ \emph {et~al.}(2025)\citenamefont {Wang}, \citenamefont {Vasseur}, \citenamefont {Trebst}, \citenamefont {Ludwig},\ and\ \citenamefont {Zhu}}]{Wang25selfdual}%
  \BibitemOpen
  \bibfield  {author} {\bibinfo {author} {\bibfnamefont {Q.}~\bibnamefont {Wang}}, \bibinfo {author} {\bibfnamefont {R.}~\bibnamefont {Vasseur}}, \bibinfo {author} {\bibfnamefont {S.}~\bibnamefont {Trebst}}, \bibinfo {author} {\bibfnamefont {A.~W.~W.}\ \bibnamefont {Ludwig}},\ and\ \bibinfo {author} {\bibfnamefont {G.-Y.}\ \bibnamefont {Zhu}},\ }\href@noop {} {\bibinfo {title} {{Decoherence-induced self-dual criticality in topological states of matter}}} (\bibinfo {year} {2025}),\ \Eprint {https://arxiv.org/abs/2502.14034} {arXiv:2502.14034 [quant-ph]} \BibitemShut {NoStop}%
\bibitem [{Note2()}]{Note2}%
  \BibitemOpen
  \bibinfo {note} {In agreement with independent work by Fava et al.~\cite {Nahum23nlsmFreeFermion}.}\BibitemShut {Stop}%
\bibitem [{\citenamefont {Eckstein}\ \emph {et~al.}(2024)\citenamefont {Eckstein}, \citenamefont {Han}, \citenamefont {Trebst},\ and\ \citenamefont {Zhu}}]{Eckstein2024}%
  \BibitemOpen
  \bibfield  {author} {\bibinfo {author} {\bibfnamefont {F.}~\bibnamefont {Eckstein}}, \bibinfo {author} {\bibfnamefont {B.}~\bibnamefont {Han}}, \bibinfo {author} {\bibfnamefont {S.}~\bibnamefont {Trebst}},\ and\ \bibinfo {author} {\bibfnamefont {G.-Y.}\ \bibnamefont {Zhu}},\ }\bibfield  {title} {\bibinfo {title} {{Robust Teleportation of a Surface Code and Cascade of Topological Quantum Phase Transitions}},\ }\href {https://doi.org/10.1103/PRXQuantum.5.040313} {\bibfield  {journal} {\bibinfo  {journal} {PRX Quantum}\ }\textbf {\bibinfo {volume} {5}},\ \bibinfo {pages} {040313} (\bibinfo {year} {2024})}\BibitemShut {NoStop}%
\bibitem [{\citenamefont {Eckstein}\ \emph {et~al.}(2025)\citenamefont {Eckstein}, \citenamefont {Han}, \citenamefont {Trebst},\ and\ \citenamefont {Zhu}}]{Eckstein2025}%
  \BibitemOpen
  \bibfield  {author} {\bibinfo {author} {\bibfnamefont {F.}~\bibnamefont {Eckstein}}, \bibinfo {author} {\bibfnamefont {B.}~\bibnamefont {Han}}, \bibinfo {author} {\bibfnamefont {S.}~\bibnamefont {Trebst}},\ and\ \bibinfo {author} {\bibfnamefont {G.-Y.}\ \bibnamefont {Zhu}},\ }\href@noop {} {\bibinfo {title} {{Learning transitions of topological surface codes}}} (\bibinfo {year} {2025}),\ \Eprint {https://arxiv.org/abs/2512.19786} {arXiv:2512.19786 [quant-ph]} \BibitemShut {NoStop}%
\bibitem [{\citenamefont {Behrends}\ \emph {et~al.}(2024)\citenamefont {Behrends}, \citenamefont {Venn},\ and\ \citenamefont {B{\'e}ri}}]{Beri24coherenterror}%
  \BibitemOpen
  \bibfield  {author} {\bibinfo {author} {\bibfnamefont {J.}~\bibnamefont {Behrends}}, \bibinfo {author} {\bibfnamefont {F.}~\bibnamefont {Venn}},\ and\ \bibinfo {author} {\bibfnamefont {B.}~\bibnamefont {B{\'e}ri}},\ }\bibfield  {title} {\bibinfo {title} {{Surface codes, quantum circuits, and entanglement phases}},\ }\href {https://doi.org/10.1103/PhysRevResearch.6.013137} {\bibfield  {journal} {\bibinfo  {journal} {Phys. Rev. Res.}\ }\textbf {\bibinfo {volume} {6}},\ \bibinfo {pages} {013137} (\bibinfo {year} {2024})}\BibitemShut {NoStop}%
\bibitem [{\citenamefont {{Yan}}\ \emph {et~al.}(2026)\citenamefont {{Yan}}, \citenamefont {{Bao}},\ and\ \citenamefont {{Vijay}}}]{Yan2026sigma}%
  \BibitemOpen
  \bibfield  {author} {\bibinfo {author} {\bibfnamefont {S.~W.}\ \bibnamefont {{Yan}}}, \bibinfo {author} {\bibfnamefont {Y.}~\bibnamefont {{Bao}}},\ and\ \bibinfo {author} {\bibfnamefont {S.}~\bibnamefont {{Vijay}}},\ }\href@noop {} {\bibinfo {title} {{Non-linear Sigma Model for the Surface Code with Coherent Errors}}} (\bibinfo {year} {2026}),\ \Eprint {https://arxiv.org/abs/2603.25665} {arXiv:2603.25665 [cond-mat.stat-mech]} \BibitemShut {NoStop}%
\bibitem [{\citenamefont {{Yang}}\ \emph {et~al.}(2026)\citenamefont {{Yang}}, \citenamefont {{Ludwig}},\ and\ \citenamefont {{Jian}}}]{Yang2026decoding}%
  \BibitemOpen
  \bibfield  {author} {\bibinfo {author} {\bibfnamefont {Z.}~\bibnamefont {{Yang}}}, \bibinfo {author} {\bibfnamefont {A.~W.~W.}\ \bibnamefont {{Ludwig}}},\ and\ \bibinfo {author} {\bibfnamefont {C.-M.}\ \bibnamefont {{Jian}}},\ }\href@noop {} {\bibinfo {title} {{Decoding coherent errors in toric codes on honeycomb and square lattices: duality to Majorana monitored dynamics and symmetry classes}}} (\bibinfo {year} {2026}),\ \Eprint {https://arxiv.org/abs/2604.08650} {arXiv:2604.08650 [cond-mat.stat-mech]} \BibitemShut {NoStop}%
\bibitem [{\citenamefont {Negari}\ \emph {et~al.}(2024)\citenamefont {Negari}, \citenamefont {Sahu},\ and\ \citenamefont {Hsieh}}]{Negari2024}%
  \BibitemOpen
  \bibfield  {author} {\bibinfo {author} {\bibfnamefont {A.-R.}\ \bibnamefont {Negari}}, \bibinfo {author} {\bibfnamefont {S.}~\bibnamefont {Sahu}},\ and\ \bibinfo {author} {\bibfnamefont {T.~H.}\ \bibnamefont {Hsieh}},\ }\bibfield  {title} {\bibinfo {title} {{Measurement-induced phase transitions in the toric code}},\ }\href {https://doi.org/10.1103/PhysRevB.109.125148} {\bibfield  {journal} {\bibinfo  {journal} {Phys. Rev. B}\ }\textbf {\bibinfo {volume} {109}},\ \bibinfo {pages} {125148} (\bibinfo {year} {2024})}\BibitemShut {NoStop}%
\bibitem [{\citenamefont {P\"utz}\ \emph {et~al.}(2025)\citenamefont {P\"utz}, \citenamefont {Vasseur}, \citenamefont {Ludwig}, \citenamefont {Trebst},\ and\ \citenamefont {Zhu}}]{Putz25nishimori}%
  \BibitemOpen
  \bibfield  {author} {\bibinfo {author} {\bibfnamefont {M.}~\bibnamefont {P\"utz}}, \bibinfo {author} {\bibfnamefont {R.}~\bibnamefont {Vasseur}}, \bibinfo {author} {\bibfnamefont {A.~W.~W.}\ \bibnamefont {Ludwig}}, \bibinfo {author} {\bibfnamefont {S.}~\bibnamefont {Trebst}},\ and\ \bibinfo {author} {\bibfnamefont {G.-Y.}\ \bibnamefont {Zhu}},\ }\bibfield  {title} {\bibinfo {title} {{Flow to Nishimori Universality in Weakly Monitored Quantum Circuits with Qubit Loss}},\ }\href {https://doi.org/10.1103/ygfz-crvp} {\bibfield  {journal} {\bibinfo  {journal} {PRX Quantum}\ }\textbf {\bibinfo {volume} {6}},\ \bibinfo {pages} {040372} (\bibinfo {year} {2025})}\BibitemShut {NoStop}%
\bibitem [{\citenamefont {Lee}\ and\ \citenamefont {Yoshida}(2026)}]{Yoshida2026}%
  \BibitemOpen
  \bibfield  {author} {\bibinfo {author} {\bibfnamefont {D.}~\bibnamefont {Lee}}\ and\ \bibinfo {author} {\bibfnamefont {B.}~\bibnamefont {Yoshida}},\ }\bibfield  {title} {\bibinfo {title} {{Randomly Monitored Quantum Codes}},\ }\href {https://doi.org/10.1103/6s65-ylvw} {\bibfield  {journal} {\bibinfo  {journal} {PRX Quantum}\ }\textbf {\bibinfo {volume} {7}},\ \bibinfo {pages} {020352} (\bibinfo {year} {2026})}\BibitemShut {NoStop}%
\bibitem [{\citenamefont {P\"utz}\ \emph {et~al.}(2026)\citenamefont {P\"utz}, \citenamefont {Garratt}, \citenamefont {Nishimori}, \citenamefont {Trebst},\ and\ \citenamefont {Zhu}}]{Learning_transition_guoyi}%
  \BibitemOpen
  \bibfield  {author} {\bibinfo {author} {\bibfnamefont {M.}~\bibnamefont {P\"utz}}, \bibinfo {author} {\bibfnamefont {S.~J.}\ \bibnamefont {Garratt}}, \bibinfo {author} {\bibfnamefont {H.}~\bibnamefont {Nishimori}}, \bibinfo {author} {\bibfnamefont {S.}~\bibnamefont {Trebst}},\ and\ \bibinfo {author} {\bibfnamefont {G.-Y.}\ \bibnamefont {Zhu}},\ }\bibfield  {title} {\bibinfo {title} {{Learning Transitions in Classical Ising Models and Deformed Toric Codes}},\ }\href {https://doi.org/10.1103/4dwm-kn11} {\bibfield  {journal} {\bibinfo  {journal} {Phys. Rev. Lett.}\ }\textbf {\bibinfo {volume} {136}},\ \bibinfo {pages} {190402} (\bibinfo {year} {2026})}\BibitemShut {NoStop}%
\bibitem [{\citenamefont {Patil}\ \emph {et~al.}(2026{\natexlab{a}})\citenamefont {Patil}, \citenamefont {Pütz}, \citenamefont {Trebst}, \citenamefont {Zhu},\ and\ \citenamefont {Ludwig}}]{patil2026highernishimoricriticalityexact}%
  \BibitemOpen
  \bibfield  {author} {\bibinfo {author} {\bibfnamefont {R.~A.}\ \bibnamefont {Patil}}, \bibinfo {author} {\bibfnamefont {M.}~\bibnamefont {Pütz}}, \bibinfo {author} {\bibfnamefont {S.}~\bibnamefont {Trebst}}, \bibinfo {author} {\bibfnamefont {G.-Y.}\ \bibnamefont {Zhu}},\ and\ \bibinfo {author} {\bibfnamefont {A.~W.~W.}\ \bibnamefont {Ludwig}},\ }\href@noop {} {\bibinfo {title} {{Higher Nishimori Criticality and Exact Results at the Learning Transition of Deformed Toric Codes}}} (\bibinfo {year} {2026}{\natexlab{a}}),\ \Eprint {https://arxiv.org/abs/2604.06324} {arXiv:2604.06324 [cond-mat.stat-mech]} \BibitemShut {NoStop}%
\bibitem [{\citenamefont {Patil}\ \emph {et~al.}(2026{\natexlab{b}})\citenamefont {Patil}, \citenamefont {Pütz}, \citenamefont {Mukherjee}, \citenamefont {Zhu}, \citenamefont {Trebst},\ and\ \citenamefont {Ludwig}}]{patil2026pottsZq}%
  \BibitemOpen
  \bibfield  {author} {\bibinfo {author} {\bibfnamefont {R.~A.}\ \bibnamefont {Patil}}, \bibinfo {author} {\bibfnamefont {M.}~\bibnamefont {Pütz}}, \bibinfo {author} {\bibfnamefont {R.}~\bibnamefont {Mukherjee}}, \bibinfo {author} {\bibfnamefont {G.-Y.}\ \bibnamefont {Zhu}}, \bibinfo {author} {\bibfnamefont {S.}~\bibnamefont {Trebst}},\ and\ \bibinfo {author} {\bibfnamefont {A.~W.~W.}\ \bibnamefont {Ludwig}},\ }\href@noop {} {\bibinfo {title} {{Learning Potts Models and $Z_3$ Toric Codes: Higher and Ordinary Nishimori Criticality}}} (\bibinfo {year} {2026}{\natexlab{b}}),\ \Eprint {https://arxiv.org/abs/2608.20268} {arXiv:2608.20268 [cond-mat.stat-mech]} \BibitemShut {NoStop}%
\bibitem [{\citenamefont {Mukherjee}\ and\ \citenamefont {Trebst}(2026)}]{Mukherjee26zqthreshold}%
  \BibitemOpen
  \bibfield  {author} {\bibinfo {author} {\bibfnamefont {R.}~\bibnamefont {Mukherjee}}\ and\ \bibinfo {author} {\bibfnamefont {S.}~\bibnamefont {Trebst}},\ }\href@noop {} {\bibinfo {title} {{Nishimori Threshold Estimation for Bayesian Inference and $\mathbb{Z}_q$ Surface Code Decoding}}} (\bibinfo {year} {2026}),\ \Eprint {https://arxiv.org/abs/2607.18374} {arXiv:2607.18374 [quant-ph]} \BibitemShut {NoStop}%
\bibitem [{\citenamefont {Vijay}\ and\ \citenamefont {Lee}(2025)}]{VijayLee26infoCritical}%
  \BibitemOpen
  \bibfield  {author} {\bibinfo {author} {\bibfnamefont {A.}~\bibnamefont {Vijay}}\ and\ \bibinfo {author} {\bibfnamefont {J.~Y.}\ \bibnamefont {Lee}},\ }\href@noop {} {\bibinfo {title} {{Information Critical Phases under Decoherence}}} (\bibinfo {year} {2025}),\ \Eprint {https://arxiv.org/abs/2512.22121} {arXiv:2512.22121 [quant-ph]} \BibitemShut {NoStop}%
\bibitem [{\citenamefont {Isakov}\ \emph {et~al.}(2011)\citenamefont {Isakov}, \citenamefont {Fendley}, \citenamefont {Ludwig}, \citenamefont {Trebst},\ and\ \citenamefont {Troyer}}]{Isakov_2011}%
  \BibitemOpen
  \bibfield  {author} {\bibinfo {author} {\bibfnamefont {S.~V.}\ \bibnamefont {Isakov}}, \bibinfo {author} {\bibfnamefont {P.}~\bibnamefont {Fendley}}, \bibinfo {author} {\bibfnamefont {A.~W.~W.}\ \bibnamefont {Ludwig}}, \bibinfo {author} {\bibfnamefont {S.}~\bibnamefont {Trebst}},\ and\ \bibinfo {author} {\bibfnamefont {M.}~\bibnamefont {Troyer}},\ }\bibfield  {title} {\bibinfo {title} {{Dynamics at and near conformal quantum critical points}},\ }\href {https://doi.org/10.1103/PhysRevB.83.125114} {\bibfield  {journal} {\bibinfo  {journal} {Phys. Rev. B}\ }\textbf {\bibinfo {volume} {83}},\ \bibinfo {pages} {125114} (\bibinfo {year} {2011})}\BibitemShut {NoStop}%
\bibitem [{\citenamefont {{Castelnovo}}\ \emph {et~al.}(2010)\citenamefont {{Castelnovo}}, \citenamefont {{Trebst}},\ and\ \citenamefont {{Troyer}}}]{Troyer10topocrit}%
  \BibitemOpen
  \bibfield  {author} {\bibinfo {author} {\bibfnamefont {C.}~\bibnamefont {{Castelnovo}}}, \bibinfo {author} {\bibfnamefont {S.}~\bibnamefont {{Trebst}}},\ and\ \bibinfo {author} {\bibfnamefont {M.}~\bibnamefont {{Troyer}}},\ }\bibfield  {title} {\bibinfo {title} {{Topological Order and Quantum Criticality}},\ }in\ \href {https://doi.org/10.1201/b10273-14} {\emph {\bibinfo {booktitle} {Understanding Quantum Phase Transitions}}}\ (\bibinfo  {publisher} {Taylor \& Francis},\ \bibinfo {year} {2010})\ pp.\ \bibinfo {pages} {167--192},\ \Eprint {https://arxiv.org/abs/0912.3272} {arXiv:0912.3272} \BibitemShut {NoStop}%
\bibitem [{\citenamefont {Zhu}\ and\ \citenamefont {Zhang}(2019)}]{Zhu19}%
  \BibitemOpen
  \bibfield  {author} {\bibinfo {author} {\bibfnamefont {G.-Y.}\ \bibnamefont {Zhu}}\ and\ \bibinfo {author} {\bibfnamefont {G.-M.}\ \bibnamefont {Zhang}},\ }\bibfield  {title} {\bibinfo {title} {{Gapless Coulomb State Emerging from a Self-Dual Topological Tensor-Network State}},\ }\href {https://doi.org/10.1103/PhysRevLett.122.176401} {\bibfield  {journal} {\bibinfo  {journal} {Phys. Rev. Lett.}\ }\textbf {\bibinfo {volume} {122}},\ \bibinfo {pages} {176401} (\bibinfo {year} {2019})}\BibitemShut {NoStop}%
\bibitem [{\citenamefont {Xu}\ \emph {et~al.}(2020)\citenamefont {Xu}, \citenamefont {Zhang},\ and\ \citenamefont {Zhang}}]{Zhang20deformFib}%
  \BibitemOpen
  \bibfield  {author} {\bibinfo {author} {\bibfnamefont {W.-T.}\ \bibnamefont {Xu}}, \bibinfo {author} {\bibfnamefont {Q.}~\bibnamefont {Zhang}},\ and\ \bibinfo {author} {\bibfnamefont {G.-M.}\ \bibnamefont {Zhang}},\ }\bibfield  {title} {\bibinfo {title} {Tensor network approach to phase transitions of a non-abelian topological phase},\ }\href {https://doi.org/10.1103/PhysRevLett.124.130603} {\bibfield  {journal} {\bibinfo  {journal} {Phys. Rev. Lett.}\ }\textbf {\bibinfo {volume} {124}},\ \bibinfo {pages} {130603} (\bibinfo {year} {2020})}\BibitemShut {NoStop}%
\bibitem [{\citenamefont {Zhi}\ \emph {et~al.}(2026)\citenamefont {Zhi}, \citenamefont {Han}, \citenamefont {Vasseur}, \citenamefont {Ludwig}, \citenamefont {Trebst},\ and\ \citenamefont {Zhu}}]{zenodo}%
  \BibitemOpen
  \bibfield  {author} {\bibinfo {author} {\bibfnamefont {Z.}~\bibnamefont {Zhi}}, \bibinfo {author} {\bibfnamefont {B.}~\bibnamefont {Han}}, \bibinfo {author} {\bibfnamefont {R.}~\bibnamefont {Vasseur}}, \bibinfo {author} {\bibfnamefont {A.~W.~W.}\ \bibnamefont {Ludwig}}, \bibinfo {author} {\bibfnamefont {S.}~\bibnamefont {Trebst}},\ and\ \bibinfo {author} {\bibfnamefont {G.-Y.}\ \bibnamefont {Zhu}},\ }\bibfield  {title} {\bibinfo {title} {{Data for ``Measurement-induced criticality for non-Abelian anyons: The monitored golden chain"}},\ }\bibfield  {journal} {\bibinfo  {journal} {Zenodo}\ }\href {https://doi.org/10.5281/zenodo.22842959} {10.5281/zenodo.22842959} (\bibinfo {year} {2026})\BibitemShut {NoStop}%
\bibitem [{\citenamefont {Fishman}\ \emph {et~al.}(2022)\citenamefont {Fishman}, \citenamefont {White},\ and\ \citenamefont {Stoudenmire}}]{itensor}%
  \BibitemOpen
  \bibfield  {author} {\bibinfo {author} {\bibfnamefont {M.}~\bibnamefont {Fishman}}, \bibinfo {author} {\bibfnamefont {S.~R.}\ \bibnamefont {White}},\ and\ \bibinfo {author} {\bibfnamefont {E.~M.}\ \bibnamefont {Stoudenmire}},\ }\bibfield  {title} {\bibinfo {title} {{The \mbox{ITensor} Software Library for Tensor Network Calculations}},\ }\href {https://doi.org/10.21468/SciPostPhysCodeb.4} {\bibfield  {journal} {\bibinfo  {journal} {SciPost Phys. Codebases}\ ,\ \bibinfo {pages} {4}} (\bibinfo {year} {2022})}\BibitemShut {NoStop}%
\bibitem [{\citenamefont {Nahum}\ and\ \citenamefont {Skinner}(2020)}]{Nahum20freefermion}%
  \BibitemOpen
  \bibfield  {author} {\bibinfo {author} {\bibfnamefont {A.}~\bibnamefont {Nahum}}\ and\ \bibinfo {author} {\bibfnamefont {B.}~\bibnamefont {Skinner}},\ }\bibfield  {title} {\bibinfo {title} {{Entanglement and dynamics of diffusion-annihilation processes with Majorana defects}},\ }\href {https://doi.org/10.1103/PhysRevResearch.2.023288} {\bibfield  {journal} {\bibinfo  {journal} {Phys. Rev. Res.}\ }\textbf {\bibinfo {volume} {2}},\ \bibinfo {pages} {023288} (\bibinfo {year} {2020})}\BibitemShut {NoStop}%
\bibitem [{\citenamefont {Lang}\ and\ \citenamefont {B\"uchler}(2020)}]{Lang_2020}%
  \BibitemOpen
  \bibfield  {author} {\bibinfo {author} {\bibfnamefont {N.}~\bibnamefont {Lang}}\ and\ \bibinfo {author} {\bibfnamefont {H.~P.}\ \bibnamefont {B\"uchler}},\ }\bibfield  {title} {\bibinfo {title} {{Entanglement transition in the projective transverse field Ising model}},\ }\href {https://doi.org/10.1103/PhysRevB.102.094204} {\bibfield  {journal} {\bibinfo  {journal} {Phys. Rev. B}\ }\textbf {\bibinfo {volume} {102}},\ \bibinfo {pages} {094204} (\bibinfo {year} {2020})}\BibitemShut {NoStop}%
\bibitem [{\citenamefont {Kumar}\ \emph {et~al.}(2024)\citenamefont {Kumar}, \citenamefont {Aziz}, \citenamefont {Chakraborty}, \citenamefont {Ludwig}, \citenamefont {Gopalakrishnan}, \citenamefont {Pixley},\ and\ \citenamefont {Vasseur}}]{Vasseur24transfer}%
  \BibitemOpen
  \bibfield  {author} {\bibinfo {author} {\bibfnamefont {A.}~\bibnamefont {Kumar}}, \bibinfo {author} {\bibfnamefont {K.}~\bibnamefont {Aziz}}, \bibinfo {author} {\bibfnamefont {A.}~\bibnamefont {Chakraborty}}, \bibinfo {author} {\bibfnamefont {A.~W.~W.}\ \bibnamefont {Ludwig}}, \bibinfo {author} {\bibfnamefont {S.}~\bibnamefont {Gopalakrishnan}}, \bibinfo {author} {\bibfnamefont {J.~H.}\ \bibnamefont {Pixley}},\ and\ \bibinfo {author} {\bibfnamefont {R.}~\bibnamefont {Vasseur}},\ }\bibfield  {title} {\bibinfo {title} {{Boundary transfer matrix spectrum of measurement-induced transitions}},\ }\href {https://doi.org/10.1103/PhysRevB.109.014303} {\bibfield  {journal} {\bibinfo  {journal} {Phys. Rev. B}\ }\textbf {\bibinfo {volume} {109}},\ \bibinfo {pages} {014303} (\bibinfo {year} {2024})}\BibitemShut {NoStop}%
\bibitem [{\citenamefont {Fidkowski}\ \emph {et~al.}(2009)\citenamefont {Fidkowski}, \citenamefont {Freedman}, \citenamefont {Nayak}, \citenamefont {Walker},\ and\ \citenamefont {Wang}}]{Fidkowski09nonabelions}%
  \BibitemOpen
  \bibfield  {author} {\bibinfo {author} {\bibfnamefont {L.}~\bibnamefont {Fidkowski}}, \bibinfo {author} {\bibfnamefont {M.}~\bibnamefont {Freedman}}, \bibinfo {author} {\bibfnamefont {C.}~\bibnamefont {Nayak}}, \bibinfo {author} {\bibfnamefont {K.}~\bibnamefont {Walker}},\ and\ \bibinfo {author} {\bibfnamefont {Z.}~\bibnamefont {Wang}},\ }\bibfield  {title} {\bibinfo {title} {{From String Nets to Nonabelions}},\ }\href {https://doi.org/10.1007/s00220-009-0757-9} {\bibfield  {journal} {\bibinfo  {journal} {Communications in Mathematical Physics}\ }\textbf {\bibinfo {volume} {287}},\ \bibinfo {pages} {805} (\bibinfo {year} {2009})}\BibitemShut {NoStop}%
\bibitem [{\citenamefont {Fendley}(2008)}]{Fendley08quantumloopsnets}%
  \BibitemOpen
  \bibfield  {author} {\bibinfo {author} {\bibfnamefont {P.}~\bibnamefont {Fendley}},\ }\bibfield  {title} {\bibinfo {title} {{Topological Order from Quantum Loops and Nets}},\ }\href {https://doi.org/10.1016/j.aop.2008.04.011} {\bibfield  {journal} {\bibinfo  {journal} {Annals of Physics}\ }\textbf {\bibinfo {volume} {323}},\ \bibinfo {pages} {3113} (\bibinfo {year} {2008})}\BibitemShut {NoStop}%
\bibitem [{\citenamefont {Lin}\ and\ \citenamefont {Zou}(2021)}]{Zou21fibonacci}%
  \BibitemOpen
  \bibfield  {author} {\bibinfo {author} {\bibfnamefont {C.-J.}\ \bibnamefont {Lin}}\ and\ \bibinfo {author} {\bibfnamefont {L.}~\bibnamefont {Zou}},\ }\bibfield  {title} {\bibinfo {title} {{Reaction-diffusion dynamics in a Fibonacci chain: Interplay between classical and quantum behavior}},\ }\href {https://doi.org/10.1103/PhysRevB.103.174305} {\bibfield  {journal} {\bibinfo  {journal} {Phys. Rev. B}\ }\textbf {\bibinfo {volume} {103}},\ \bibinfo {pages} {174305} (\bibinfo {year} {2021})}\BibitemShut {NoStop}%
\bibitem [{\citenamefont {Fava}\ \emph {et~al.}(2023)\citenamefont {Fava}, \citenamefont {Piroli}, \citenamefont {Swann}, \citenamefont {Bernard},\ and\ \citenamefont {Nahum}}]{Nahum23nlsmFreeFermion}%
  \BibitemOpen
  \bibfield  {author} {\bibinfo {author} {\bibfnamefont {M.}~\bibnamefont {Fava}}, \bibinfo {author} {\bibfnamefont {L.}~\bibnamefont {Piroli}}, \bibinfo {author} {\bibfnamefont {T.}~\bibnamefont {Swann}}, \bibinfo {author} {\bibfnamefont {D.}~\bibnamefont {Bernard}},\ and\ \bibinfo {author} {\bibfnamefont {A.}~\bibnamefont {Nahum}},\ }\bibfield  {title} {\bibinfo {title} {{Nonlinear Sigma Models for Monitored Dynamics of Free Fermions}},\ }\href {https://doi.org/10.1103/PhysRevX.13.041045} {\bibfield  {journal} {\bibinfo  {journal} {Phys. Rev. X}\ }\textbf {\bibinfo {volume} {13}},\ \bibinfo {pages} {041045} (\bibinfo {year} {2023})}\BibitemShut {NoStop}%
\end{thebibliography}%

\clearpage
\onecolumngrid
\begin{center}
    {\large\textbf{Supplemental Material for ``Measurement-induced criticality for non-Abelian anyons: The monitored golden chain''}}\\[1em]
    Zhaohui Zhi,$^1$ Bo Han,$^2$ Romain Vasseur,$^{3,4}$ Andreas W.~W.~Ludwig,$^5$ Simon Trebst,$^2$ and Guo-Yi Zhu$^{1}$\\[0.8em]
    {\normalsize
    $^1$The Hong Kong University of Science and Technology (Guangzhou), Nansha, Guangzhou, 511453, Guangdong, China\\
    $^2$Institute for Theoretical Physics, University of Cologne, Z\"ulpicher Straße 77, 50937 Cologne, Germany\\
    $^3$Department of Theoretical Physics, University of Geneva, 24 quai Ernest-Ansermet, 1211 Gen\`eve, Switzerland\\
    $^4$Geneva Quantum Center, University of Geneva\\
    $^5$Department of Physics, University of California, Santa Barbara, CA 93106, USA\\[0.4em]
    }
\end{center}

\vspace{1em}
\twocolumngrid

\setcounter{figure}{0}
\renewcommand{\thefigure}{S\arabic{figure}}
\renewcommand{\theHfigure}{S\arabic{figure}}
\setcounter{equation}{0}
\renewcommand{\theequation}{S\arabic{equation}}
\renewcommand{\theHequation}{S\arabic{equation}}
\setcounter{table}{0}
\renewcommand{\thetable}{S\arabic{table}}
\renewcommand{\theHtable}{S\arabic{table}}
\setcounter{section}{0}
\renewcommand{\thesection}{S\arabic{section}}
\renewcommand{\theHsection}{S\arabic{section}}
\setcounter{secnumdepth}{2}

\section{From a measurement trajectory to the RSOS tensor network}
\label{sec:rsos_tensor_network}

Here we derive a constrained tensor network directly from the fusion measurement trajectory as in Fig.~1 of the main text. 
We first fix our convention for the $F$-symbols. The recoupling tensor $(F^{abc}_{d})^{e}_{f}$ relates the two fusion orders:
\begin{equation}
\begin{tikzpicture}[
    baseline={(current bounding box.center)},
    x=0.85cm,
    y=0.6cm,
    spine/.style={line width=0.9pt},
    anyon/.style={line width=0.8pt},
    vertex/.style={circle,fill=black,inner sep=1.1pt}
]
    \draw[spine] (0.3,0) -- (2.7,0);
    \foreach \j/\l in {1/b, 2/c}{
        \draw[anyon] (\j,0) -- (\j,1);
        \node[vertex] at (\j,0) {};
        \node[above] at (\j,1) {\scriptsize$\l$};
    }
    \node[below] at (1.5,0) {\scriptsize$e$};
    \node[right] at (2.7,0) {\scriptsize$d$};
    \node[left] at (0.3,0) {\scriptsize$a$};
\end{tikzpicture}
=
\sum_f \left(F^{abc}_{d}\right)^{e}_{f}
\begin{tikzpicture}[
    baseline={(current bounding box.center)},
    x=0.85cm,
    y=0.6cm,
    spine/.style={line width=0.9pt},
    anyon/.style={line width=0.8pt},
    vertex/.style={circle,fill=black,inner sep=1.1pt}
]
    \draw[spine] (0.5,0) -- (2.5,0);
    \draw[anyon] (1.5,0) -- (1.5,0.5);
    \node[vertex] at (1.5,0) {};
    \node[vertex] at (1.5,0.5) {};
    \draw[anyon] (1.5,0.5) -- (1,1);
    \draw[anyon] (1.5,0.5) -- (2,1);
    \node[above] at (1,1) {\scriptsize$b$};
    \node[above] at (2,1) {\scriptsize$c$};
    \node[right] at (1.65,0.25) {\scriptsize$f$};
    \node[left] at (0.5,0) {\scriptsize$a$};
    \node[right] at (2.5,0) {\scriptsize$d$};
\end{tikzpicture}
\ .
\label{eq:f_symbol_definition}
\end{equation}
For Fibonacci anyons we adopt the standard gauge~\cite{Trebst07goldenchain}, in which all $F$-symbols are real: every $F$-symbol with the vacuum $\mathbf{1}$ among $a,b,c$ equals $1$ whenever allowed by the fusion rules, and the only nontrivial recoupling matrix is
\begin{equation}
    F^{\tau\tau\tau}_{\tau}
    =\frac{1}{\phi}
    \begin{pmatrix}
        1 & \sqrt{\phi}\\
        \sqrt{\phi} & -1
    \end{pmatrix},
    \label{eq:fibonacci_F_matrix}
\end{equation}
with both rows and columns ordered as $(\mathbf{1},\tau)$.

Starting from the sequential fusion-tree basis
\begin{equation}
    \ket{\boldsymbol{x}}=\ket{x_0,x_1,\ldots,x_{L-1}},
    \qquad x_j\in\{\mathbf{1},\tau\},
    \label{eq:rsos_fusion_path_basis}
\end{equation}
where adjacent labels satisfy the fusion-multiplicity constraint $N_{x_{j-1}\tau}^{x_j}=1$, i.e., two neighboring spine labels cannot both be $\mathbf{1}$. Thus the $x_j$ are restricted ``heights.'' Equivalently, after resolving the alternating sublattices, these binary labels encode paths on the four-node $A_4$ adjacency graph; this is the standard fusion-tree/RSOS correspondence~\cite{Trebst07goldenchain}.
Now consider a measurement of the two $\tau$ anyons at sites $j$ and $j+1$, which can be mapped to a local projector by an $F$-move
\begin{equation}
    \raisebox{-0.5\height}{\includegraphics[width=\columnwidth]{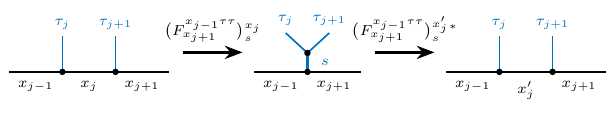}} \ ,
    \label{eq:pi_two_Fmoves}
\end{equation}
where we use blue color to highlight the edges with fixed $\tau$ label. Consequently, the local fusion projector is actually a rank-4 tensor with four indices and has matrix elements
\begin{equation}
    \left[\Pi_j^s\right]_{x'_j x_j}^{x_{j-1},x_{j+1}}
    =
    \left(F_{x_{j+1}}^{x_{j-1}\tau\tau}\right)^{x'_j *}_{s}
    \left(F_{x_{j+1}}^{x_{j-1}\tau\tau}\right)^{x_j}_{s} \ .
    \label{eq:local_fusion_projector_F}
\end{equation}
The outer spine labels $x_{j-1},x_{j+1}$ are spectators throughout: they enter only as conditions of the $F$ matrix. Only five basis states satisfy the adjacency rule in the local constrained space $(x_{j-1},x_j,x_{j+1})$. Hence $\Pi_j^s$ is a $5\times 5$ three-qubit operator acting on $(x_{j-1},x_j,x_{j+1})$, but it changes only the middle label.
We represent it as a building block of the RSOS tensor network, parameterized by the measured channel $s$,
\begin{equation}
    \includegraphics[height=2.2cm]{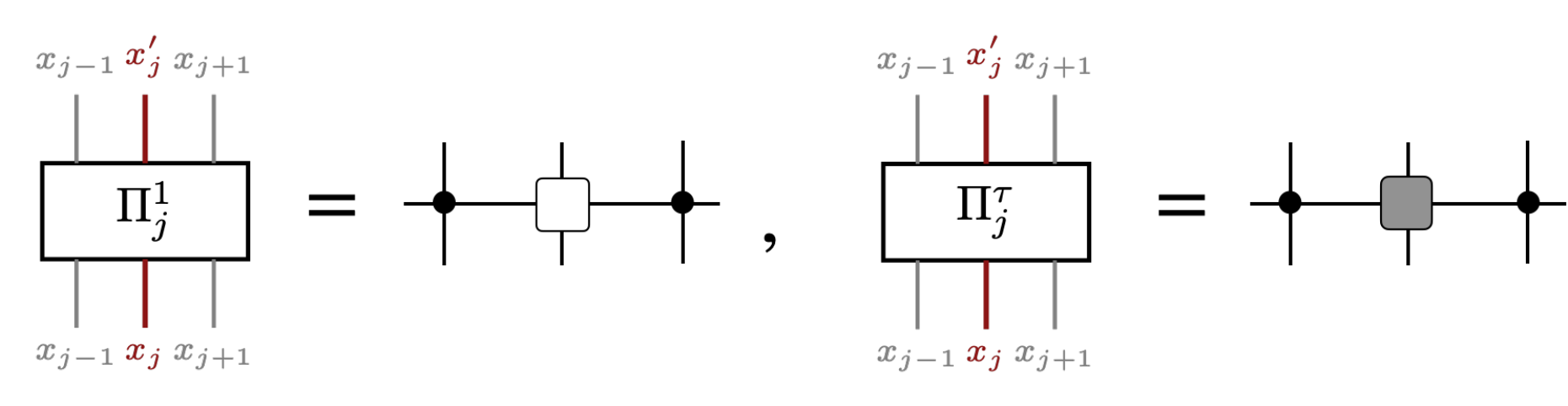} \ .
    \label{eq:pi_3qubit}
\end{equation}
This locality is the origin of a single face tensor.

Then we derive the explicit form of the projector~\cite{Trebst07goldenchain}. For $x_{j-1}=x_{j+1}=\tau$, both $x_j$ and the measured channel $s$ can take the values $\mathbf{1},\tau$, giving
\begin{equation}
    [\Pi_j^{\mathbf{1}}]^{\tau,\tau}
    =
    \begin{pmatrix}
        \phi^{-2} & \phi^{-3/2}\\
        \phi^{-3/2} & \phi^{-1}
    \end{pmatrix}
    = \mathbbm{1}-[\Pi_j^\tau]^{\tau,\tau}\ .
    \label{eq:fibonacci_projector_block}
\end{equation}
The remaining admissible triples are one dimensional
\begin{equation}
    [\Pi_j^{\mathbf{1}}]^{\mathbf{1},\mathbf{1}}=1,
    \qquad
    [\Pi_j^{\mathbf{1}}]^{\mathbf{1},\tau}
    =[\Pi_j^{\mathbf{1}}]^{\tau,\mathbf{1}}=0.
    \label{eq:fibonacci_projector_scalar_blocks}
\end{equation}
We express these matrices in terms of Pauli matrices $X$ and $Z$ acting on the binary space spanned by $x_j\in \{\mathbf{1}, \tau\}$: defining $\sigma^{\mu\nu} = 2[\Pi_j^{\mathbf{1}}]^{\mu\nu} - \mathbbm{1}$,
explicitly $\sigma^{\mathbf{1} \mathbf{1}}=-Z$, $\sigma^{\tau \mathbf1}=\sigma^{\mathbf1 \tau}=Z$, and $\sigma^{\tau \tau}=-\phi^{-3}Z+2\phi^{-3/2}X$, one verifies $(\sigma^{\mu\nu})^2 = \mathbbm{1}$ for all $\mu, \nu \in \{{\bf 1}, \tau \}$. In the scalar blocks of \cref{eq:fibonacci_projector_scalar_blocks} the middle label is pinned to $x_j=\tau$ by the adjacency rule, so only the eigenvalue of $\sigma^{\mu\nu}$ on $\ket{\tau}$ is physical. With the projectors $P^{\mathbf1(\tau)}=(1\pm Z)/2$ onto the neighboring three-spine indices of \cref{eq:fusion_path}, the Kraus operator of the main text can then be rewritten as
\begin{equation}
\begin{aligned}
    M_s &\propto e^{\,\eta_s\lambda\sum_{\mu\nu}  P_{j-1}^\mu \sigma^{\mu\nu}_j P_{j+1}^\nu/2} \\
    &\propto \sum_{\mu\nu}P_{j-1}^\mu\, e^{\eta_s \lambda\sigma^{\mu\nu}_j/2}\, P_{j+1}^\nu \, ,
\end{aligned}
    \label{eq:kraus_pauli_form}
\end{equation}
where $s\in\{\mathbf{1},\tau\}$ labels the anyonic outcome and its numerical encoding is $\eta_{\mathbf{1}}=+1$ and $\eta_\tau=-1$.

For an observed outcome $s$ at site $j$, define the local face weight
\begin{equation}
    W_s\!\left(
    \begin{matrix}
        x_{j-1} & x'_j\\[-2pt]
        x_j & x_{j+1}
    \end{matrix}\right)
    \equiv
    \bra{x_{j-1},x'_j,x_{j+1}}M_{s, j}
    \ket{x_{j-1},x_j,x_{j+1}}.
    \label{eq:trajectory_face_weight}
\end{equation}
And the Kraus operators can be rewritten as
\begin{align}
    M_{\mathbf{1}, j}
    &\propto \mathbbm{1}+(e^\lambda-1)\Pi_j^{\mathbf{1}}
     =\mathbbm{1}+\frac{e^\lambda-1}{\phi}e_j,
    \nonumber\\
    M_{\tau, j}
    &\propto \mathbbm{1}+(e^{-\lambda}-1)\Pi_j^{\mathbf{1}}
     =\mathbbm{1}-\frac{1-e^{-\lambda}}{\phi}e_j \ ,
    \label{eq:kraus_TL_face_operator}
\end{align}
where $e_j$ is the Temperley--Lieb algebra generator. 
The outcome therefore selects one of two local Temperley--Lieb face operators. The four labels in \cref{eq:trajectory_face_weight} are precisely the four corners of the square tensor drawn in \cref{fig:fusion_protocol}(d); a forbidden pair of neighboring heights makes the corresponding weight vanish.

To assemble the spacetime tensor network, insert a resolution of the fusion-tree identity between consecutive brickwork layers. For a record $\mathbf{s}\equiv\{s_{j,t}\}$ and fixed initial and final paths, this gives
\begin{equation}
    \begin{aligned}
        &\bra{\boldsymbol{x}^{(T)}}\mathcal{T}(\mathbf{s})
        \ket{\boldsymbol{x}^{(0)}}\\
        &\quad=
        \sum_{\{\boldsymbol{x}^{(t)}\}}
        \prod_{(j,t)\in\mathrm{brickwork}}
        W_{s_{j,t}}\!\left(
        \begin{matrix}
            x_{j-1}^{(t)} & x_j^{(t+1)}\\[-2pt]
            x_j^{(t)} & x_{j+1}^{(t)}
        \end{matrix}\right).
    \end{aligned}
    \label{eq:trajectory_RSOS_partition_function}
\end{equation}
The sums over intermediate fusion trees contract neighboring face tensors, while the Fibonacci adjacency rule restricts all internal heights. This is the RSOS form of the single-ket trajectory transfer matrix. Closing or fixing the temporal boundary gives the corresponding partition function or boundary-to-boundary amplitude. 

For a uniform (in space and time) post-selected 
measurement record $s_{j,t}=s_0$ with time translation symmetry, the transfer matrix of the entire circuit is a power of a fixed double-layer transfer matrix, 
$\mathcal{T}_{s_0}=\mathcal{T}_{s_0,\mathrm{e}}\mathcal{T}_{s_0,\mathrm{o}}$, where $\mathcal{T}_{s_0,\mathrm{e}}$ is the transfer matrix for the even layer and $\mathcal{T}_{s_0,\mathrm{o}}$ is for the odd layer, and where the measurement outcomes at all sites are fixed to the same value $s_0$.
It is this single-ket operator that admits the integrable RSOS identification. A corresponding statement applies also to the case of an arbitrary (not uniformly postselected) set $\mathbf{s}$ of measurement outcomes, which is the main focus of this paper, where the single clean RSOS partition function is replaced by a random version due to the randomness of the measurement record $\mathbf{s}$.  

Finally, for the uniform record $s_{j,t}=\mathbf{1}$, every face weight in \cref{eq:trajectory_face_weight} is a matrix element of the same local RSOS face operator,
\begin{equation}
    W_j(u)\equiv \mathbbm{1}
    +\frac{\sin u}{\sin(\pi/5-u)}e_j
    \propto M_{\mathbf{1},j},
    \label{eq:uniform_A4_RSOS_face}
\end{equation}
with $(e^\lambda-1)/\phi=\sin u/\sin(\pi/5-u)$; this is the critical $A_4$ RSOS transfer matrix associated with the tricritical-Ising CFT~\cite{Friedan1984,Belavin1984}. Here $u$ is the RSOS spectral parameter. It labels the Yang--Baxter face transfer matrices and the relative weight of the two local connectivities $\mathbbm{1}$ and $e_j$. It should be distinguished from the metric factor $\alpha$ relating circuit time to continuum Euclidean distance. The uniform $s_{j,t}=\tau$ record selects the other RSOS regime through the negative coefficient in \cref{eq:kraus_TL_face_operator}. A generic Born-sampled record spatially and temporally mixes these two admissible face weights. It is therefore a constrained, correlated random RSOS network: the disorder is not independently assigned, because the probability of each $s_{j,t}$ is generated self-consistently by the evolving many-body state.


\section{Topological Symmetry}\label{sec:Y_derivation}

\subsection{Derivation and MPO of topological symmetry $\hat Y$}
\begin{figure*}[htbp]
    \centering
    \includegraphics[width=0.8\textwidth]{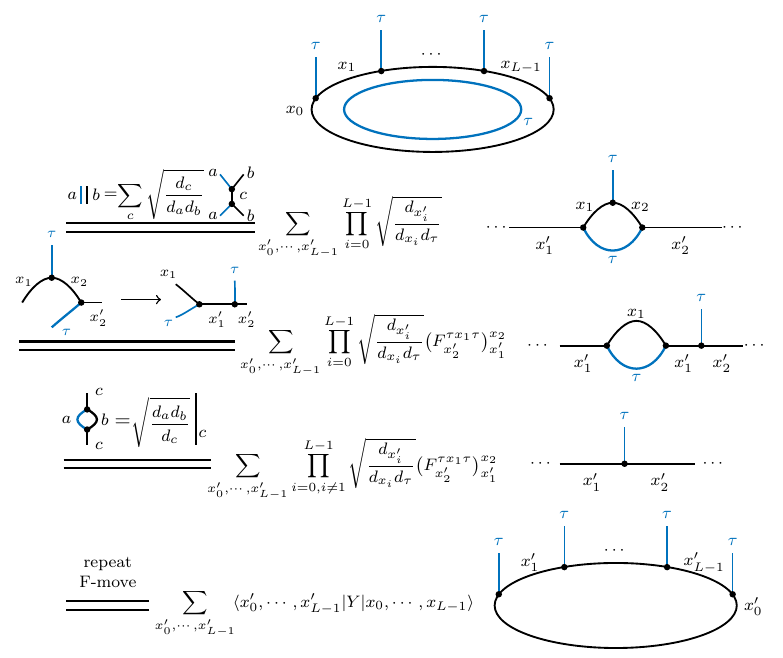}
    \caption{{\bf Deriving the matrix elements of topological symmetry operator $\hat Y$.} The periodic fusion tree describes a chain of Fibonacci anyons nucleated on a topological liquid on a torus. An auxiliary $\tau$ loop is threaded along the noncontractible cycle and absorbed into the spine by sequential $F$-moves, converting $\{x_j\}$ into $\{x'_j\}$. For $x_i=x'_i=\tau$, every $F$-move contributes $(F^{\tau\tau\tau}_{\tau})^{\tau}_{\tau}=-1/\phi$, giving $\mathfrak{y}=(-1/\phi)^L$.}
    \label{fig:Yprocess}
\end{figure*}

The construction of $\hat Y$ is a loop-insertion operation (\cref{fig:Yprocess}).  Starting from a fusion-tree basis state $\ket{\boldsymbol{x}}$, we place an auxiliary $\tau$ line along the noncontractible cycle parallel to the spine.  At site $i$, an $F$-move transfers this line through the trivalent vertex, changes the spine label from $x_i$ to $x'_i$, and contributes one $F$ symbol.  After the auxiliary line has been moved once around the ring, it closes on itself and the remaining graph is the primed fusion tree.  The amplitude therefore factorizes as
\begin{equation}
    \bra{\boldsymbol{x}'}\hat Y\ket{\boldsymbol{x}}
    =
    \prod_{i=0}^{L-1}
    \bigl(F_{x'_{i+1}}^{\tau x_i\tau}\bigr)^{x_{i+1}}_{x'_i},
    \quad
    x_L=x_0,\quad x'_L=x'_0 \,.
    \label{eq:Y_loop_insertion}
\end{equation}

\begin{figure}[htbp]
    \centering
    \includegraphics[width=0.98\columnwidth]{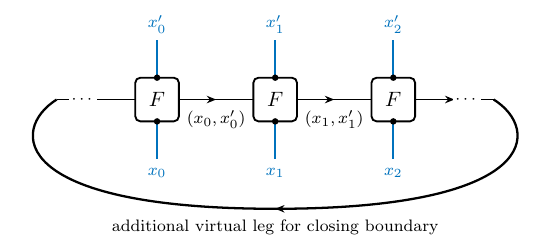}
    \caption{{\bf MPO representation of the $\hat Y$.}
        On the periodic ring, $\hat Y$ is a tensor train of site tensors $F$ [\cref{eq:Y_MPO_tensor}] with physical bra (ket) legs $x'_i$ ($x_i$). The virtual bonds carry the current fusion-channel pair $(x_i,x'_i)$; an additional virtual leg for the closing boundary carries the first-site pair $(x_0,x'_0)$, giving bond dimension independent of $L$: $\chi_Y=16$ ($\chi_Y=4$ for OBC).
    }
    \label{fig:YMPO}
\end{figure}

Note that explicit construction of the full matrix $\hat{Y}$ is not scalable to large system sizes, thus we write it into a matrix product operator (MPO) form with virtual bond dimension 4 for open chain and 16 for periodic chain,
illustrated in \cref{fig:YMPO}, where each local tensor is given by
\begin{equation}
    \hspace*{-0.18cm}\begin{tikzpicture}[
        baseline={(0,-0.12)},
        x=0.88cm,
        y=0.88cm,
        spine/.style={line width=0.9pt},
        anyon/.style={line width=0.8pt},
        vertex/.style={circle,fill=black,inner sep=1.0pt},
        lbl/.style={font=\scriptsize},
        bondarrow/.style={postaction={decorate,decoration={markings,mark=at position 0.55 with {\arrow{Stealth[length=4pt]}}}}}
    ]
        \draw[line width=0.9pt,fill=white] (-0.35,-0.35) rectangle (0.35,0.35);
        \node at (0,0) {$F$};
        \draw[anyon,brightblue] (0,0.35) -- (0,0.8);
        \draw[anyon,brightblue] (0,-0.35) -- (0,-0.8);
        \draw[spine,bondarrow] (-2.0,0) -- (-0.35,0);
        \draw[spine,bondarrow] (0.35,0) -- (2.0,0);
        \node[vertex] at (0,0.35) {};
        \node[vertex] at (0,-0.35) {};
        \node[lbl,brightblue,above=0pt] at (0,0.8) {$x'_i$};
        \node[lbl,brightblue,below=0pt] at (0,-0.8) {$x_i$};
        \node[lbl,below=1pt] at (-1.25,0) {$(x_{i-1},x'_{i-1})$};
        \node[lbl,below=1pt] at (1.25,0) {$(x_i,x'_i)$};
    \end{tikzpicture}
    =
    \begin{pmatrix}
        0 & 0 & 0 & 0\\
        0 & 0 & 1 & 1\\
        0 & \phi^{-1} & 0 & \phi^{-1/2}\\
        0 & \phi^{-1/2} & 1 & -\phi^{-1}
    \end{pmatrix},
    \label{eq:Y_MPO_tensor}
\end{equation}
where the physical legs are determined by the virtual legs and thus the tensor is compactly represented as a 4-by-4 matrix with row index being $(x_{i-1}, x_{i-1}')$ and column index being $(x_i, x_i')$ respectively, 
in the ordered pair basis $(\mathbf{1},\mathbf{1}),(\mathbf{1},\tau),(\tau,\mathbf{1}),(\tau,\tau)$.  

\subsection{Basis state is a superposition of sectors}

The topological symmetry operator $\hat Y$ obeys $\hat Y^2=\mathbbm{1}+\hat Y$, as follows by merging the auxiliary $\tau$ loop into the periodic fusion tree using consecutive $F$-moves in \cref{fig:Yprocess}. Because it measures the total anyon flux through the ring, it commutes with every local fusion channel measurement. The roots of $\lambda^2=\lambda+1$ give eigenvalues $\lambda_{\mathbf{1}}=\phi$ and $\lambda_\tau=-\phi^{-1}$ in global topological charge sectors $y=\mathbf{1}$ and $y=\tau$, respectively. These eigenvalues can also be obtained from the modular $S$ matrix ratios $S_{y\tau}/S_{y\mathbf{1}}$.

The corresponding sector projectors can be written as 
\begin{equation}\label{eq:projectors}
\begin{aligned}
\hat{\Pi}_{\mathbf{1}}
&= \frac{\hat Y - \lambda_{\tau}\mathbbm{1}}{\lambda_{\mathbf{1}} - \lambda_{\tau}}
= \frac{\hat Y + \phi^{-1}}{\phi + \phi^{-1}},\\
\hat{\Pi}_{\tau}
&= \frac{\lambda_{\mathbf{1}}\mathbbm{1} - \hat Y}{\lambda_{\mathbf{1}} - \lambda_{\tau}}
= \frac{\phi - \hat Y}{\phi + \phi^{-1}}.
\end{aligned}
\end{equation}
Note that these projectors also represent the unnormalized density matrices of the maximally dephased mixed states preserving the strong topological symmetry; they should not be confused with the fusion channel projectors $\Pi_j^a$ in \cref{eq:local_fusion_projector_F}.
Then any computational basis state $\bm{x}$ can be decomposed as
\begin{equation*}
\begin{split}
\ket{\bm{x}} = \hat{\Pi}_{\mathbf{1}}\ket{\bm{x}}+\hat{\Pi}_\tau\ket{\bm{x}} = \sqrt{w_{\mathbf{1}}}\ket{\bm{x}_{\mathbf{1}}}+\sqrt{w_\tau}\ket{\bm{x}_{\tau}} \\
\end{split} 
\end{equation*}
where $w_y=\bra{\bm{x}}\hat{\Pi}_y\ket{\bm{x}}$ is the weight of the basis state $\bm{x}$ in flux sector $y$, and $\ket{\bm{x}_y}=\hat{\Pi}_y\ket{\bm{x}}/\sqrt{w_y}$ is its normalized projection onto that sector.
Notice that the $\hat{Y}$-projected states $\ket{\bm{x}_y}$ are linear combinations of the fusion-tree basis states $\ket{\bm{x}}$.

Using \cref{eq:projectors}, we can explicitly compute the weights $w_{\mathbf{1}}$ and $w_\tau$ as
\begin{equation}\label{eq:initial_state_weights}
\begin{split}
w_{\mathbf{1}} & = \frac{1}{\phi + \phi^{-1}}(\bra{\bm{x}}\hat Y\ket{\bm{x}} + \phi^{-1}) \\
w_\tau &= \bra{\bm{x}}\hat{\Pi}_\tau\ket{\bm{x}} = \frac{1}{\phi + \phi^{-1}}(\phi - \bra{\bm{x}}\hat Y\ket{\bm{x}}) \\
\end{split} 
\end{equation}
We only need to compute diagonal matrix elements of $\bra{\bm{x}}\hat Y\ket{\bm{x}} = \prod_{i=0}^{L-1}\bigl(F^{\tau x_i \tau}_{x_{i+1}}\bigr)^{x_{i+1}}_{x_i}$. 
    Here $\bigl(F^{\tau x_i \tau}_{x_{i+1}}\bigr)^{x_{i+1}}_{x_i}$ can be interpreted as the overlap between the following two configurations, taking $i=1$ as an example
\begin{equation}
    \bigl(F^{\tau x_1\tau}_{x_2}\bigr)^{x_2}_{x_1}
    =
    \left\langle
    \begin{tikzpicture}[
        baseline={(current bounding box.center)},
        x=0.72cm,
        y=0.72cm,
        spine/.style={line width=0.9pt},
        anyon/.style={line width=0.8pt},
        loop/.style={line width=1.1pt,brightblue},
        vertex/.style={circle,fill=black,inner sep=1.1pt},
        lbl/.style={font=\scriptsize}
    ]
        \draw[loop] (0,0) -- (0.7,0);
        \node[lbl,below=1pt] at (0.35,0) {$\tau$};
        \draw[spine] (0.7,0) -- (0.7,0.55);
        \node[lbl,above=0pt] at (0.7,0.55) {$x_1$};
        \draw[spine] (0.7,0) -- (1.5,0);
        \node[lbl,below=1pt] at (1.1,0) {$x_2$};
        \draw[anyon,brightblue] (1.5,0) -- (1.5,0.55);
        \node[lbl,brightblue,above=0pt] at (1.5,0.55) {$\tau$};
        \draw[spine] (1.5,0) -- (2.3,0);
        \node[lbl,below=1pt] at (1.9,0) {$x_2$};
        \node[vertex] at (0.7,0) {};
        \node[vertex] at (1.5,0) {};
    \end{tikzpicture}
    \middle|
    \begin{tikzpicture}[
        baseline={(current bounding box.center)},
        x=0.72cm,
        y=0.72cm,
        spine/.style={line width=0.9pt},
        anyon/.style={line width=0.8pt},
        loop/.style={line width=1.1pt,brightblue},
        vertex/.style={circle,fill=black,inner sep=1.1pt},
        lbl/.style={font=\scriptsize}
    ]
        \draw[loop] (0,0) -- (0.7,0);
        \node[lbl,below=1pt] at (0.35,0) {$\tau$};
        \draw[spine] (0.7,0) -- (1.5,0);
        \node[lbl,below=1pt] at (1.1,0) {$x_2$};
        \draw[spine] (0.7,0) -- (0.7,0.3);
        \node[lbl,right=1pt] at (0.7,0.15) {$x_1$};
        \draw[spine] (0.7,0.3) -- (0.3,0.55);
        \node[lbl,above=0pt] at (0.3,0.55) {$x_1$};
        \draw[anyon,brightblue] (0.7,0.3) -- (1.1,0.55);
        \node[lbl,brightblue,above=0pt] at (1.1,0.55) {$\tau$};
        \node[vertex] at (0.7,0) {};
        \node[vertex] at (0.7,0.3) {};
    \end{tikzpicture}
    \right\rangle
    =
    -\frac{1}{\phi}\delta_{x_1, \tau}\delta_{x_{2}, \tau}
    \ .
    \label{eq:Fmove}
\end{equation}
The fusion rule of Fibonacci anyons constrains that $x_1=x_2=\tau$ for nonzero element.
Thus $\bra{\bm{x}}\hat Y\ket{\bm{x}}=\mathfrak{y}\delta_{\bm{x},\tau\tau\tau\cdots}$, with $\mathfrak{y}=(-1/\phi)^L$. Only the all-$\tau$ configuration therefore contributes to $\operatorname{Tr}\hat{Y}=\mathfrak{y}$, as shown graphically in \cref{fig:Yprocess,eq:Fmove}.

\subsection{Topological charge sharpening}
Since every Kraus operator commutes with $\hat Y$, the
monitoring dynamics cannot convert a state in the $y=\mathbf{1}$ flux sector into one in the $y=\tau$ sector, or vice versa.  Nevertheless, the probabilities
of the same local measurement record are generically different in the two sectors.  A trajectory can therefore progressively learn the value of the
conserved global charge.  Topological charge sharpening is precisely this quantum-nondemolition readout: the charge is not changed by the measurements, while our uncertainty about it is reduced by the accumulated record.

To make this information flow explicit, we entangle the two flux sectors of the initial state in \cref{eq:initial_fusion_path} with a reference qubit $R$.
We prepare
\begin{equation}
    \ket{\Psi_0}
    =\sqrt{w_\tau}\ket{0}_R\ket{\psi_{0,\tau}}
    +\sqrt{w_{\mathbf{1}}}\ket{1}_R\ket{\psi_{0,\mathbf{1}}}.
    \label{eq:charge_reference_state}
\end{equation}
For the all-$\tau$ fusion tree used throughout this work,
\begin{equation}
    w_{\mathbf{1}}
    =\frac{1+\phi\mathfrak{y}}{1+\phi^2},\qquad
    w_\tau=1-w_{\mathbf{1}},\qquad
    \mathfrak{y}=(-\phi^{-1})^L,
    \label{eq:charge_prior}
\end{equation}
so that $w_{\mathbf{1}}\to (d_\mathbf{1}/D)^2 = (1+\phi^2)^{-1}=0.276393\ldots$
as $L\to\infty$.  After a measurement record $\mathbf{s}_t$, the reduced reference density matrix is diagonal in this basis,
\begin{align}
    \rho_R(\mathbf{s}_t)
    &=p_\tau(\mathbf{s}_t)\ketbra{0}{0}
      +p_{\mathbf{1}}(\mathbf{s}_t)\ketbra{1}{1}, \\
    p_y(\mathbf{s}_t)
    &=\frac{w_y
    \norm{\mathcal{T}(\mathbf{s}_t)\ket{\psi_{0,y}}}^2}
    {\sum_{y'=\mathbf{1},\tau}w_{y'}
    \norm{\mathcal{T}(\mathbf{s}_t)\ket{\psi_{0,y'}}}^2}.
    \label{eq:charge_posterior}
\end{align}
Thus the monitored quantum trajectory implements an ordinary Bayesian update of a hidden, conserved topological label.  Equivalently, its log-likelihood
ratio is
\begin{equation}
    \ln\frac{p_{\mathbf{1}}}{p_\tau}
    =\ln\frac{w_{\mathbf{1}}}{w_\tau}
    +\ln\frac{\norm{\mathcal{T}(\mathbf{s}_t)\ket{\psi_{0,\mathbf{1}}}}^2}
    {\norm{\mathcal{T}(\mathbf{s}_t)\ket{\psi_{0,\tau}}}^2}.
    \label{eq:charge_log_likelihood}
\end{equation}
Every outcome adds a small amount of evidence to the second term.  Although no single local measurement resolves the nonlocal flux, the spacetime record eventually drives the log-likelihood ratio to either sign of infinity, which means that the system is ``purified'' to a definite topological charge. Such a purification process can also be visualized through final-state Bloch vectors, which give a trajectory-resolved view. The reference qubit then points to one of two antipodal points on the Bloch sphere, which represent purification into the conserved global topological charge sectors $y=\mathbf{1}$ and $y=\tau$, respectively.

We quantify the remaining uncertainty by the reference qubit entropy
\begin{equation}
    S_R(\mathbf{s}_t)
    \equiv S(\rho_R)
    =-p_{\mathbf{1}}\ln p_{\mathbf{1}}-p_\tau\ln p_\tau.
    \label{eq:charge_entropy}
\end{equation}
Initially it is the binary entropy of the sector prior with value $- \sum_{y} w_{y}\ln w_{y} \sim 0.59$ (not $\ln2$).  It vanishes only when a trajectory has selected a definite
flux sector.

A Born trajectory is less strongly biased at each step than a post-selected one, and therefore sharpens more slowly, but both require $O(L)$ measurements to achieve purification. This is physically natural for a nonlocal loop charge: local outcomes acquire sector information only through correlations propagated around the periodic chain.
The reference qubit entropy decays after a time of order $L$; the observed linear scaling $t^\#\propto L$ is compatible with the dynamical exponent $z=1$ and the emergent spacetime symmetry inferred independently from the Lyapunov spectrum and correlation matching.

\subsection{Learning global topological charge}
Charge sharpening has an operational classical counterpart.  An observer who is given only the measurement record can infer the conserved topological sector without directly measuring $\hat Y$.  We quantify this learnability by a binary classification decoder on the measurement record $\mathbf{s}$, as shown in End Matter. 

In \Cref{fig:topological_charge_decoding_dynamics}, we generate $10^4$ independent labeled test quantum trajectories in sector $y^*\in\{\mathbf{1},\tau\}$. For $y^*= \mathbf{1}$, we prepare the golden-chain ground state, which is at the tricritical Ising critical point, and the Potts model ground states for $y^*=\tau$. Note that for $L\in 6\mathbb{Z}_{>0}$, we need to select a fusion-tree basis state and project it to the $y=\tau$ sector, because the Potts model ground states are in the $y=\mathbf{1}$ sector now.
Then we sequentially measure every local outcome according to Born rule. $y^*$ is retained solely to score its prediction $\tilde y$.

During decoding, the decoder constructs $K=100$ orthonormal random trial states $\ket{\psi_{y,k}}$, $k=1,\ldots,K$, in each sector $y\in\{\mathbf{1},\tau\}$. For each fixed $k$, the same quantum trajectory $\mathbf{s}$ is applied to the two trial states $\ket{\psi_{\mathbf{1},k}}$ and $\ket{\psi_{\tau,k}}$, yielding  
the state-resolved decoder free energy
\begin{equation}
    \begin{aligned}
    F_{y,k} = -\ln P(\mathbf{s}|y,k)= -\ln\norm{\mathcal{T}(\mathbf{s})\ket{\psi_{y,k}}}^2,
    \end{aligned} \,
    \label{eq:decoding_free_energy}
\end{equation}
the maximum-likelihood decoder directly assigns
\begin{equation}
    \begin{aligned}
    \tilde y_k(\mathbf{s})
    =\operatorname*{arg\,max}_{y\in\{\mathbf{1},\tau\}} P(\mathbf{s}|y,k)
    \end{aligned} \ ,
    \label{eq:ml_decoder}
\end{equation}
where $\arg\max_y(f(y))$ is the function that returns the argument $y$ value that maximizes the objective function $f(y)$.
The time-dependent average performance of the decoder is
\begin{equation}
    P_{\mathrm{success}}(t)
    =\mathbb E_{y^*} \mathbb E_{\mathbf{s}_{t}|y^*} \mathbb E_{k}
      \delta_{y^*,\tilde y_k(\mathbf{s}_{t})},
    \label{eq:seedwise_decoding_success}
\end{equation}
where $\mathbb E_{y^*}$ gives equal weight to the two prepared charge sectors, $\mathbb E_{\mathbf{s}_t|y^*}$ averages over test Born measurement records in sector $y^*$ at time $t$, and $\mathbb E_k$ uniformly averages over the $K$ trial-state indices, measuring the average performance of decoder.
Here $\mathbf{s}_{t}$ denotes the entire measurement record that the decoder has received until time $t$.
Thus we plot the decoding error probability $1 - P_{\mathrm{success}}(t)$ in \cref{fig:topological_charge_decoding_dynamics}. 
The decoder does not use any information about the initial state or the
probability distribution $w_y$ for the two sectors (the ``prior distribution'' of the two sectors). 
It only receives the record $\mathbf{s}_t$ and uses unbiased trial states $\ket{\psi_{y,k}}$, and must infer the sector from the record alone. The decoder is therefore a classical algorithm that learns the conserved topological charge from the measurement record, without any direct access to the quantum system.


\section{Measurement statistics}
We first consider Born ensembles in which the conserved topological charge has already sharpened to $y=\mathbf{1}$ or $y=\tau$. For linear observables of the density matrix, we can trace out the measurement record, unraveling the trajectories into a mixed state that evolves under the Born-averaged channel
\begin{equation}
    \Phi_\gamma(\rho)=\sum_{\mathbf{s}}\mathcal{T}(\mathbf{s})\rho \mathcal{T}(\mathbf{s})^\dagger.
    \label{eq:born_channel}
\end{equation}
Since every local Kraus operator commutes with $\hat{\Pi}_y$, this channel
cannot transfer weight between the two sectors due to the strong topological symmetry.  Moreover, unitality gives
\begin{equation}
    \Phi_\gamma(\hat{\Pi}_y)
    =\hat{\Pi}_y\sum_{\mathbf{s}}\mathcal{T}(\mathbf{s})\mathcal{T}(\mathbf{s})^\dagger
    =\hat{\Pi}_y.
    \label{eq:sector_fixed_point}
\end{equation}
Assuming equilibration within each symmetry block, the sector-resolved stationary state is thus
\begin{equation}
    \begin{aligned}
        \rho_y^\star&=\frac{\hat{\Pi}_y}{D_y},
        &D_y&=\operatorname{Tr}\hat{\Pi}_y \ .
    \end{aligned}
    \label{eq:stationary_density_matrix}
\end{equation}
Here $D_y=\operatorname{Tr}\hat{\Pi}_y$ are the sector dimensions. Note the Hilbert space dimension of the Fibonacci anyon chain is the number of binary strings of length $L$ on a ring with no adjacent $\mathbf{1}$s, namely the Lucas number
$\mathrm{Tr}\,\mathbbm{1}=\mathcal{L}_L$, which is related to the Fibonacci numbers as $\mathcal{L}_n=F_{n+1}+F_{n-1}$ with $F_0=0$, $F_1=1$, and $F_{n+1}=F_n+F_{n-1}$. The number of paths in which the $L$ anyons fuse globally to $\mathbf{1}$ is $D_{\mathbf{1}}=F_{L-1}$, and correspondingly $D_\tau=F_{L+1}$ for global fusion to $\tau$. These dimensions are also consistent with $\operatorname{Tr}\hat{Y}=(-1/\phi)^L$.

It remains to compare the local fusion probabilities in these two sharpened sectors.  Removing a selected pair that fuses to $\mathbf{1}$ leaves a periodic fusion tree of length $L-2$.  Together with $\operatorname{Tr}\hat{Y}=\mathfrak{y}$, this gives
\begin{align}
    \operatorname{Tr}(\Pi^{\mathbf{1}})&=\mathcal{L}_{L-2},
    &\operatorname{Tr}(\Pi^\tau)&=\mathcal{L}_{L-1},
    \label{eq:trace_pi_one}\\
    \operatorname{Tr}(\Pi^{\mathbf{1}}\hat{Y})&=\mathfrak{y}\phi^2,
    &\operatorname{Tr}(\Pi^\tau \hat{Y})&=-\mathfrak{y}\phi.
    \label{eq:trace_pi_one_y}
\end{align}
Substitution into the sector projectors in \cref{eq:projectors} gives the exact finite-size fusion probabilities in specific sectors,
\begin{align}
    \left\langle\Pi^\tau\right\rangle_{\mathbf{1}}
    &=\frac{\phi^{-1}\mathcal{L}_{L-1}-\phi\mathfrak{y}}
            {(\phi+\phi^{-1})F_{L-1}}
      =\frac{F_{L-2}}{F_{L-1}}
      \xrightarrow{L\to\infty}\frac{1}{\phi},\nonumber\\
    \left\langle\Pi^\tau\right\rangle_\tau
    &=\frac{\phi\mathcal{L}_{L-1}+\phi\mathfrak{y}}
            {(\phi+\phi^{-1})F_{L+1}}
      =\frac{F_L}{F_{L+1}}
      \xrightarrow{L\to\infty}\frac{1}{\phi}.
    \label{eq:finite_size_fusion_probability}
\end{align}
Thus both sharpened sectors also have $\langle\Pi^{\mathbf{1}}\rangle_y\to\phi^{-2}$, with only exponentially small, sector-dependent finite-size corrections, which means that we can not distinguish which sector we are in from measurement outcome density directly. The $\tau$ readout POVM,
\begin{equation}
    M_\tau^\dagger M_\tau
    =\frac{1+\gamma}{2}\Pi^\tau
     +\frac{1-\gamma}{2}\Pi^{\mathbf{1}},
    \label{eq:tau_kraus_povm}
\end{equation}
has the sector-independent thermodynamic probability
\begin{align}
    P_{\tau\times\tau\to\tau}^{(y)}
    &=\frac{1+\gamma}{2}\left\langle\Pi^\tau\right\rangle_y
      +\frac{1-\gamma}{2}\left\langle\Pi^{\mathbf{1}}\right\rangle_y \nonumber\\
    &\longrightarrow
      \frac{1+\gamma}{2\phi}+\frac{1-\gamma}{2\phi^2},
    \qquad y=\mathbf{1},\tau.
    \label{eq:stationary_measurement_outcome_probability}
\end{align}
In particular, the projective limit gives $P_{\tau\times\tau\to\tau}^{(y)}\to1/\phi$, as observed in \cref{fig:outcome_probability}.

A local one-point probability therefore {\sl cannot} distinguish the two sectors in the thermodynamic limit.  This does not preclude charge sharpening: the global charge is encoded in the correlations of the extended spacetime record, rather than in its mean outcome density.

Finally, consider a state in mixed sectors.  After Born averaging, its stationary diagonal part is
\begin{equation}
    \rho^\star=\sum_{y=\mathbf{1},\tau}w_y\rho_y^\star,
    \qquad
    w_y=\operatorname{Tr}(\hat{\Pi}_y\rho_0),
\end{equation}
and linearity gives $P_{\tau\times\tau\to\tau}=\sum_yw_yP_{\tau\times\tau\to\tau}^{(y)}$.  Hence the mixed-sector probability has the same thermodynamic limit, independently of the weights $w_y$; at finite $L$ it retains only the exponentially small weighted corrections in \cref{eq:finite_size_fusion_probability}.  The same conclusion applies to an initial coherent superposition of the two sectors, since a local fusion projector commutes with $\hat{Y}$ and therefore has no matrix elements between distinct $y$ blocks.

\section{Spacetime anisotropy}
\label{sec:anisotropy}
\begin{figure}[htbp]
    \centering
    \includegraphics[width = 0.6\columnwidth]{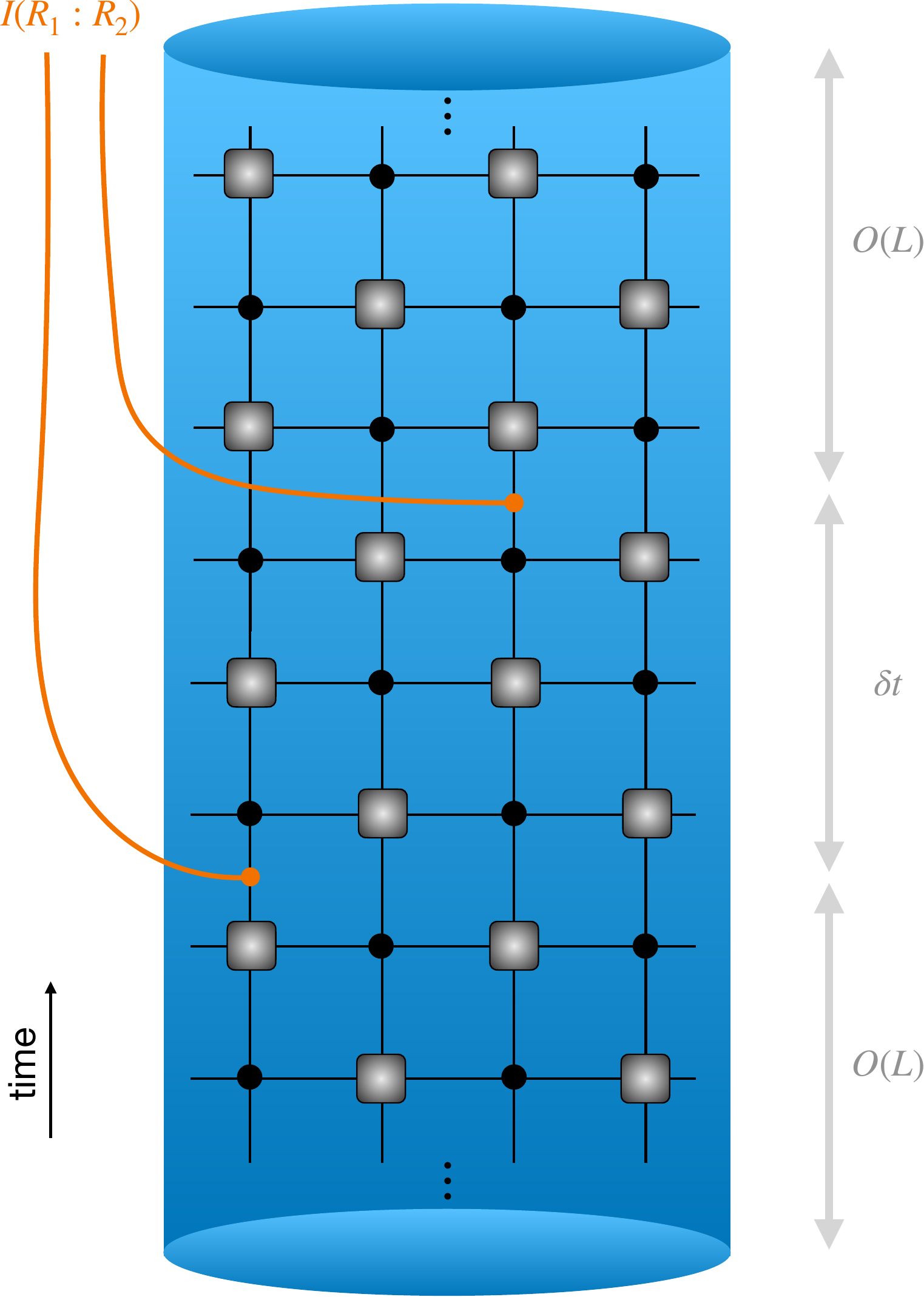}
    \caption{{\bf Monitored circuits probing spatiotemporal correlation} by injecting reference qubits $R_1,R_2$ to the bulk of the spacetime separated by spatial distance $L/2$ {\it or} temporal distance $\delta t$. We keep them away from the initial and final temporal boundary by $O(L)$ temporal distances.
    }
    \label{reference_circuits}
\end{figure}

CFT requires a spacetime-symmetric continuum, but a lattice model with a finite time step is anisotropic, with spatial and temporal scales related by $L=\alpha t$. Only $S_c/TL$ is directly accessible, so the raw Casimir slope of Shannon entropy of the measurement records only measures the composite $\alpha c$. And extracting the true $c$ requires evaluating the sound velocity $\alpha$.
Here we follow Ref.~\cite{Pixley22MIPTCFT,Zabalo_2020,Gullans20scalabledecoder, Vasseur24transfer} to insert two probes at different spacetime locations of the circuit to evaluate $\alpha$ by comparing the spatial and temporal correlations.
CFT predicts a primary field $\varphi$ satisfies $\expval{\varphi(\vb{x}_i)\varphi(\vb{x}_j)}_{\mathbb C}=\abs{\vb{x}_i-\vb{x}_j}^{-2\Delta}$ 
 in the complex plane $\mathbb C$, with scaling dimension $\Delta$. We can map 
the system to a cylinder of circumference $L$ and take the ratio of temporal and spatial correlators $I(\delta x, \delta t)$ to     cancel $\Delta$,
\begin{equation}\label{eq:spatial_temporal_ratio}
    \frac{I(0, \delta t)}{I(\delta x = L/2, 0)} = \left(\frac{2e^{\pi \alpha \delta t/L}}{e^{2\pi \alpha \delta t/L}-1}\right)^{2\Delta} \ ,
\end{equation}
so that tuning 
 $\delta t$ until the ratio equals $1$ gives 
\begin{equation}
    \alpha = \frac{L}{\pi\delta t}\ln(1+\sqrt2) \ .
    \label{eq:alpha}
\end{equation}
Numerically the correlators are read off from the mutual information $I(R_i:R_j)=S_{R_i}+S_{R_j}-S_{R_iR_j}$ of reference qubits inserted at the two spacetime points (\cref{reference_circuits}). The reference qubits are inserted to form a Bell pair with probed qubit $\ket{R \psi} = \frac{1}{\sqrt{2}}(\ket{0 \phi_0 } + \ket{1 \phi_1 })$. Here $\ket{\phi_{0(1)}}$ means that we pick up the computational basis containing $0(1)$ at given site, then we attach an additional $0(1)$ as reference qubit to form Bell pair. 

The isotropic points follow by matching the single-ket gates in \cref{eq:kraus_TL_face_operator} to the critical $A_4$ RSOS face operator in \cref{eq:uniform_A4_RSOS_face}, with $e_j=\phi\Pi_j^{\mathbf{1}}$ and $x(u)=\sin u/\sin(\pi/5-u)$.
The tricritical Ising and $\mathbb Z_3$ parafermion anisotropy angles are $\vartheta=5u$ and $\vartheta=10|u|/3$, respectively~\cite{BianchiniErcolessiPearceRavanini15,KimPearce87anisotropy}. Accounting for the diagonal-to-diagonal embedding of one odd--even brickwork period fixes the isotropic circuit angle to $\vartheta=\arctan(4/3)$. For the uniform $\mathbf{s}=\mathbf{1}$ trajectory, which realizes the tricritical-Ising $A_4$ RSOS model, the anisotropy at arbitrary $0\leq\gamma<1$ is
\begin{equation}
    \begin{aligned}
        x_{\mathbf{1}}(\gamma)
        &=\frac{e^{\lambda}-1}{\phi}
          =\frac{\sqrt{(1+\gamma)/(1-\gamma)}-1}{\phi},\\
        u_{\mathbf{1}}(\gamma)
        &=\arctan\!\left[
            \frac{x_{\mathbf{1}}(\gamma)\sin(\pi/5)}
                 {1+x_{\mathbf{1}}(\gamma)\cos(\pi/5)}
          \right],\\
        \alpha_{\mathbf{1}}(\gamma)
        &=\frac{\tan[5u_{\mathbf{1}}(\gamma)/2]}
                {\tan(\vartheta/2)},
        \qquad \vartheta=\arctan\frac{4}{3}.
    \end{aligned}
    \label{eq:rsos_anisotropy_gamma}
\end{equation}
This expression makes explicit the singular projective limit $\alpha_{\mathbf{1}}(\gamma)\to\infty$ as $\gamma\to1^-$. The ferromagnetic counterpart $\alpha_{\tau}(\gamma)$ follows from the same matching with $x_{\tau}(\gamma)=(e^{-\lambda}-1)/\phi$ and $u_{\tau}(\gamma)=\arctan\!\bigl[x_{\tau}(\gamma)\sin(\pi/5)/(1+x_{\tau}(\gamma)\cos(\pi/5))\bigr]$, but with the $\mathbb{Z}_3$-parafermion anisotropy angle $\vartheta=10|u|/3$, i.e., $\alpha_{\tau}(\gamma)=\tan[10|u_{\tau}(\gamma)|/6]/\tan(\vartheta/2)$. A comparison of $\alpha$ for the two post-selected cases and the Born ensemble is shown in \cref{fig:anisotropy_comparison}. 

The brickwork-isotropic measurement strengths for the two uniform trajectories are obtained by imposing the corresponding isotropic RSOS angles,
\begin{equation*}
    u_{\mathbf{1}}=\frac{\vartheta}5,\qquad
    u_{\tau}=-\frac{3\vartheta}{10},\qquad
    x_s=x(u_s),
\end{equation*}
and matching the coefficients of $e_j$. This gives
\begin{equation}
    \begin{aligned}
        \gamma_{\rm iso}^{\tau\times\tau\to \mathbf{1}}
        &=\frac{(1+\phi x_{\mathbf{1}})^2-1}{(1+\phi x_{\mathbf{1}})^2+1}
        \simeq0.4841738,\\
        \gamma_{\rm iso}^{\tau\times\tau\to\tau}
        &=\frac{1-(1+\phi x_{\tau})^2}{1+(1+\phi x_{\tau})^2}
        \simeq0.6809541.
    \end{aligned}
    \label{eq:postselected_isotropic_points}
\end{equation}
\begin{figure}[htbp]
    \centering
    \includegraphics[width=.49\columnwidth]{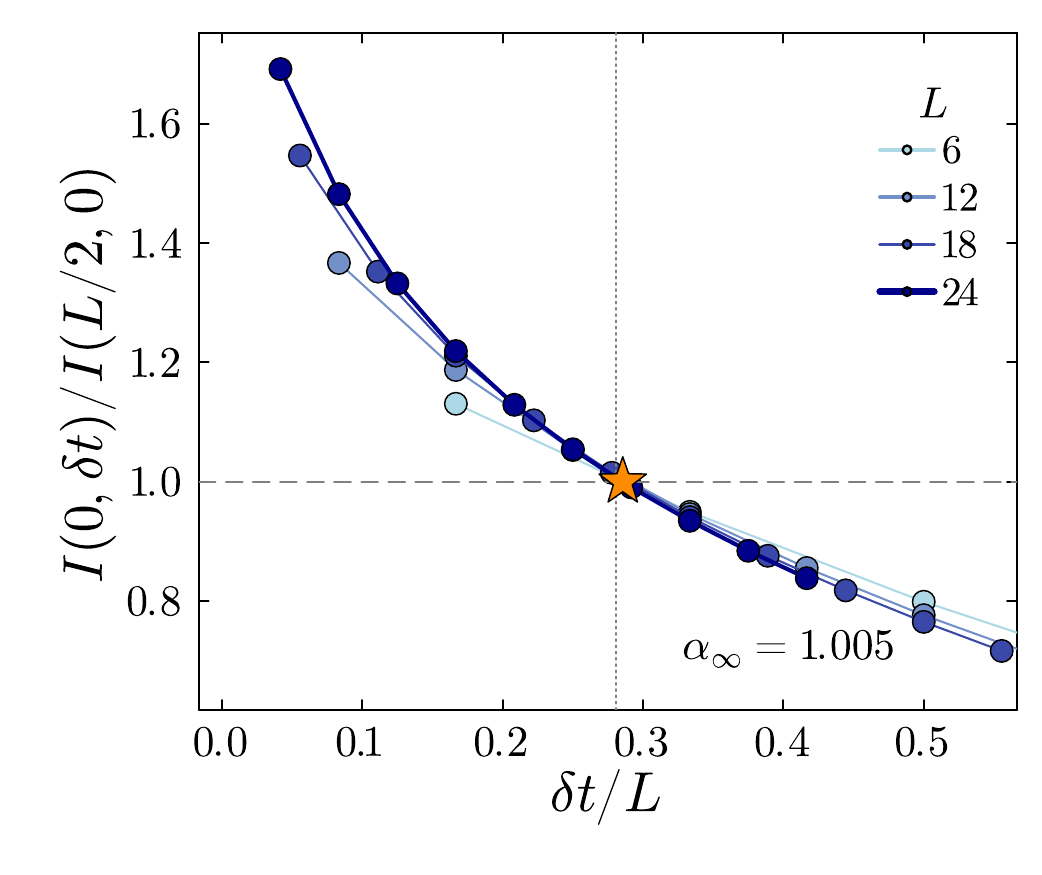}
    \includegraphics[width=.49\columnwidth]{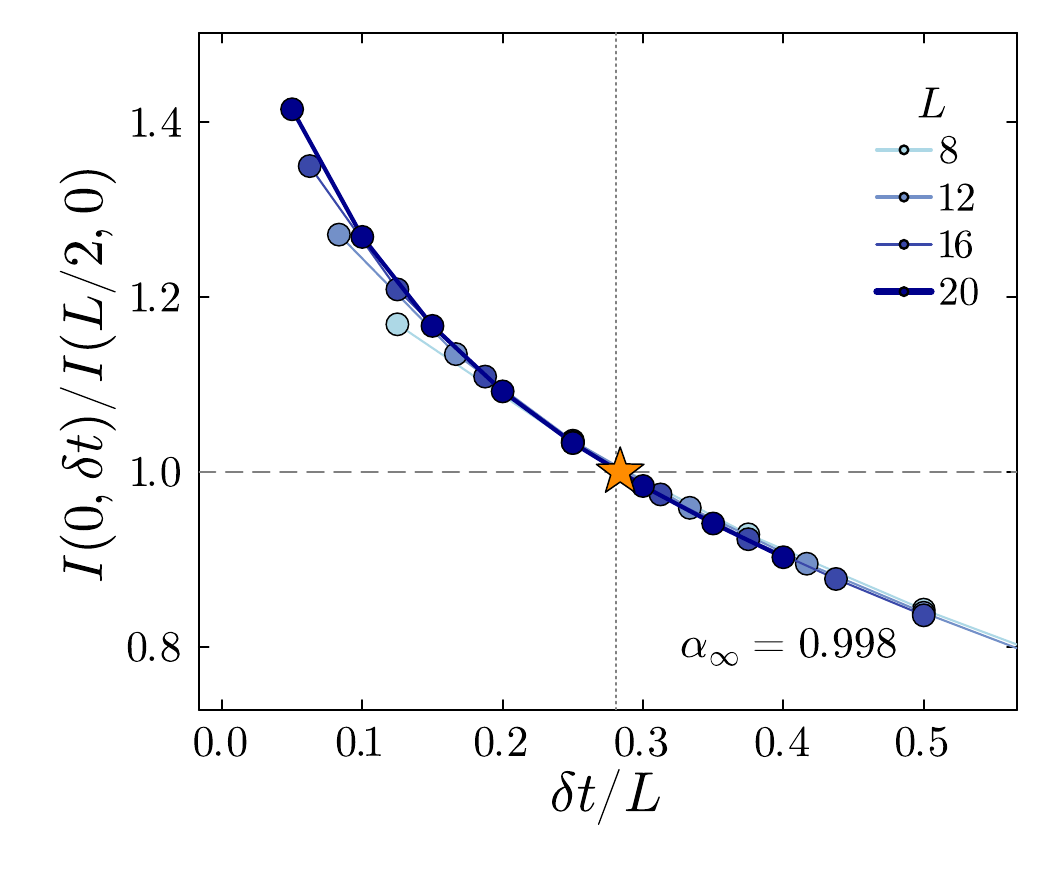}
    \caption{\textbf{Benchmark spacetime anisotropy for post-selected cases}.
    (a) Uniform $\mathbf{s}=\tau$ post-selection at the measurement strength $\gamma\simeq0.6809541$ analytically expected to be isotropic in spacetime, for $L=6,12,18,24$. (b) Uniform $\mathbf{s}=\mathbf{1}$ post-selection at isotropic point $\gamma\simeq0.4841738$, for $L=8,12,16,20$. The orange star marks the numerically interpolated crossing point at $\alpha=1$ for the largest size, which roughly agrees with the analytical value $\delta t^*$ obtained by setting $\alpha=1$ in \cref{eq:alpha} (shown as vertical dashed lines). The extrapolations give $\alpha_\infty=1.005(11)$ and $0.9977(20)$ in (a) and (b), respectively.
    }
    \label{fig:fibo_postselection}
\end{figure}
The reference qubit correlation-function crossings in \cref{fig:fibo_postselection} verify these predictions. Fits over $L\ge12$ give $\alpha_\infty=1.005(11)$ for $\mathbf{s}=\tau$ ($\tau\times\tau\to\tau$) and $\alpha_\infty=0.9977(20)$ for $\mathbf{s}=\mathbf{1}$ ($\tau\times\tau\to\mathbf{1}$). Note that in addition to the universal Casimir term, the next-leading finite-size corrections to the free-energy density are $L^{-18/5}$ and $L^{-4}$, contributed by the leading irrelevant primary fields. Including these corrections gives $\alpha c=0.798724$ for $\mathbf{s}=\tau$ and $0.699914$ for $\mathbf{s}=\mathbf{1}$, and hence $c=0.795(8)$ and $0.7015(14)$, consistent with $4/5$ and $7/10$, respectively. For the $\mathbf{s}=\tau$ parafermion CFT, both the free-energy density and the anisotropy are computed using sizes $L\in6\mathbb Z_{>0}$.

In the Born scenario, the monitored Ising anyon chain shares a common isotropic point with the clean monitored Ising circuit. However, this correspondence does not extend to the Born ensemble of the Fibonacci chain. In the Ising measurement circuit, spacetime rotation maps the string-net configuration associated with each Born trajectory to another trajectory with the same Born weight, and the Abelian anyon loop does not branch under rotation. By contrast, the two Fibonacci fusion outcomes define inequivalent channels with unequal quantum dimension. Their non-Abelian nature allows the string-net loop gas to branch, precluding an analogous trajectory-to-trajectory pairing. The Born ensemble is therefore not constrained to share the lattice metric of either uniformly post-selected ensemble; its spacetime anisotropy must be determined independently and exhibits a distinct emergent isotropic point. At $\gamma=0.95$, the Born-sampled value extrapolates to $\alpha=2.24(8)$ from $L=14$--$28$ (\cref{fig:spacetimecorr}), giving $c_{\rm mix}=2.60(2)/2.24(8)=1.16(4)$.

Motivated by the two exact uniform-trajectory RSOS metrics, we parametrize the Born anisotropy by
\begin{equation}
    \alpha_{\rm B}(\gamma)
    =\frac{A\,\alpha_{\mathbf{1}}(\gamma)\alpha_{\tau}(\gamma)}
    {1+B\,\alpha_{\mathbf{1}}(\gamma)\alpha_{\tau}(\gamma)}.
    \label{eq:born_anisotropy_crossover}
\end{equation}
Here $\alpha_{\mathbf{1}}$ and $\alpha_{\tau}$ are the exact AFM and FM expressions defined above. A weighted fit gives $A=0.3624(25)$ and $B=0.0676(40)$. This phenomenological form reproduces $\alpha_{\rm B}\propto\gamma^2$ in the weak-measurement limit and approaches the finite value $A/B$ in the projective limit. Solving $\alpha_{\rm B}(\gamma)=1$ gives
\begin{equation}
    \gamma_{\rm iso}^{\rm B}=0.8142(20),
\end{equation}
which is numerically compatible with the compact conjectural value
\begin{equation}
    \gamma_{\rm iso}^{\rm B}\stackrel{?}{=}\frac{1}{3}+\ln\phi
    =0.814545\ldots\,.
\end{equation}
We emphasize that, unlike $\alpha_{\mathbf{1}}$ and $\alpha_{\tau}$, neither \cref{eq:born_anisotropy_crossover} nor the proposed compact value is known to be exact; an analytic derivation for the Born ensemble is presently lacking.

\begin{figure}[htbp]
    \centering
    \includegraphics[width=0.8\columnwidth]{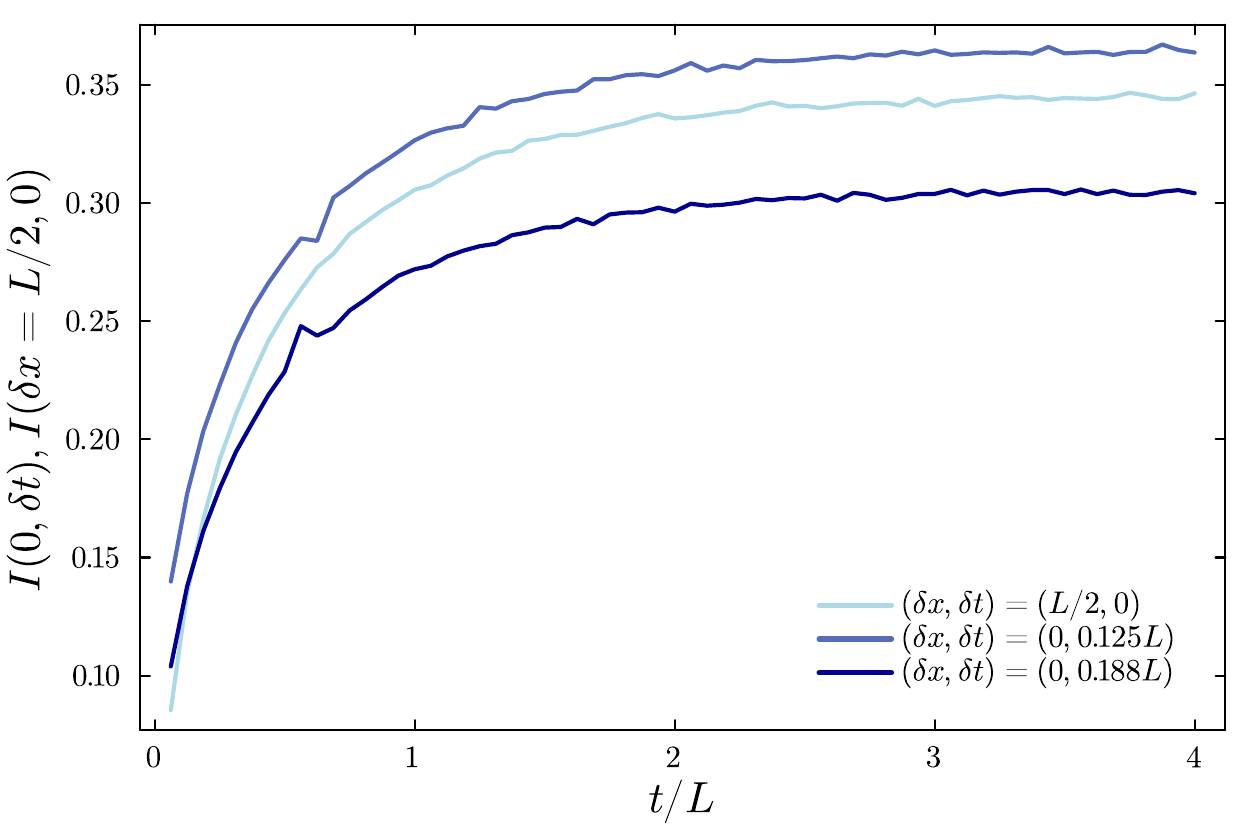}
    \caption{{\bf Dynamics of the mutual information of reference qubits} at $\gamma=0.95$. Temporal and spatial correlation dynamics over evolution time $t$ for various $\delta t$. Here we use $L=16$ and average over 100000 Born trajectories.
    }
    \label{fig:fibo_anisotropy_born}
\end{figure}



\section{Shannon entropy density}
\label{sec:shannon}

Besides the universal Casimir term, the record Shannon entropy contains a nonuniversal volume-law part (cf.\ \cref{eq:casimir}), $S_c/(\alpha T)=\epsilon_0L-\pi c_{\rm eff}/(6L)$. The bulk coefficient $\epsilon_0$ is the Shannon entropy density per spacetime unit area $\alpha Lt$. The combination
\begin{equation*}
    \alpha\epsilon_0 = \lim_{L\to\infty}\frac{S_c}{LT}
\end{equation*}
is the Shannon entropy produced per bit or per measurement gate ($L$ gates per period). It quantifies the uncertainty in the record per gate and is bounded above by $\ln2$ for a binary readout.

\begin{figure}[htb!]
    \centering
    \includegraphics[width=0.85\columnwidth]{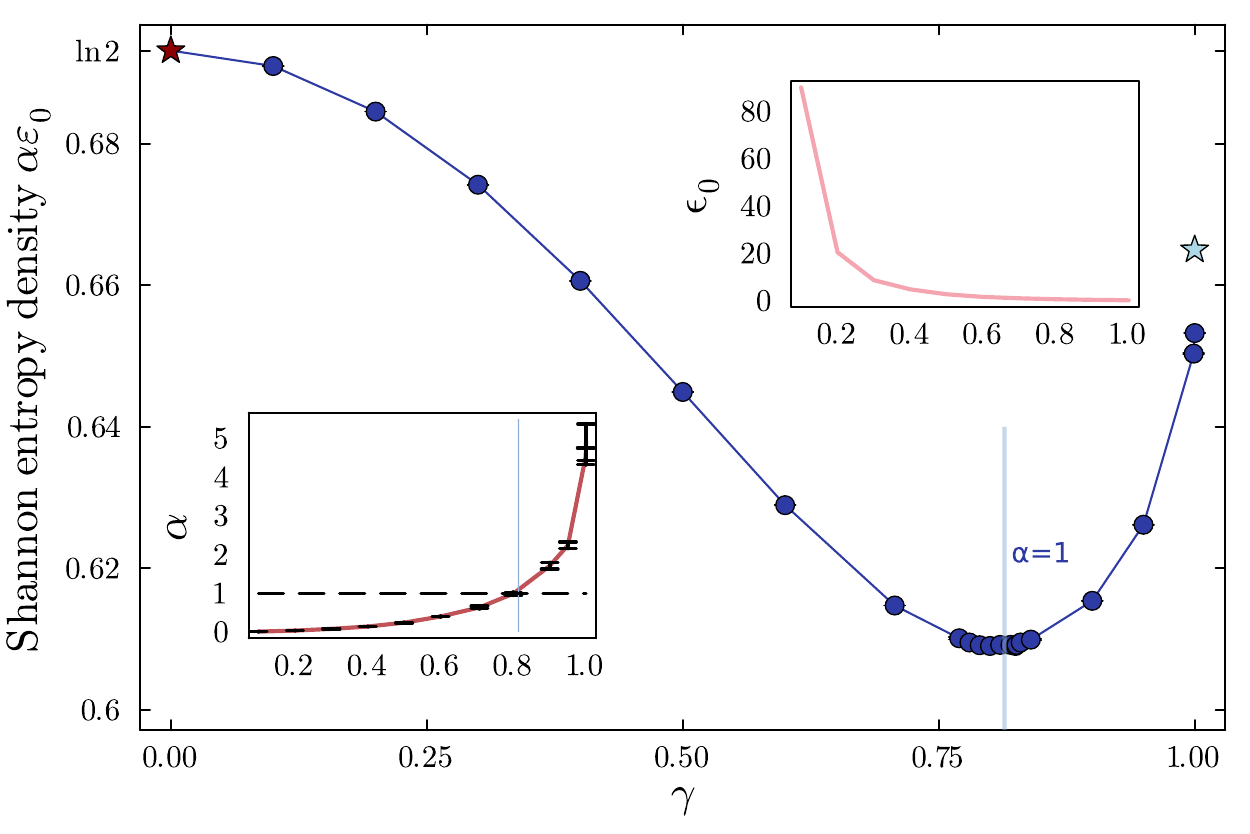}
    \caption{\textbf{Shannon entropy density and sound velocity} versus measurement strength $\gamma$. The bulk entropy density per bit, $\alpha\epsilon_0$, is obtained from the large-$L$ limit of $S_c/(LT)$. Red and blue stars mark values $(\gamma,\alpha\epsilon_0)=(0,\ln2)$ and $(1,(\phi+2)\ln\phi/\phi^2)$, respectively.
    The blue ribbon indicates the $\alpha=1$ region near $\gamma\approx0.814$, determined from the fit of $\alpha(\gamma)$ (lower-left inset), where $\alpha\epsilon_0$ is also minimal. Insets show $\epsilon_0$ and $\alpha$ separately.
    }
    \label{fig:bulkcoef}
\end{figure}

Two limiting values provide useful benchmarks. At $\gamma=0$ ($\lambda=0$), the Kraus operators are proportional to the identity, $M_{\mathbf{1}}=M_\tau=\mathbbm 1/\sqrt2$. The outcomes are state-independent fair coin flips, so $\alpha\epsilon_0=\ln2$ (red star in \cref{fig:bulkcoef}). At $\gamma=1$, treating the outcomes as independent and identically distributed with $P(\tau\times\tau\to\tau)=1/\phi$ and $P(\tau\times\tau\to\mathbf{1})=1/\phi^2$ gives
\begin{equation*}
    \alpha\epsilon_0 = -\frac{1}{\phi}\ln\frac{1}{\phi}-\frac{1}{\phi^2}\ln\frac{1}{\phi^2} = \frac{(\phi+2)\ln\phi}{\phi^2}
\end{equation*}
(blue star in \cref{fig:bulkcoef}), which is ruled out by our numerical results, corroborating the spacetime correlation of the measurement record. The blue ribbon shows the $\alpha=1$ region determined from the Born anisotropy fit of \cref{sec:anisotropy}, where $\alpha\epsilon_0$ approximately attains its minimum. The insets show the bare $\epsilon_0$ and $\alpha$ that enter $\alpha\epsilon_0$.

\section{Casimir fits across measurement strengths}
\label{sec:casimir_fit_windows}

\begin{figure*}[p]
    \centering
    \includegraphics[height=0.87\textheight]{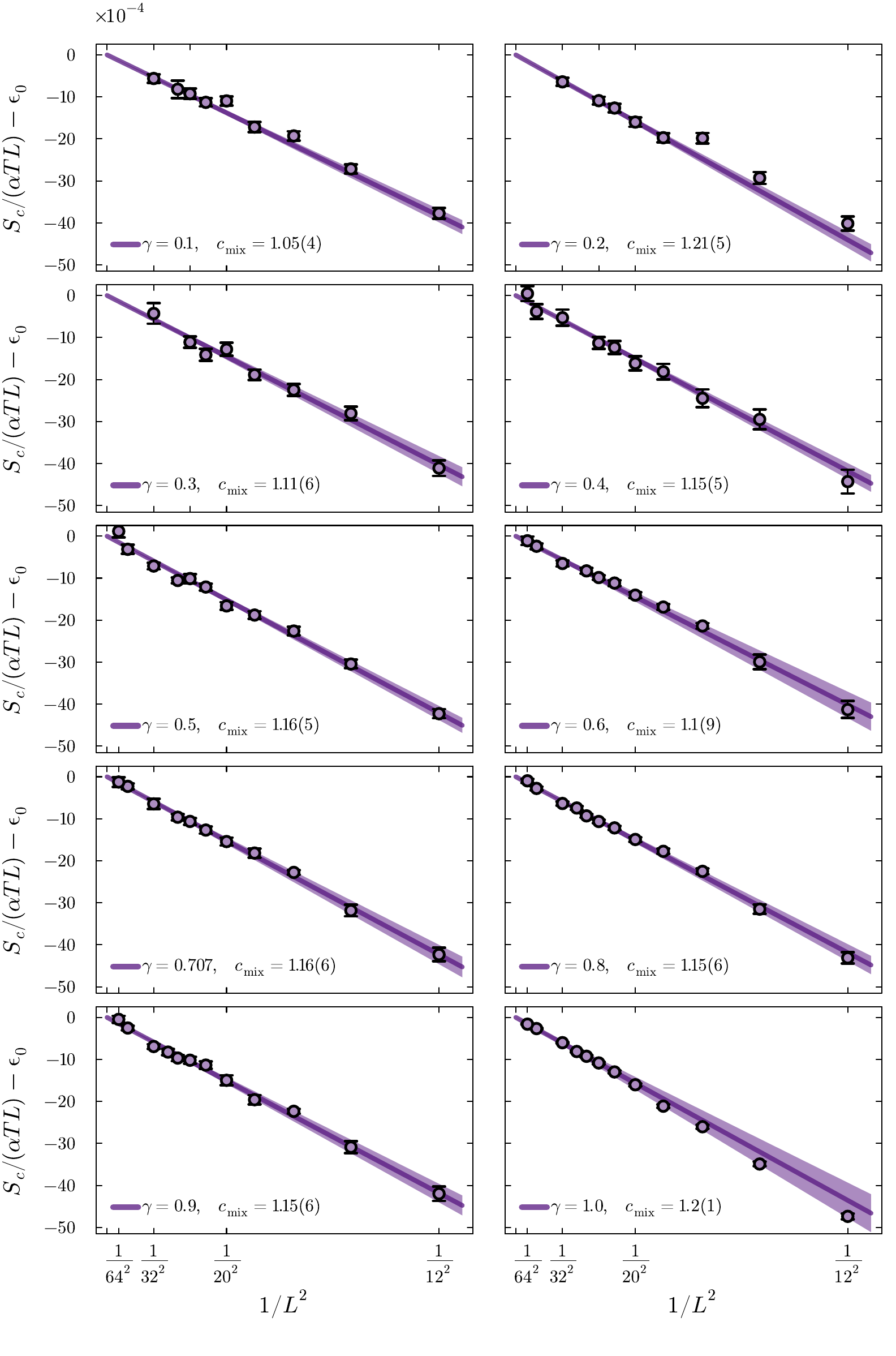}
    \caption{\textbf{Casimir fits across measurement strengths} for the mixed sector. 
    From top to bottom and left to right, the panels show
    $S_c/(\alpha TL)-\epsilon_0$ versus $1/L^2$. The measurement strengths
    and fitted sizes are $\gamma=0.1$ ($L=18, 22, 24, 26, 32$);
    $\gamma=0.2$ ($L=18$--$24,32$); $\gamma=0.3$ ($L=12$--$24,32$);
    $\gamma=0.4$ ($L=14$--$24,32,48,64$);
    $\gamma=0.5$ ($L=12$--$26,32,48,64$);
    $\gamma=0.6$ ($L=24$--$26,32,48,64$);
    $\gamma=0.707$ ($L=12$--$26,32,48,64$);
    $\gamma=0.8$ ($L=20$--$28,32,48,64$);
    $\gamma=0.9$ ($L=12$--$28,32,48,64$); and
    $\gamma=1$ ($L=22$--$28,32,48,64$).
    For measurement strengths with appreciable finite-size corrections, the
    smallest system sizes are discarded from the fit. Purple lines show the
    best fits, and the ribbons indicate their uncertainties.
    }
    \label{fig:casimir_fits_all_gamma}
\end{figure*}

To expose the finite-size windows underlying the critical scaling of the Shannon entropy in the mixed sector (Fig.~7 in the End Matter),
\cref{fig:casimir_fits_all_gamma} shows the subtracted Shannon entropy density
for every measurement strength used in the central-charge extraction, apart from the $\gamma=0.95$ example already shown in Fig.~2(b) in the main text. 
The fit uses the size windows listed in the caption of
\cref{fig:casimir_fits_all_gamma}. The data at $\gamma=0.1,0.2,0.3$ extend to $L=32$, whereas the remaining panels extend to $L=64$. The smaller accessible sizes in the weak-measurement regime reflect its substantially higher computational cost.

The weak-measurement regime is numerically demanding for two related reasons.
First, as $\gamma\to0$, each measurement layer produces only a weak update of the conditioned state, so equilibration requires increasingly long cylinders. For example, the aspect ratio used here decreases from $T/L=700$ at $\gamma=0.1$ to $T/L=150$, $40$, and $2$ at $\gamma=0.2$, $0.3$, and $0.95$, respectively [\cref{tab:aspect_ratio}]. The cost of obtaining an independent steady-state trajectory therefore grows rapidly toward small $\gamma$, limiting the accessible system sizes.

Second, the strong spacetime anisotropy requires very long temporal separations to match the spatial correlator; for example, the $\gamma=0.1$ calculation requires $\delta t\sim10^3$. The universal finite-size correction 
is also parametrically small because the raw Casimir slope measures $\alpha c$. Converting this slope to $c_{\rm eff}$ divides both the signal and its statistical uncertainty by $\alpha$. At $\gamma=0.1$ we find $\alpha=0.00767(21)$; taking $c_{\rm mix}\simeq1.2$ and $L=32$, the unrescaled universal contribution is only
$\pi\alpha c_{\rm mix}/(6L^2)\simeq4.7\times10^{-6}$. Resolving such a small correction against the nonuniversal bulk term requires substantially more trajectories than at moderate $\gamma$.

The opposite, near-projective regime is difficult for a complementary reason. Since $\delta t^*/L=\ln(1+\sqrt2)/(\pi\alpha)$, a large $\alpha$ pushes the correlation crossing below the discrete temporal resolution unless $L$ is correspondingly increased. The choice $\gamma=0.95$ in Fig.~2(b)(c) 
is therefore a practical compromise: the Casimir coefficient $\alpha c$ is sufficiently large to resolve accurately, while the correlation crossing remains accessible for the available sizes.

Compared to Ising monitored dynamics, the Fibonacci anyon model is more numerically demanding because of its larger local Hilbert space and the nontrivial topological constraints. The accessible system sizes are therefore smaller, and the finite-size windows for the Casimir fits are correspondingly narrower. Nevertheless, the extracted central charges are consistent across the measurement strengths, as shown in Fig.~7 in the End Matter. 

\clearpage
\onecolumngrid
\section{Computational resources}
\label{sec:resources}
Our simulations run on the AMD EPYC 7763 and Intel Xeon Platinum 8358P clusters at HKUST(GZ). \Cref{tab:computational_resources} lists the sampling parameters for each method and task, and \cref{tab:aspect_ratio} the circuit aspect ratio (temporal periods / spatial size) across $\gamma$.

\begin{table}[!ht]
\centering
\begin{tabular}{c|cc|ccc}
\toprule
method & \multicolumn{2}{c|}{MPS evolution} & \multicolumn{3}{c}{exact state evolution} \\
\midrule
Task & bulk monitored & Lyapunov spectrum & bulk monitored & Lyapunov spectrum & spacetime correlation \\
Observable & $S_c/T$, $S_{\rm vN}$ &$\lambda_n$ & $S_c/T$, $S_{\rm vN}$ &$\lambda_n$ & $I(R_i:R_j)$ \\
\midrule
largest system size $L$ & 128 & 36 & 28 & 28 & 28 \\
\# of data points $(\gamma, L)$ & 30 & 10 & 148 & 23 & $\sim$1300 \\
\# of samples / data point & $\sim$10,000 & 5,000 & $\sim$50,000 & 20,000 & 10,000 \\
total core hours & 3.5 {\rm million} & 0.4 {\rm million} & 1.2 {\rm million} & 0.2 {\rm million} & 1.5 {\rm million} \\
total file size for $\{\mathbf{s}\}$ & 200 GB & 5 GB & 1 TB & 60 GB & 1.5 TB \\
\bottomrule
\end{tabular}
\caption{\textbf{Computational resources.} For each method and task: the largest system size, number of data points (measurement strengths $\gamma$ across sizes $L$), samples per point, total core hours, and data storage. MPS evolution reaches the Born ensemble up to $L=128$ ($\chi\le400$), from which $c_{\rm eff}$ and $c_{\rm ent}$ are extracted (Fig.~2, Fig.~3) for 3 sectors; exact evolution ($L\le28$) covers the bulk dynamics and the reference qubit circuits used to determine $\alpha$ via $I(R_i:R_j)$ (\cref{reference_circuits,fig:fibo_anisotropy_born}).}
\label{tab:computational_resources}
\vspace{1.5em}

\centering
\begin{tabular}{c|ccccccccccc}
\toprule
$\gamma$ & 0.1 & 0.2 & 0.3 & 0.4 & 0.5 & 0.6 & 0.707 & 0.8 & 0.9 & 0.95 & 1.0 \\
\midrule
MPS ($L\ge 32$) & 700 & 150 & 40 & 28 & 20 & 12 & 10 & 7 & 3 & 2 & 2 \\
Exact ($L\le 28$) & 1250 & 250 & 100 & 50 & 40 & 22.5 & 17.5 & 12.5 & 4 & 4 & 2 \\
\bottomrule
\end{tabular}
\caption{{\bf Aspect ratio $T/L$} (temporal periods / spatial size) for the monitored dynamics at different measurement strengths $\gamma$, set by the equilibration time (shorter for stronger measurement).}
\label{tab:aspect_ratio}
\end{table}
\twocolumngrid

\onecolumngrid

\end{document}